\documentclass[5p,twocolumn,preprint,times]{elsarticle}

\usepackage{silence}
\usepackage{natbib}
\usepackage{graphicx}
\usepackage{tikz}
\usepackage{tabularx}
\usepackage[english]{babel}
\usetikzlibrary{shapes,arrows}
\usepackage{array}
\usetikzlibrary{shapes.geometric}
\newcolumntype{L}[1]{>{\raggedright\let\newline\\\arraybackslash\hspace{0pt}}m{#1}}
\newcolumntype{C}[1]{>{\centering\let\newline\\\arraybackslash\hspace{0pt}}m{#1}}
\newcolumntype{R}[1]{>{\raggedleft\let\newline\\\arraybackslash\hspace{0pt}}m{#1}}
\usepackage{enumerate}
\usepackage{dcolumn}
\usepackage{xcolor}
\usepackage{bigints}
\usepackage{upgreek}

\usepackage{comment}

\usepackage[colorlinks=true,urlcolor=blue]{hyperref} 

\usepackage{xspace}
\usepackage{booktabs}
\usepackage{multirow}
\usepackage{adjustbox}
\usepackage{relsize}
\usepackage{colortbl}
\usepackage{steinmetz}
\usepackage{scalerel}
\usepackage{float}
\usepackage{caption}

\usepackage[switch]{lineno}

\usepackage[
per-mode=symbol,
range-phrase=\text{--},
range-units=single, 
product-units = power,
]{siunitx}
\DeclareSIUnit{\rpm}{RPM}
\DeclareSIUnit{\inch}{in}
\DeclareSIUnit{\feet}{ft}
\DeclareSIUnit{\pixel}{px}
\DeclareSIUnit{\rad}{rad}

\journal{Aerospace Science and Technology}

\usepackage{nomencl}
\usepackage{multicol}
\usepackage{etoolbox}
\usepackage{framed}

\makenomenclature

\renewcommand*\nompreamble{\begin{multicols}{2}}
	\renewcommand*\nompostamble{\end{multicols}}

\renewcommand\nomgroup[1]{%
	\item[\bfseries
		\ifstrequal{#1}{G}{Greek symbols}{%
			\ifstrequal{#1}{D}{Dimensionless numbers}{%
				\ifstrequal{#1}{A}{Abbreviations}{%
					\ifstrequal{#1}{R}{Roman symbols}{%
                        \ifstrequal{#1}{L}{List of operators}{%
    						  \ifstrequal{#1}{S}{Subscripts and superscripts}{}}}}}}%
	]%
}

\usepackage{dblfloatfix}

\begin{document}
	
	\begin{frontmatter}
		
		\title{On the aerodynamics of an accelerating and decelerating propeller in hover at low Reynolds numbers}
		
		\author[1,2]{\texorpdfstring{M. Ratz\corref{cor1}}{M. Ratz}}
		\ead{manuel.ratz@vki.ac.be}
		\author[1]{C. Schram}
		\author[3]{A. Sciacchitano}
		\author[1,2,4]{M. A. Mendez}
		\cortext[cor1]{Corresponding author}
		\address[1]{Environmental and Applied Fluid Dynamics, von Karman Institute for Fluid Dynamics, Waterloosesteenweg 72, Sint-Genesius-Rode, Belgium}
		\address[2]{Aero-Thermo-Mechanics Laboratory, École Polytechnique de Bruxelles, Université Libre de Bruxelles, Av. Franklin Roosevelt 50, Brussels, 1050, Belgium}
		\address[3]{Faculty of Aerospace Engineering, Delft University of Technology, Kluyverweg 1, 2629 HS Delft, the Netherlands}
		\address[4]{Experimental Aerodynamics and Propulsion Lab, Universidad Carlos III de Madrid, Av. de la Universidad 30, 28911 Leganés, Spain}
		
		\date{\\textsf{today}}
		
		\begin{abstract}
			Small unmanned aerial vehicles commonly regulate propeller thrust through rapid changes in rotational speed, yet it remains unclear how the inflow, the wake, and the near-surface velocity respond to these changes at low Reynolds numbers. A commercial two-bladed propeller was tested in hover with custom blade holders at three additional pitch offsets relative to the original blade geometry, $\Theta = 0$, $5$, and \SI{10}{\degree}, and accelerated and decelerated between 3000 and \SI{5000}{\rpm} with open-loop step responses. The resulting nondimensional acceleration and deceleration numbers, below 0.011 and 0.007, suggest negligible acceleration effects. Phase-locked particle image velocimetry (PIV) at 40, 60, and \SI{80}{\percent} rotor radius and in the meridian plane was combined with load measurements. The thrust coefficient responds quasi-steadily at $\Theta = 0^\circ$, but shows hysteresis at the higher pitch offsets, with lower values during acceleration and higher values during deceleration than under steady conditions. This trend opposes expectations from dynamic inflow theory. The effect is largest at $\Theta = 5^\circ$, with deviations from the steady values of up to \SI{8}{\percent} during acceleration and \SI{5}{\percent} during deceleration, compared with \SI{4}{\percent} and \SI{3}{\percent} at $\Theta = 10^\circ$. PIV reveals that the rotor-disk inflow, blade deformation, and sectional circulation remain largely quasi-steady, whereas velocity fluctuations, blade-wake vorticity, and tip-vortex circulation vary strongly, with the largest deviations at low rotational speeds rather than at peak acceleration or deceleration.
		\end{abstract}
		
		
	\end{frontmatter}

	\begin{table*}[htb!]  
		\begin{framed}
			\nomenclature[Ra0]{$a_\textrm{blade}$}{local blade acceleration, $\frac{\textrm{d}}{\textrm{d}t} u_\textrm{blade}$, ${\unit{\meter\per\second^2}}$}
			\nomenclature[RC0]{$c$}{propeller chord length, \unit{\meter}}
			\nomenclature[RC2]{$c_x$}{propeller chord length along the $x-$axis, \unit{\meter}}
			\nomenclature[RC1]{$c_s$}{propeller chord length along the $\xi-$axis, \unit{\meter}}
			\nomenclature[RC4]{$C_l$}{lift coefficient}
			\nomenclature[RC7]{$C_T$}{thrust coefficient, $T / (\rho \pi R^2 (\Omega R)^2)$}
			\nomenclature[RC7]{$C_p$}{pressure coefficient, $(p - p_\infty) / (0.5 \rho_\infty u_\textrm{blade}^2)$}
			\nomenclature[RD0]{$D$}{propeller diameter, $\unit{\meter}$}
			\nomenclature[Rn1]{$n$}{rotational speed, $\unit{\rpm}$}
			\nomenclature[RM]{$M$}{rotor revolution count}
			\nomenclature[RR0]{$r$}{radial distance from the axis of rotation,  ${\unit{\meter}}$}
			\nomenclature[RR2]{$R$}{propeller radius,  ${\unit{\meter}}$}
			\nomenclature[RRe]{$\textrm{Re}_c$}{chord-based Reynolds number, $c\,u_\textrm{blade} / \nu$}
			\nomenclature[Rt0]{$t$}{time coordinate,  ${\unit{\second}}$}
			\nomenclature[RT]{$T$}{propeller thrust, \unit{\newton}}
			\nomenclature[Ru00]{$(u, v)$}{velocities in Cartesian coordinates, \si{\meter\per\second}}
			\nomenclature[Ru1]{$u_\textrm{blade}$}{local blade velocity, $\Omega r$,  ${\unit{\meter\per\second}}$}
			\nomenclature[Ru2]{$u_\textrm{tip}$}{blade tip velocity, $\Omega R$,  ${\unit{\meter\per\second}}$}
			\nomenclature[Ru5]{$U$}{velocity magnitude, $\sqrt{u^2 + v^2}$,  ${\unit{\meter\per\second}}$}
			\nomenclature[RX1]{$(x, y)$}{spanwise Cartesian coordinate system, ${\unit{\meter}}$}
			\nomenclature[RX2]{$(X, Y)$}{slipstream Cartesian coordinate system, ${\unit{\meter}}$}
			
			\nomenclature[G$\alpha0$]{$\alpha$}{angle of attack, \unit{\degree}}
			\nomenclature[G$\alpha1$]{$\alpha_0$}{zero-lift angle of attack, \unit{\degree}}
			\nomenclature[G$\Theta$0]{$\Theta_0$}{geometric propeller twist angle, \unit{\degree}}
			\nomenclature[G$\Theta$1]{$\Theta$}{additional propeller pitch offset, \unit{\degree}}
			\nomenclature[G$\theta$2]{$\theta$}{propeller twist angle due to loading, \unit{\degree}}
			\nomenclature[G$\delta_y$]{$\delta_y$}{sectional vertical bending, \unit{\meter}}
			\nomenclature[G$\nu$]{$\nu$}{kinematic viscosity of air, \unit{\meter^2\per\second}}
			\nomenclature[G$\lambda_0$]{$\lambda$}{inflow into the rotor disk}
			\nomenclature[G$\lambda_1$]{$\lambda(r)$}{local inflow into the rotor disk, $v_i(r)/ (\Omega r)$}
			\nomenclature[G$\gamma$1]{$\Gamma_v$}{tip-vortex circulation, \unit{\meter^2\per\second}}
			\nomenclature[G$\gamma$0]{$\Gamma_s$}{sectional airfoil circulation, \unit{\meter^2\per\second}}
			\nomenclature[G$\rho$0]{$\rho$}{density of air, \unit{\kilogram\per\meter^3}}
			\nomenclature[G$\Psi$]{$\Psi$}{propeller phase angle, \unit{\degree}}
			\nomenclature[G$\xi$]{$(\xi, \eta)$}{spanwise surface-normal coordinate system, \unit{\meter}}
			\nomenclature[G$\tau$]{$\tau$}{nondimensional time}
			\nomenclature[G$\zeta$]{$\zeta$}{tip vortex age, \unit{\degree}}
			\nomenclature[G$\Omega$]{$\Omega$}{propeller angular speed, $2\pi n/60$, \unit{\rad\per\second}}
			\nomenclature[G$\omega$]{$\omega_z$}{out-of-plane vorticity, \unit{\per\second}}
			
			
			
			\nomenclature[L$a$]{$\langle q \rangle$}{ensemble average operator}
            \nomenclature[L$a$]{$ q^\prime$}{fluctuation component, $q - \langle q \rangle$}
			
			\nomenclature[ABEMT]{BEMT}{blade element momentum theory}
			\nomenclature[AFOV]{FOV}{field of view}
			\nomenclature[APIV]{PIV}{particle image velocimetry}
			\nomenclature[AUAV]{UAV}{unmanned aerial vehicle}
			\nomenclature[AURANS]{URANS}{unsteady Reynolds-averaged Navier--Stokes}
			\nomenclature[ALSB]{LSB}{laminar separation bubble}
			\nomenclature[AESC]{ESC}{electronic speed controller}
			\printnomenclature
		\end{framed}
	\end{table*}
	
	\section{Introduction}
	\label{sec:1}
	
	Interest in multi-rotor unmanned aerial vehicles (UAVs) operating at low Reynolds numbers ($\mathcal{O}(10^4$--$10^5)$) has increased steadily in the last decade due to their use in many fields \cite{Hassanalian2017, Ayamga2021}, including pesticide control \cite{Faical2014}, agriculture \cite{Shouji2021}, earth observation \cite{Lucieer2014}, civil engineering \cite{Tatum2017, Greenwood2019}, wildlife monitoring \cite{Goebel2015,Hodgson2016}, parcel delivery \cite{Tavana2017}, disaster management \cite{Restas2015}, and firefighting \cite{Viegas2022}, to mention but a few. In these applications, UAVs often encounter complex and unsteady flow environments, including high turbulence levels \cite{Zhang2024} and strong gusts \cite{Park2023}, requiring continuous thrust changes. Typically, the thrust is controlled by varying the rotational speed, since the motors have high electrical efficiency over a large range of rotational speeds \cite{Moore2010, McDonald2014}, making transient operation common. As an alternative to fixed-pitch propellers, variable-pitch propellers have also been explored to improve control authority \cite{Cutler2015, Pavel2022} and propulsive efficiency \cite{Sheng2016}. However, the broader operating envelope can place substantial portions of the blade close to or beyond stall. The present work focuses on rotational speed transients at elevated pitch angles, spanning conditions from laminar separation with reattachment to fully separated flow.
	
	A blade operating near stall challenges every level of the modeling hierarchy. Low-order methods based on blade element momentum theory (BEMT) \cite{Bramwell2001, Leishman2016} rely on steady sectional airfoil polars, typically obtained from potential flow theory \cite{Porter2016} or coupled with boundary-layer formulations \cite{Drela1989, Cutler2015}. Dynamic inflow models \cite{Pitt1981, Peters1991, Peters1995} account for the lag in the induced velocity during changing rotational speed, but their predictions remain sensitive to the chosen model \cite{Sekula2022} and can differ from dynamics identified in experiments \cite{Asper2024}. In the conditions investigated in this work, rotational speed transients question the use of steady sectional polars, while separation and stall challenge the viscous-inviscid models used to generate them. Similar limitations arise in vortex-particle methods coupled with lifting-line theory \cite{Valentin2024,Grava2026a}: although these methods resolve the propeller slipstream and naturally account for dynamic inflow, they still rely on steady sectional polars and neglect changing sectional aerodynamics due to rotation. Higher fidelity methods, such as unsteady Reynolds-averaged Navier--Stokes (URANS) simulations, fully resolve the rotor during transients (see e.g., \cite{Mortimer2020a}), but do not remove modeling difficulties associated with low Reynolds numbers. The flow may separate, transition, and reattach, forming a laminar separation bubble (LSB) \cite{Carmichael1981, Lissaman1983, Mueller2003}, requiring a dedicated transition model \cite{Menter2006} adapted to rotating geometries \cite{Pisharoti2023}. Yet, even for simple two-dimensional airfoils, URANS does not consistently outperform viscous-inviscid approaches such as XFOIL \cite{Morgado2016}.
	
	On the experimental side, data largely concern airfoils in steady conditions (e.g., \cite{Burgmann2008, Boutilier2012}), where airfoil performance responds strongly to the Reynolds number \cite{Winslow2018} and free-stream turbulence \cite{Wang2014, Yu2024b}. Far fewer studies have investigated transient conditions. Among them, rapid free-stream accelerations and decelerations were shown to shift the LSB opposite to quasi-steady expectations \cite{Ellsworth1991}, while accelerations from rest showed continued development of the LSB after reaching quasi-steady conditions \cite{Dierl2024}. In both cases, the airfoil did not stall, and the acceleration levels exceeded values typically reached by conventional fixed-pitch rotors, being more representative of torque-modulated concepts \cite{Paulos2018,Bernardes2023,Flores2024}. For smaller, quasi-steady free-stream changes, Toppings and Yarusevych \cite{Toppings2024, Toppings2025} observed significant hysteresis in the lift of a low-Reynolds-number airfoil between reattached and separated flow states. They relate this to the formation and bursting of an LSB and measure different timescales for a 2D airfoil and a finite wing. However, extrapolating these insights to propellers is not straightforward because rotation delays stall \cite{Himmelskamp1945} and flattens the LSB \cite{Khedr2025}, requiring measurements on a full rotor (e.g., \cite{Mulleners2012b,Schwermer2019}). This motivates studying accelerating rotors rather than accelerating airfoils.
	
	Few studies investigated transient propeller aerodynamics at low Reynolds numbers. Aerodynamic loads of a small-scale, two-bladed rotor in forward flight were characterized during a change of rotational speed \cite{Sekula2022}, both experimentally and with a time-marching solution using different inflow models in commercial software, resulting in varying vibratory loads. Grava et al. \cite{Grava2026a} studied rotor speed ramps with an unsteady vortex particle method. Thrust and torque coefficients traced hysteresis loops around steady values, attributed to dynamic inflow, as the model uses steady polars, agnostic to LSB and rotation effects. The tip vortex trajectory and circulation magnitude likewise showed transient behavior \cite{Grava2026a}. Chance et al. \cite{Chance2026} measured a five-bladed rotor under harmonic speed modulation. Cycle-mean thrust decreased by up to \SI{8}{\percent}, while phase-averaged loads showed only small phase delays at the highest modulation frequencies. Phase-averaged particle image velocimetry (PIV, \cite{Raffel2018}) indicated higher normalized downstream velocity during deceleration, with unchanged upstream flow. Ratz et al. \cite{Ratz2026b} examined open-loop step responses of a two-bladed propeller and observed a hysteresis loop in the thrust coefficient, while the flow over the blade remained attached and quasi-steady. Transient differences appeared mainly upstream and varied across the propeller span. All the aforementioned studies used a single moderate fixed pitch with attached flow; transient behavior at higher pitch leading to separation during accelerations/decelerations remains unmeasured.
	
	\begin{figure*}[t]
		\centering
		\includegraphics[width=\linewidth,clip,trim=0cm 4.9cm 0cm 0cm]{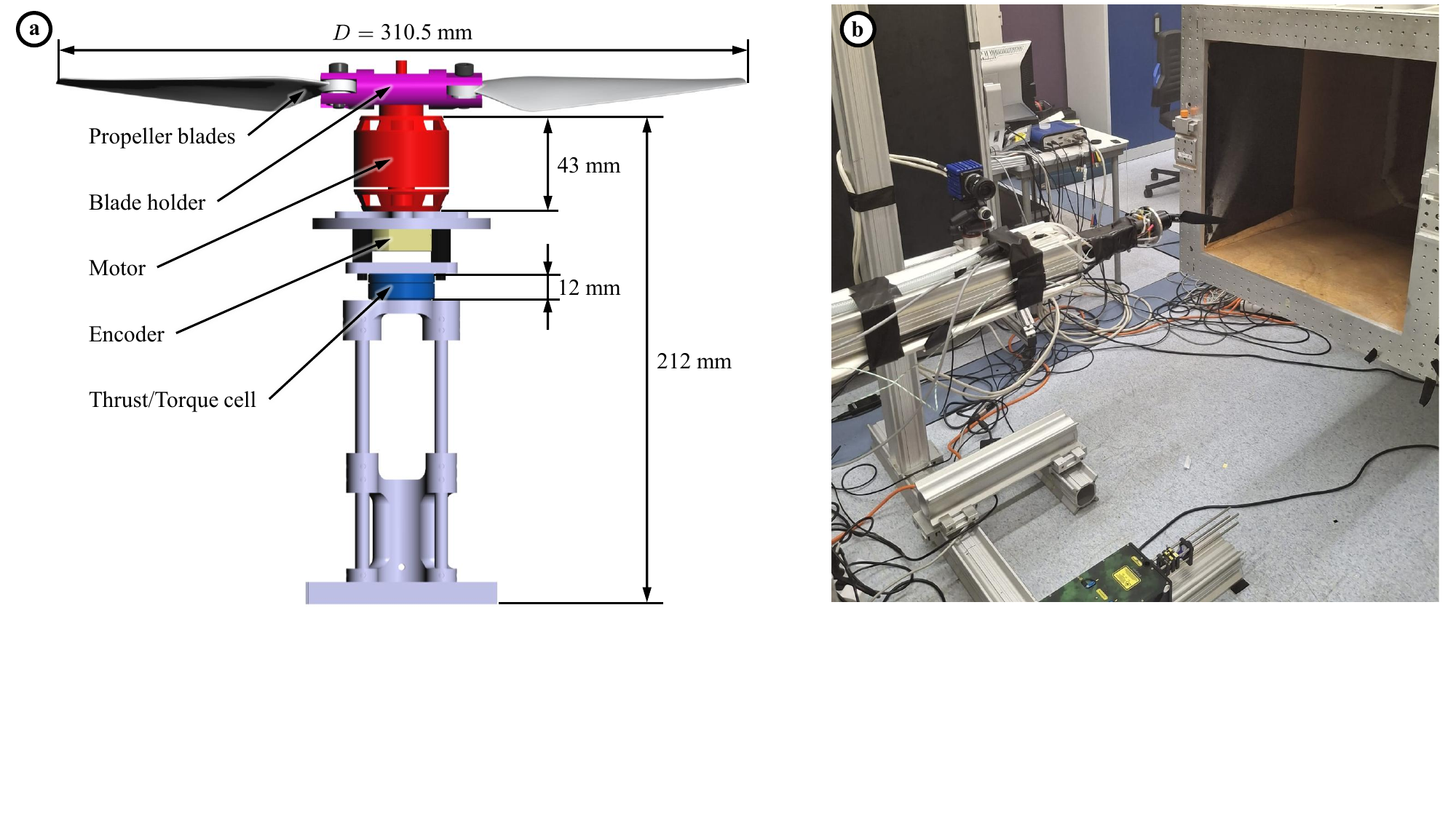}
		\caption{Propeller drivetrain. (a) Blade holder (purple), drivetrain with the motor (red), capacitive encoder (yellow), and thrust and torque cell (blue). (b) Setup mounted in the W-tunnel at TU Delft.}
		\label{fig:01_exp_setup}
	\end{figure*}

    This work addresses this gap using load measurements and phase-locked PIV on a two-bladed propeller in hover, subjected to open-loop step changes between 3000 and \SI{5000}{\rpm} at pitch offsets $\Theta = 0$, 5, and \SI{10}{\degree}. The nondimensional accelerations remain within the quasi-steady range of 2D airfoils, while the different pitch offsets and spanwise locations cover attached flow, laminar separation with reattachment, and fully separated flow. Load cell measurements show hysteresis in the thrust coefficient at high pitch offsets, while the response remains quasi-steady at $\Theta = \SI{0}{\degree}$. The sense of this hysteresis opposes classic dynamic inflow theory. The flow field is characterized with phase-locked PIV in two regions: the slipstream in the meridian plane, resolving tip vortex trajectory, strength, and rotor-disk inflow; and three spanwise stations at 40, 60, and \SI{80}{\percent} of the rotor radius, characterizing flow separation over the blade and in the blade wake. The PIV measurements capture transient flow dynamics that escape quasi-steady techniques like temperature-sensitive paint \cite{Jaroslawski2022, Li2024b} and oil flow visualization \cite{Grande2021, Grande2022}. The near-wall resolution does not support a reconstruction of the LSB topology, so we characterize the transient response using sectional circulation, shear layer velocity fluctuations, and blade-wake vorticity.
	
	The remainder of the paper is organized as follows. Section~\ref{sec:2} describes the rotor test rig, PIV measurement setup, and test procedure, while data processing is described in Sec.~\ref{sec:3}. Section~\ref{sec:5} presents the results, starting with the load measurements, which show hysteresis. Potential origins of the hysteresis are then discussed in the context of the blade deformation as well as the spanwise and slipstream PIV. Finally, Sec.~\ref{sec:6} closes with conclusions and perspectives.

	\section{Materials and methods}
	\label{sec:2}
	
	Measurements were conducted in the W-tunnel at TU Delft. All data were acquired in hovering conditions, with the wind tunnel switched off during all acquisitions. It was switched on to transport the seeding and redistribute it in the test room between test runs. The propeller was characterized under two different conditions, herein referred to as steady (constant rotational speed) and transient (changing rotational speed).
	
	\subsection{Propeller test rig}
    \label{sec:2_1}
	
	Figure~\ref{fig:01_exp_setup} shows the propeller drivetrain and the setup mounted in the wind tunnel. The propeller was located at least $4D$ from the walls and floor in the radial direction and more than $20D$ from downstream boundaries, with $D$ the propeller diameter. The setup consisted of a T-Motor AT3520 550KV brushless DC motor, which was connected to a T-Motor AT75A 2--6S electronic speed controller (ESC) and powered by two parallel RUZIZAO bench power supplies delivering \SI{22}{\volt} and a peak current of \SI{20}{\ampere}. Below the motor plate was a CUI AMT11 capacitive encoder with 2048 pulses per revolution, used to compute the rotational speed $n$. The thrust $T$ was measured with a FUTEK QMA150 custom torque and thrust biaxial sensor; torque was not considered further because it showed no significant hysteresis. The thrust channel was calibrated \textit{in situ} with weights corresponding to the maximum measured thrust. 
	
	The load cell was excited with \SI{10}{\volt} DC. Its rated capacity is \SI{100}{\newton}, and nonlinearity and hysteresis stay within $\pm \SI{0.5}{\percent}$ of the rated output. The sensor operates between \num{-51} and \SI{93}{\celsius}, with temperature compensation between \num{16} and \SI{71}{\celsius}. To mitigate thermal effects, compressed air was directed onto the motor stator, and the motor and load cell temperatures were monitored with type-K thermocouples.
	
	Motor speed was regulated with a hardware-timed pulse-width-modulated signal at \SI{500}{\hertz}, provided by a Raspberry Pi 5. The thrust signal was conditioned with a FUTEK IAA100 analog amplifier, and the signals were sampled by an MCC 128 DAQ HAT with 16-bit resolution at \SI{10}{\kilo\hertz}. Thermocouple signals were acquired with an MCC 134 DAQ HAT with 24-bit resolution at \SI{1}{\hertz}. The encoder pulses were sampled with an Arduino GIGA R1 WiFi, and the rotational speed was computed based on the number of pulses counted every \SI{2}{\milli\second}, yielding an effective frequency of \SI{500}{\hertz}. This Arduino handled the rotational speed alone; a second Arduino performed the phase-locking (see Sec.~\ref{sec:2_4}).
	
	\subsection{Propeller blade geometry}
	\label{sec:2_2}
	
	\begin{figure*}[t]
		\centering
		\includegraphics[width=\linewidth]{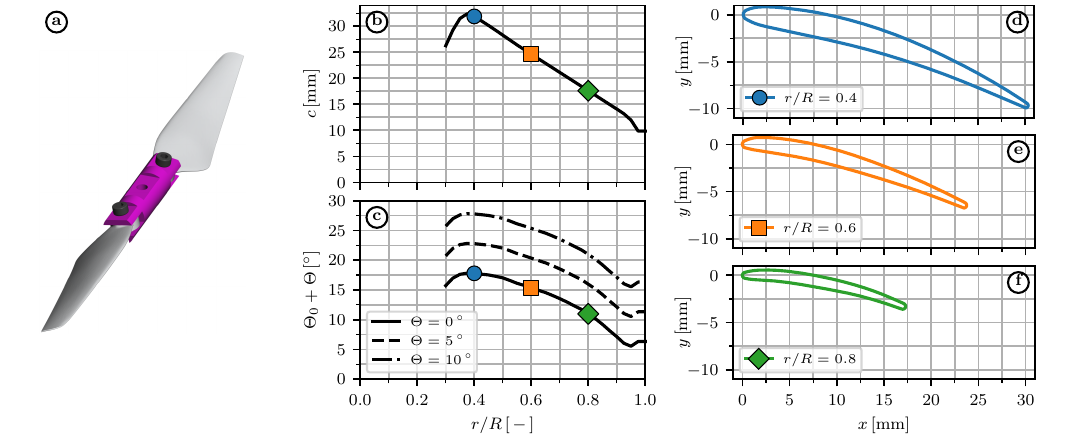}
		\caption{Propeller CAD geometry (a) and the chord (b) and twist (c) distributions over the normalized propeller radius. Propeller profiles extracted at $r/R=0.4,\,0.6,\,0.8$ for $\Theta = \SI{0}{\degree}$ (d--f) in a Cartesian coordinate system centered at the leading edge $(x,\,y)$. Markers in subfigures (b,c) correspond to the colors of the extracted slices in (d--f).}
		\label{fig:02_twist_chord}
	\end{figure*}

    Blades from Holybro X500 V2 1045 propellers were mounted on three sets of CNC-machined aluminum holders, imposing additional pitch offsets $\Theta = 0$, 5, and \SI{10}{\degree} relative to the built-in twist $\Theta_0$. The holder sets had comparable mass and rotational inertia, and each resulted in a two-bladed propeller.  The original propeller diameter and nominal pitch were $D_B = \SI{254}{\milli\meter}$ and \SI{114.3}{\milli\meter}, respectively, and the blade holders increased the diameter to \SI{310.5}{\milli\meter}. Figure~\ref{fig:02_twist_chord}(a) shows the blade and holder geometry, reconstructed from a 3D scan of the original blade. Figures~\ref{fig:02_twist_chord}(b,c) show the chord $c$ and total twist $\Theta_0+\Theta$ as functions of normalized radius $r/R$, with $r$ the radial position and $R$ the propeller radius. The chord distribution is unchanged across pitch offsets, while the twist is shifted by \SI{5}{\degree} increments. Figures~\ref{fig:02_twist_chord}(d--f) show the blade sections at $r/R=0.4$, 0.6, and 0.8 for $\Theta=\SI{0}{\degree}$, corresponding to the three spanwise PIV measurement locations.
	
	\begin{table*}[hbt!]
		\caption{\label{tab:piv_params} Parameters of the PIV system.}
		\centering
		\begin{tabular}{lcccc}
			\hline
			\hline\\[-7pt]
			$r/R$ & 0.0 & 0.4 & 0.6 & 0.8\\
			Camera & \multicolumn{4}{c}{LaVision Imager sCMOS} \\
			Pixel size & \multicolumn{4}{c}{\qtyproduct{6.5 x 6.5}{\micro\meter}} \\
			Sensor size & \multicolumn{4}{c}{$\SI{2560}{\pixel} \times \SI{2160}{\pixel}$} \\
			\multirow{2}{*}{Field of view} & \qtyproduct{317.8 x 268.1}{\milli\meter} & \multicolumn{3}{c}{\qtyproduct{51.8 x 43.7}{\milli\meter}} \\
			& $2.05\,R \times 1.73\,R$ & $1.62\,c \times 1.37\,c$ & $2.10\,c \times 1.77\,c$ & $2.94\,c \times 2.48\,c$ \\
			Objective &  AF Nikkor \SI{35}{\milli\meter} & \multicolumn{3}{c}{AF Micro-Nikkor \SI{200}{\milli\meter}} \\
			Aperture $f\#$ & $5.6$ & \multicolumn{3}{c}{$11$} \\
			Magnification & 0.052 & \multicolumn{3}{c}{0.321} \\
			Imaging resolution & \SI{8.1}{\pixel\per\milli\meter} & \multicolumn{3}{c}{\SI{49.4}{\pixel\per\milli\meter}}  \\
			Laser pulse separation (steady) & \qtyrange{84}{200}{\micro\second} & \qtyrange{12}{23}{\micro\second} & \qtyrange{8}{15}{\micro\second} & \qtyrange{6}{11}{\micro\second} \\
			Laser pulse separation (transient) & \SI{84}{\micro\second} & \SI{12}{\micro\second} & \SI{8}{\micro\second} & \SI{6}{\micro\second} \\
			Laser source & \multicolumn{4}{c}{Quantel Evergreen EVG00200 Nd:YAG} \\
			Maximum laser frequency & \multicolumn{4}{c}{\SI{15}{\hertz}} \\
			Laser sheet thickness & \multicolumn{4}{c}{\SI{1.5}{\milli\meter}} \\
			Acquisition software & \multicolumn{4}{c}{DaVis 8.4} \\
			Particle seeding & \multicolumn{4}{c}{SAFEX water-glycol droplets, \SI{1}{\micro\meter}} \\
			& & & & \\[-7pt]
			\hline
			\hline
		\end{tabular}
	\end{table*}
	
	\subsection{Particle image velocimetry setups}
    \label{sec:2_3}

\begin{figure*}[t]
		\centering
		\includegraphics[width=\linewidth,clip,trim=0cm 5.5cm 0cm 0cm]{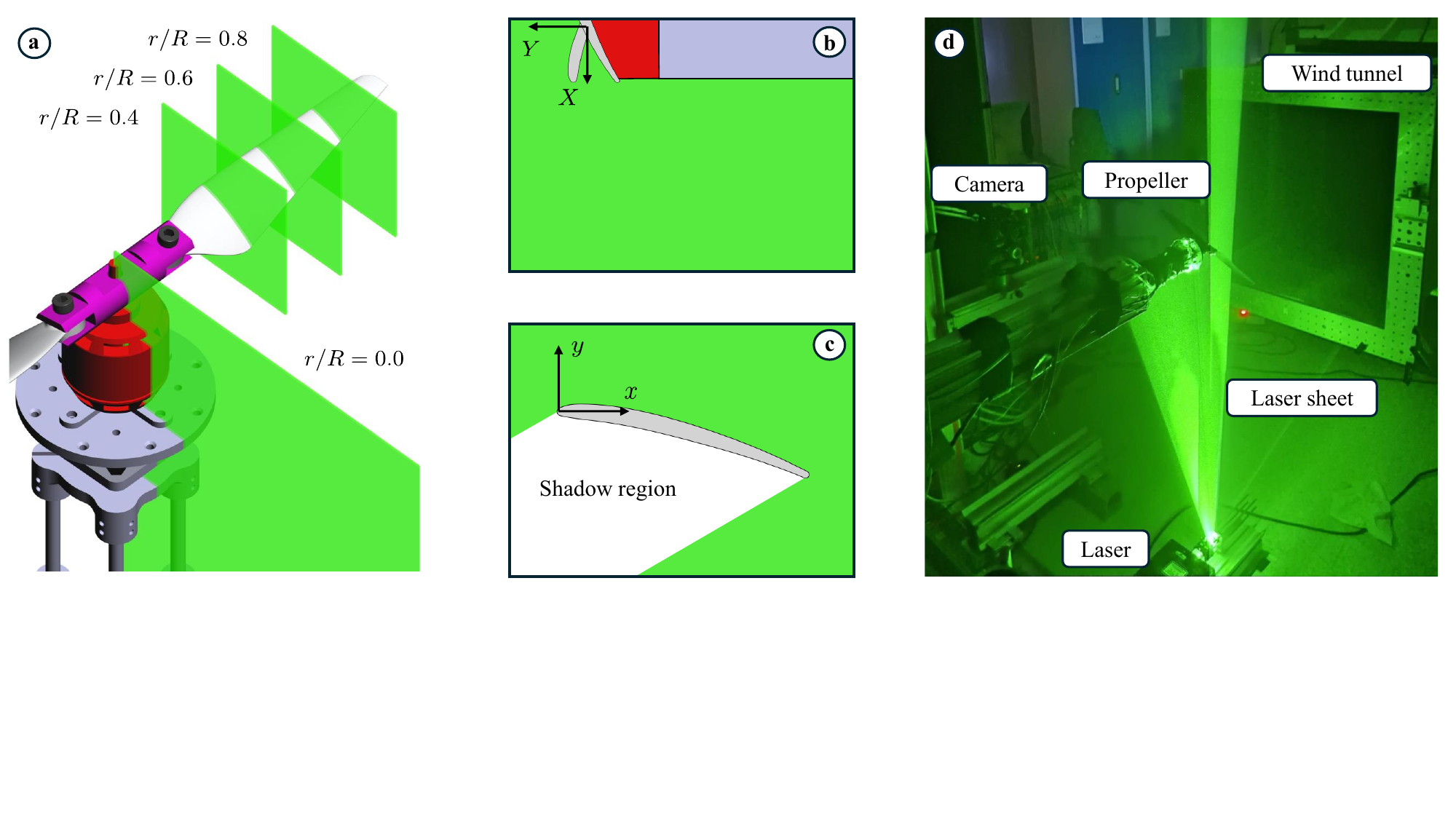}
		\caption{PIV measurement setup. (a) Positions of the laser sheet for the four different measurement locations. Field of view for the slipstream measurements with coordinate system $(X,\,Y)$, centered in the hub (b), and the spanwise measurements with coordinate system $(x,\,y)$, centered in the leading edge (c). Also sketched is the shadow region where the blade blocks laser light. (d) Picture of the planar PIV setup for $r/R=0.0$.}
		\label{fig:03_piv_setup}
	\end{figure*}

    Two regions were surveyed with 2D-2C planar phase-locked PIV: the slipstream in the meridian plane at $r/R = 0$ and different span locations at $r/R=0.4,\,\,0.6,$ and 0.8. Their laser-sheet positions and fields of view (FOVs) are shown in Fig.~\ref{fig:03_piv_setup}(a--c), together with the coordinate systems used throughout the paper. The meridian-plane measurements characterize rotor inflow and tip-vortex shedding, while the spanwise measurements resolve the blade flow and wake. Because the blade blocks the laser sheet, measurements were restricted to the suction side, where the flow separates and forms an LSB. Figure~\ref{fig:03_piv_setup}(d) shows the meridian-plane setup in the wind tunnel.
	
	For both setups, the flow was seeded using a SAFEX Twin Fog generator with SAFEX-Inside-Nebelfluid, a water--glycol mixture of \SI{1}{\micro\meter} median particle diameter. To reach uniform concentration, the laboratory was filled for \SI{30}{\minute} before the first test, and particles were added intermittently. The particles were illuminated by a Quantel Evergreen EVG00200 Nd:YAG laser, expanded into a laser sheet using spherical and cylindrical lenses. For the spanwise measurements, the blades were coated with thin layers of matte black paint to reduce surface reflections. The added mass was negligible. The illuminated droplets were imaged using a low-speed LaVision Imager sCMOS camera with a resolution of $\SI{2560}{\pixel} \times \SI{2160}{\pixel}$ and a pixel pitch of \SI{6.5}{\micro\meter}. For the slipstream measurements, an AF Nikkor \SI{35}{\milli\meter} objective was used at $f\# = 5.6$, resulting in an FOV of \qtyproduct{317.8 x 268.1}{\milli\meter}. For the spanwise measurements, a small winglet on the propeller obstructed the view of the outer spanwise sections, so the camera was equipped with a Scheimpflug mount with an angle of less than \SI{5}{\degree}. An AF Micro-Nikkor \SI{200}{\milli\meter} lens at $f\#=11$ resulted in an FOV of \qtyproduct{51.8 x 43.7}{\milli\meter}.
	
	All recordings were acquired with DaVis 8.4 from LaVision, and a full overview of the acquisition parameters for the different test cases is provided in Table~\ref{tab:piv_params}.

	\subsection{Test procedure}
	\label{sec:2_4}
	
	\begin{figure}[t]
		\centering
		\includegraphics[width=\linewidth]{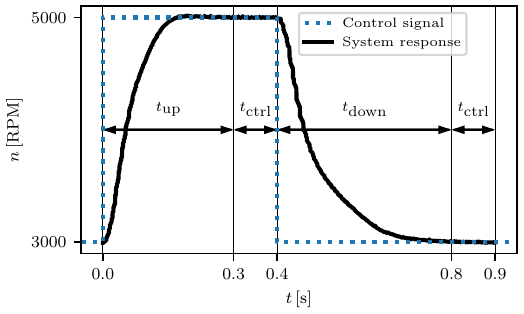}
		\caption{Typical step cycle of the step up and down response in the rotational speed for $\Theta = \SI{0}{\degree}$. During $t_\textrm{up}$ and $t_\textrm{down}$ the system is in open loop and during $t_\textrm{ctrl}$ in closed loop. The control signal corresponds to minimum and maximum duty cycles, which map to 3000 and \SI{5000}{\rpm}.}
		\label{fig:step_response_time_history}
	\end{figure}
    
	For the steady measurements, data were acquired at discrete rotational speeds, set with a PID controller. For the transient measurements, the propeller was subjected to open-loop step changes between two duty cycle commands, corresponding to minimum and maximum rotational speed. Accelerations (step up) and decelerations (step down) were investigated within the same cycle, as illustrated in Fig.~\ref{fig:step_response_time_history}. The sequence started at $t=\SI{0}{\second}$ with a step up command to the ESC, followed by $t_\textrm{up} = \SI{0.3}{\second}$ of open-loop response. A PID controller then corrected setpoint drift for $t_\textrm{ctrl} = \SI{0.1}{\second}$ before the open-loop step down, which lasted $t_\textrm{down} = \SI{0.4}{\second}$ and was followed by another setpoint drift correction during $t_\textrm{ctrl} = \SI{0.1}{\second}$. The drift remained within \SI{1}{\percent} of the target, meaning only small corrections.
	
	For the load measurements, synchronization between the acquisition devices caused the first \SI{5}{\milli\second} of each step to be unavailable. This interval is comparable to the ESC dead time before the rotational speed begins to change and was therefore excluded from the analysis. The dead time was determined from the rotational speed signal acquired in parallel with the PIV triggering. PIV data were acquired in the same manner: steady conditions at constant rotational speed and transient conditions with step responses. Hardware limitations ruled out time-resolved measurements, so low-speed phase-locked PIV was used. The encoder index, generated once per revolution, was sampled by a second Arduino and triggered the PIV system. It was not possible to align index pulse and blade position because of a minimum delay between the synchronizer trigger and the camera image. Even in fast response mode, the camera required \SI{1}{\milli\second} between receiving a trigger and capturing an image, corresponding to a phase lag $\Delta\Psi = \SI{30}{\degree}$ at maximum speed. Compensating for the lag by aligning the encoder for every test proved impractical because of varying speed in transient conditions. Instead, the encoder was offset by approximately \SI{120}{\degree} from the desired blade position in the camera FOV, ensuring positive trigger delays across all rotational speeds. In steady conditions, the delay was set individually for each rotational speed in the acquisition software. In transient conditions, the second Arduino tracked the number of revolutions since the step command and assigned delay times pre-computed from the mean rotational speed. The method thus relies on cycle-to-cycle repeatability, quantified in \ref{app:4}.

    The maximum laser frequency of \SI{15}{\hertz} restricted the acquisition to every sixth revolution, so the transients were sampled over repeated step responses. In each repetition, the Arduino triggered the PIV system on every sixth revolution after the step at $t=0$, starting from revolution $k = 0,\dots,5$. The step was repeated at fixed $M$ until enough samples had been acquired, and $M$ was then incremented. Because each triggering point occurred at a different rotational speed for one value of $k$, the pulse separation time was kept constant in transient conditions and adapted to the speed in steady conditions (see Table~\ref{tab:piv_params}).
	
\begin{figure}[t]
		\centering
		\includegraphics[width=\linewidth]{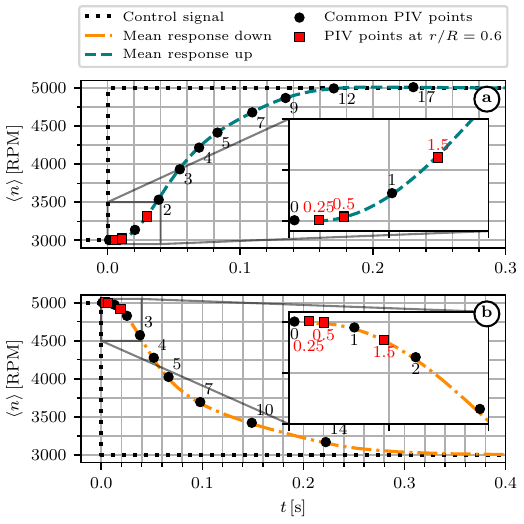}
		\caption{Rotational speed during step up (a) and step down (b) at $\Theta = \SI{0}{\degree}$, together with the PIV phase-locking points for $r/R=0.6$. The dotted line gives the commanded control signal and the dashed (a) and dash-dotted (b) lines the ensemble average of the load acquisitions. Markers denote phase-locking points with annotations corresponding to the number of revolutions since the start of the transient. Black dots (integers) correspond to points common to measurement planes and red squares (fractions) to the additional points at $r/R=0.6$. The insets show the first \SI{40}{\milli\second} of the steps.}
		\label{fig:phaselock_methods}
	\end{figure}

    Figure~\ref{fig:phaselock_methods} shows the phase-locking points for $\Theta = \SI{0}{\degree}$. Marker annotations correspond to the number of revolutions since the start of the transients. Two point sets are distinguished: those common to all laser sheet positions (black dots, integers) and additional points acquired only at $r/R=0.6$ (red squares, fractions). The latter resolve changes during timescales shorter than one revolution at the onset of acceleration and deceleration. The 10 to \SI{90}{\percent} rise time spanned approximately seven revolutions, while the 90 to \SI{10}{\percent} fall time spanned approximately eleven revolutions. The other pitch offsets had similar orders of magnitude, given in the next section.
	
	\subsection{Test matrix}
    \label{sec:2_5}
	
	The steady measurements covered fixed rotational speeds from  2750 to \SI{5250}{\rpm} in steps of \SI{250}{\rpm}, the step responses a range from 3000 to \SI{5000}{\rpm}. The blade tip Mach number stays below 0.25 throughout, so the flow is treated as incompressible. For the loads, steady data were acquired for \SI{30}{\second} at \SI{500}{\hertz} for the rotational speed and at \SI{10}{\kilo\hertz} for the thrust. The transient data were recorded over $t_\textrm{up} = \SI{0.3}{\second}$ and $t_\textrm{down}=\SI{0.4}{\second}$, with 500 repetitions for each step response.

    \begin{table*}[t]
    \caption{\label{tab:exp_params} Test matrix and propeller operating conditions in transient conditions.}
    \centering
    \begin{tabular}{lccccccccc}
        \hline
        \hline\\[-0.35cm]
        $\Theta$ $[\si{\degree}]$ & $r/R\,[-]$ & \# PIV Samples & $\langle \tau \rangle_{10-90}$ [$-$] & $\langle \tau \rangle_{90-10}$ [$-$] & $\langle M \rangle_{10-90}$ & $\langle M \rangle_{90-10}$ & $\langle \textrm{Re}_c \rangle$ $\left[\times 10^{4}\right]$ & \multicolumn{2}{c}{$\vert c \hspace{0.03cm} \langle a \rangle_\textrm{blade} / \langle u\rangle_\textrm{blade}^2 \vert$ [$-$]} \\
        & & & Step up & Step down & Step up & Step down & & Step up & Step down\\[0.1cm]
        \hline\\[-0.35cm]
        \multirow{4}{*}{0} & $0.0$ & 250 & $-$ & $-$ & \multirow{4}{*}{6.9} & \multirow{4}{*}{11.2} & $-$ & $-$ & $-$ \\
        & $0.4$ & 200 & 84 & 137 & & & $4.13 \,\dots\, 6.88$ & $\leq 0.011$ & $\leq 0.006$ \\
        & $0.6$ & 500 & 163 & 266 & & & $4.79 \,\dots\, 7.98$ & $\leq 0.006$ & $\leq 0.003$ \\
        & $0.8$ & 200 & 304 & 495 & & & $4.56 \,\dots\, 7.60$ & $\leq 0.003$ & $\leq 0.002$ \\[0.25cm]
        
        \multirow{4}{*}{5} & $0.0$ & 250 & $-$ & $-$ & \multirow{4}{*}{6.4} & \multirow{4}{*}{8.5} & $-$ & $-$ & $-$ \\
        & $0.4$ & 200 & 79 & 104 & & & $4.13 \,\dots\, 6.88$ & $\leq 0.011$ & $\leq 0.006$ \\
        & $0.6$ & 500 & 153 & 201 & & & $4.79 \,\dots\, 7.98$ & $\leq 0.006$ & $\leq 0.003$ \\
        & $0.8$ & 200 & 285 & 375 & & & $4.56 \,\dots\, 7.60$ & $\leq 0.003$ & $\leq 0.002$ \\[0.25cm]
        
        \multirow{4}{*}{10} & $0.0$ & 250 & $-$ & $-$ & \multirow{4}{*}{6.1} & \multirow{4}{*}{7.0} & $-$ & $-$ & $-$ \\
        & $0.4$ & 200 & 75 & 86 & & & $4.13 \,\dots\, 6.88$ & $\leq 0.010$ & $\leq 0.007$ \\
        & $0.6$ & 500 & 146 & 167 & & & $4.79 \,\dots\, 7.98$ & $\leq 0.005$ & $\leq 0.003$ \\
        & $0.8$ & 200 & 272 & 312 & & & $4.56 \,\dots\, 7.60$ & $\leq 0.003$ & $\leq 0.002$ \\[0.1cm]
        \hline
        \hline
    \end{tabular}
\end{table*}
	
	For the PIV, steady data were acquired using the same rotational speeds with $250,\,200,\,500$, and $200$ samples at $r/R=0.0,\,0.4,\,0.6$, and $0.8$, respectively. In transient conditions, the same number of samples was recorded at every phase-locking point. The meridian plane measurements at $r/R=0.0$ were conducted at a phase angle of $\Psi=\SI{0}{\degree}$, midway between the two blades (see Fig.~\ref{fig:03_piv_setup}a). Additionally, data at $\Psi = \SI{90}{\degree}$  were used to define the blade and rotor disk position from the laser reflection. Since no PIV fields were extracted, only 50 samples were taken per point.
	
	A summary of the full test matrix and operating conditions is provided in Table~\ref{tab:exp_params}. Given are the 10 to \SI{90}{\percent} rise and 90 to \SI{10}{\percent} fall times, where the nondimensional time is defined as
    \begin{align}
        \langle \tau \rangle (t) = \frac{1}{c} \int_{0}^{t} \langle u \rangle_\textrm{blade}(t^\prime) \, \textrm{d}t^\prime\,,
    \end{align}
    where $\langle u \rangle_\textrm{blade} = \langle \Omega \rangle r$ with $\langle \Omega \rangle = 2 \pi \langle n \rangle / 60$ in \si{\radian\per\second} and $\langle \, \cdot\, \rangle$ denotes an ensemble average across all repetitions. Therefore, $\langle \tau \rangle$ equals the distance traveled by the local blade section in chord lengths. The corresponding number of revolutions $\langle M \rangle$ is also given. As expected, higher pitch offsets result in quicker decelerations because of higher drag, while the step up response remains similar, as the motor has sufficient torque.

    The local conditions at each spanwise PIV plane are characterized by the chord-based Reynolds number
    \begin{align}
        \langle \textrm{Re}_c \rangle = \frac{\langle u \rangle_\textrm{blade} \,c}{\nu}\,,
    \end{align} with $\nu$ the kinematic viscosity of air, obtained from Sutherland's law at the recorded room temperature and pressure. The maximum nondimensional acceleration/deceleration
    \begin{align}
        \left\vert \frac{c\,\langle a \rangle_\textrm{blade}}{\langle u \rangle_\textrm{blade}^2} \right\vert\,,
    \end{align} with $\langle a \rangle_\textrm{blade}$ the blade acceleration, does not exceed 0.011 in all cases, indicating negligible acceleration effects. Nevertheless, measurable departures from quasi-steady behavior are found in both the loads and selected flow quantities (see Sec.~\ref{sec:5}).
    
	\section{Data processing and reduction}
	\label{sec:3}
	
	\subsection{Aerodynamic loads}
    \label{sec:3_1}

    For steady data, the load and rotational speed were time-averaged. In transient conditions, data were ensemble-averaged by averaging data obtained at the same elapsed time $t$ from the step trigger. Because vibrations affected the data even after averaging, each step response was filtered with a third-order forward-backward Butterworth low-pass filter at \SI{30}{\hertz} cut-off, chosen to retain low-frequency transients while suppressing structural vibrations. The filtered signals were ensemble-averaged to obtain $\langle T \rangle (t)$ and $\langle \Omega \rangle (t)$, where the rotational speed data were interpolated onto the thrust time base with cubic splines.
    	
	The ensemble-averaged thrust and rotational speed were then used to compute the thrust coefficient:
	\begin{align}
		\langle C_T \rangle = \frac{\langle T \rangle}{\rho \pi R^2 (\langle \Omega \rangle R)^2} \,,
	\end{align} where $\rho$ is the density of air and $\pi R^2$ the rotor disk area.
    
    A second-order polynomial fitted to the steady thrust coefficients provides a continuous baseline against which the transient thrust coefficients are expressed as a relative difference:
	\begin{align}
		\label{eq:rel_ct}
		\Delta \langle C_{T} \rangle = \frac{\langle C_T \rangle^\textrm{transient} - \langle C_T \rangle^\textrm{steady}}{\langle C_T \rangle^\textrm{steady}}\,.
	\end{align}
    
	\subsection{Velocity fields}
    \label{sec:3_2}
	
	An ensemble of image pairs was available at each rotational speed in steady conditions, and every phase-locking point in transient ones (see~Fig.~\ref{fig:phaselock_methods}). Different pre- and post-processing strategies were required for the spanwise and slipstream measurements and are described separately below.
	
	\subsubsection{Spanwise measurements}
	\label{sec:3_2_1}

    Pre-processing of the spanwise data was applied to every ensemble and had two objectives. First, removing background reflections, including both time-invariant ones (motor and hub screw) and varying local ones (blade surface flare). Second, correcting the varying horizontal blade position due to ESC jitter, vertical bending $\delta_y$, and twisting $\theta$.

    \begin{enumerate}
        \item The pixel-wise minimum background image was subtracted, eliminating time-invariant reflections.
        \item The reflection of the blade surface at the trailing edge was isolated using thresholding and morphological operations. Subpixel-accurate cross-correlation with a reference image at \SI{2750}{\rpm} gave the in-plane displacement.
        \item The images were shifted by their displacements, ensuring one common blade position across the ensemble. The images were averaged pixel-wise, and the mean blade reflection was isolated with thresholding and morphological operations. The blade orientation was obtained by fitting a rigid 2D transformation (two translations and one rotation) of the scanned blade contour to the extracted reflection. The fit minimized the distance between scanned shape and reflection, weighted by the image intensity, and also provided the average twist $\langle \theta \rangle$ relative to the reference image.
        \item The extracted airfoil position was used to define the masks: points inside the airfoil and the shadow region (Fig.~\ref{fig:03_piv_setup}c) were masked, together with a band within \SI{15}{\pixel} of the surface, where laser flare was too strong to yield reliable vectors. This second mask corresponds to $0.010\,c$, $0.012\,c$, and $0.017\,c$ for the three spans, matching the typical thickness of a reattached LSB \cite{Zhang2020a,Toppings2021,Toppings2024,Khedr2025}. Coupled with the spatial averaging inherent to PIV, the near-wall region thus cannot be resolved to capture direct LSB dynamics.
        \item Pre-processing ended with a mean normalization \cite{Raffel2018}: each image was divided by the pixel-wise mean, and a sliding mean over seven consecutive images was subtracted from the result. The filtered images were then rescaled by the global mean intensity.
    \end{enumerate}
	
	The preprocessed images were analyzed using the open-source software PaIRS \cite{Paolillo2024}, with a multi-pass scheme and window deformation \cite{Scarano2000}. Over three iterations, the window size was reduced from $\SI{128}{\pixel} \times \SI{128}{\pixel}$ to $\SI{32}{\pixel} \times \SI{32}{\pixel}$ with \SI{75}{\percent} overlap between neighboring windows. This corresponds to final vector spacings of $0.005$, $0.007$, and $0.009\,c$ at $r/R=0.4$, 0.6, and 0.8. Outliers were removed after each pass using a local median test \cite{Westerweel2005} and replaced by interpolation.
	
	\begin{figure}[t]
		\centering
		\includegraphics[width=\linewidth]{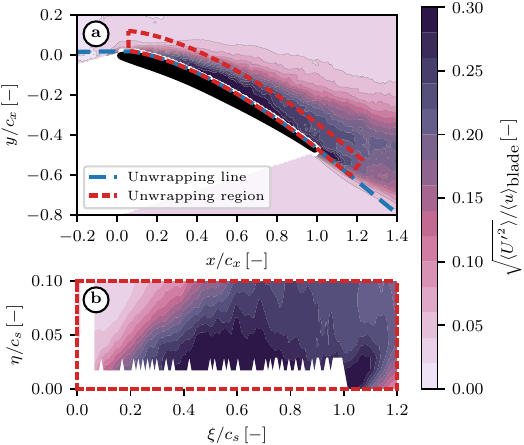}
		\caption{Definition of the unwrapping region around the airfoil for surface data analysis. (a) R.m.s. of magnitude fluctuations $\sqrt{\langle U^{\prime^2} \rangle} / \langle u \rangle_\textrm{blade}$ at \SI{5000}{\rpm} for $\Theta = \SI{10}{\degree}$ at $r/R=0.6$ in Cartesian coordinates centered at the leading edge, together with the surface line extended beyond the leading and trailing edges (blue long dashes) and the region used for unwrapping (red short dashes). (b) Magnitude r.m.s. values interpolated to the surface-normal coordinate system in the red short dashed box.}
		\label{fig:span_wake_surface_unwrapping}
	\end{figure}
	
	After cross-correlation, interrogation windows were masked within \SI{15}{\pixel} of the detected airfoil. Each instantaneous field was then rotated by the corresponding twist $\langle \theta \rangle$, both in the spatial grid and the velocity components. The full processing pipeline delivered an ensemble of velocity fields per rotational speed and phase-locking point.

    Velocity data were analyzed through ensemble averages of the velocities $\langle u \rangle$, $\langle v \rangle$, and the ensemble average of the velocity magnitude $\langle U \rangle = \langle \sqrt{u^2 + v^2} \rangle$ and the r.m.s. of its fluctuations $ \sqrt{\langle U^{\prime^2}\rangle} $, with $U^\prime = U - \langle U \rangle$. In these calculations, the horizontal flow $u$ was expressed in the blade reference frame, with the local blade velocity $\langle u \rangle_\textrm{blade}$ subtracted from the horizontal component; the blade velocity was also used for normalization. The second-order statistics are not fully converged with 500 samples at $r/R=0.6$ and 200 samples at the other two locations. Nevertheless, the differences between steady and transient conditions reported in Sec.~\ref{sec:5_3} exceed the residual scatter.

    Because the pitch offset differs between cases, the results employ three coordinate systems. The first is Cartesian $(x/c_x, y/c_x)$, centered at the leading edge and used for global comparisons of the flow field, with $c_x$ the chord projected onto the $x-$axis. The second uses the same Cartesian coordinate system, shifted to the trailing edge, used for the airfoil wake. The third is surface normal $(\xi / c_s, \eta / c_s)$, with $\xi$ the surface-tangential and $\eta$ the normal coordinate, and $c_s$ is the chord length along the $\xi-$axis, used for the upper surface flow. Figure~\ref{fig:span_wake_surface_unwrapping}(a) shows the r.m.s. of magnitude fluctuations $\sqrt{\langle U^{\prime^2} \rangle} / \langle u \rangle_\textrm{blade}$ for $\Theta = \SI{10}{\degree}$ at $r/R=0.6$. Displayed are the region surrounding the upper surface and the unwrapping line, which is the detected airfoil position extended beyond both edges. Data inside this region were mapped to the surface-normal coordinates with bilinear interpolation and are depicted in Fig.~\ref{fig:span_wake_surface_unwrapping}(b).

    In addition to the mean and r.m.s. velocity magnitude, three further quantities are considered. First, the pressure field is estimated by integrating the incompressible Navier--Stokes equations based on the PIV data. The integration uses a finite marching algorithm \cite{Oudheusden2013} with Dirichlet boundary conditions based on isentropic flow at the upper boundary and Neumann boundary conditions on the remaining boundaries. The surface pressure is obtained via linear extrapolation of the pressure field to the suction side surface. Results are given in terms of the ensemble-averaged pressure coefficient, normalized with respect to the blade velocity:
    \begin{align}
        \langle C_p \rangle = \frac{\langle p \rangle - \langle p \rangle_\infty}{\frac{1}{2}\rho_\infty \langle u\rangle_\textrm{blade}^2}\,.
    \end{align} 

    Second, the normalized sectional airfoil circulation $ \langle \Gamma_s \rangle / (\langle u \rangle_\textrm{{blade}}\,c_x)$ is estimated from a line integral of the mean flow around a control volume excluding the laser shadow. Third, the instantaneous wake vorticity $\omega_z$ is computed via finite differences. The absolute vorticity is spatially averaged in the region $x/c_x > 0.05,\,y/c_x < 0.4$ and then averaged across the ensemble to obtain the mean absolute wake vorticity $\langle \overline{\vert\omega_z \vert} \rangle c_x/ \langle u \rangle_\textrm{blade}$.

    The image-based detection used for aligning and masking also provides information about the blade deformation, which has been observed to affect the power coefficient \cite{Russell2017} and vibratory loads \cite{Moehren2022}. Advanced span-resolved measurements rely on dotted targets \cite{Nowicki2016} or speckle patterns \cite{Sicard2013} through digital image correlation \cite{Sutton2009}. Yet, they require a dedicated setup and are beyond the scope of the present work; the deformation estimates obtained here are sufficient for the discussion in Sec.~\ref{sec:5_2}.
	
	\subsubsection{Slipstream measurements}
	\label{sec:3_2_2}

    The PIV images were pre-processed by masking and background removal. Motor housing and support structure were masked manually using a constant mask across all cases. A second mask covered the reflection of a motor cooling opening in a \SI{6}{\pixel}-wide band centered at $Y/R = -0.1$ from $X/R = 0$ to 0.7, where $(X/R, Y/R)$ is the coordinate system used for the slipstream data. Background reflections were then removed with proper orthogonal decomposition by removing the first two modes of each image ensemble (see \cite{Mendez2017} for details).
	
	The images were processed with the same multi-pass scheme as the spanwise measurements. Due to the smaller magnification the interrogation window size was reduced from $\SI{96}{\pixel} \times \SI{96}{\pixel}$ to $\SI{24}{\pixel} \times \SI{24}{\pixel}$ over three iterations, with \SI{75}{\percent} overlap, corresponding to a vector spacing of $0.0048\,R$. Outliers were again removed using the same local median test as the spanwise data. Ensemble averaging then gave mean velocities for every steady speed and every phase-locking point, each being normalized by the blade tip velocity $\langle u \rangle_\textrm{tip} = \langle \Omega \rangle R$.
	
	\begin{figure}[t]
		\centering
		\includegraphics[width=\linewidth]{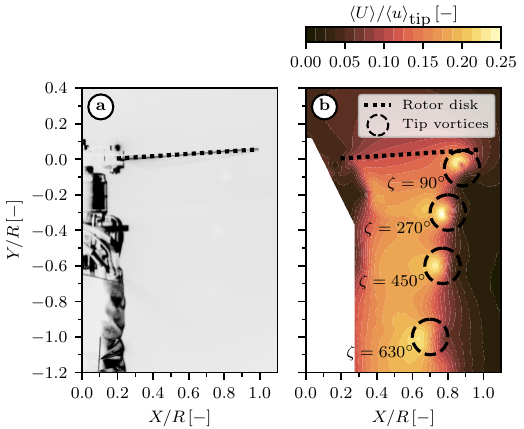}
		\caption{Definition of the rotor disk at \SI{5000}{\rpm} in steady conditions at $\Theta = \SI{5}{\degree}$. (a) Mean PIV image with inverted color scale for $\Psi = \SI{90}{\degree}$, zoomed to the propeller blade, and (b) ensemble-averaged velocity magnitude $\langle U \rangle/\langle u \rangle_\textrm{tip}$ for $\Psi = \SI{0}{\degree}$, together with four circles indicating the positions of the first four tip vortices, with $\zeta$ denoting their vortex age. The estimated rotor disk is drawn as a black dotted line.}
		\label{fig:inflow_methodology}
	\end{figure}
	
	The slipstream measurements were also used to estimate the disk inflow state by measuring the downwash velocity along the disk line between blade passages at $\Psi = \SI{0}{\degree}$ \cite{Mortimer2025, Yu2026}. The disk position was obtained from the data at $\Psi = \SI{90}{\degree}$, using the laser reflection from the blade underside. The root position was set manually, and the tip was detected via thresholding.
	
	An averaged raw image with $\Psi = \SI{90}{\degree}$ is depicted in Fig.~\ref{fig:inflow_methodology}(a) for $\Theta = \SI{5}{\degree}$ at \SI{5000}{\rpm}, with the estimated disk position drawn as a line between defined root and detected tip; no appreciable blade curvature was observed within the measurement scale. Figure~\ref{fig:inflow_methodology}(b) shows the corresponding ensemble-averaged velocity magnitude $\langle U \rangle/\langle u \rangle_\textrm{tip}$ for $\Psi = \SI{0}{\degree}$, locating the inflow plane relative to root and tip vortex. This method yielded the disk position of every ensemble for each steady rotational speed and each transient phase-locking point.

    The instantaneous inflow state was estimated by interpolating the vertical velocity along the disk line and integrating the induced velocity $v_i$, assuming azimuthal symmetry \cite{Mortimer2025,Yu2026}:
	\begin{align}
		\label{eq:inflow}
		\lambda = \frac{1}{\pi R^2} \int_0^{2\pi} \int_0^R \frac{v_i}{\langle \Omega \rangle R} \, r \, \textrm{d}r \, \textrm{d}\psi = \frac{2}{R^2} \int_0^R \frac{v_i}{\langle \Omega \rangle R} \, r \, \textrm{d}r\,.
	\end{align}
	This inflow state was estimated for every instantaneous field and then ensemble-averaged to obtain the inflow state $\langle \lambda \rangle$.

    Lastly, the tip vortices were tracked. An example of the measured tip vortices is shown in Fig.~\ref{fig:inflow_methodology}(b), where four circles denote the detected vortices in the ensemble-averaged field, each labeled with its vortex age $\zeta$, defined as the traversed azimuthal angle since blade passage. In hover, the tip vortex wanders between realizations, so ensemble averaging smears it, particularly for older vortices. Thus, the tip vortices were identified in the instantaneous snapshots.
		
	\begin{figure}[t]
		\centering
		\includegraphics[width=\linewidth]{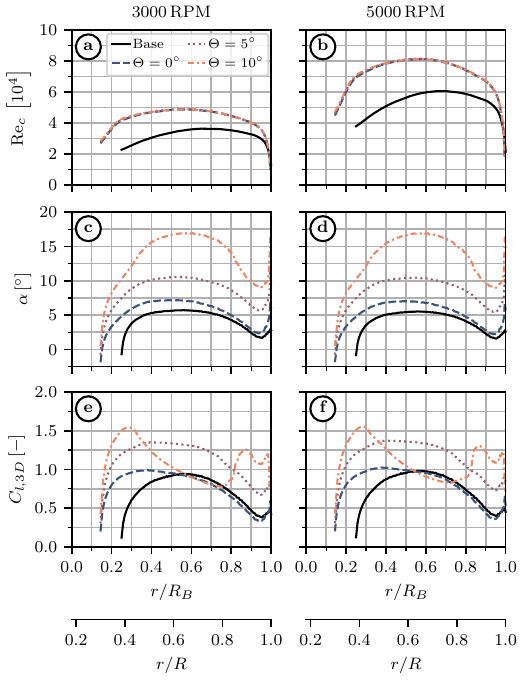}
		\caption{Results of the BEMT analysis of the original propeller (base) in comparison to the custom propeller with different pitch offsets, according to the legend. (a,b) Sectional Reynolds number, (c,d) angle of attack, and (e,f) lift coefficient over the normalized spanwise blade position at 3000 (a,c,e) and \SI{5000}{\rpm} (b,d,f). The two abscissas relate the spanwise coordinates between baseline ($R_B$) and custom propeller ($R$) as the radius changes, and the legend in subfigure (a) applies to all subfigures.}
		\label{fig:bemt}
	\end{figure}

    Following De Gregorio et al. \cite{DeGregorio2020}, tip vortices were identified using the $\Gamma_2$ criterion \cite{Graftieaux2001}, avoiding to compute velocity gradients, which are noisy because the vortex cores are void of particles.
	
	For a two-dimensional velocity field, the scalar function $\Gamma_2 \in [-1, 1]$ at a point $\boldsymbol{x}$ is defined as:
	\begin{align}
		\Gamma_2 (\boldsymbol{x}) = \frac{1}{N} \sum_{\boldsymbol{x}^\prime \in \mathcal{B}(r_\Gamma)} \frac{ \left( (\boldsymbol{x} - \boldsymbol{x}^\prime) \times \boldsymbol{u}_\textrm{HP} \right) \cdot \boldsymbol{z}}{\vert\vert (\boldsymbol{x} - \boldsymbol{x}^\prime) \vert\vert \, \vert\vert \boldsymbol{u}_\textrm{HP} \vert\vert}\,,
	\end{align} where $\boldsymbol{x}^\prime$ are all the data points contained within a disk $\mathcal{B}(r_\Gamma)$ of radius $r_\Gamma$, $N$ is the number of points within this disk, $\boldsymbol{u}_\textrm{HP}$ is the high-pass filtered velocity field, and $\boldsymbol{z}$ is the unit out-of-plane vector. The high-pass filter results in Galilean invariance, and is implemented by subtracting a Gaussian low-pass of \SI{10}{\pixel} standard deviation from each snapshot; the residual was smoothed with a second Gaussian of \SI{2}{\pixel} to suppress spurious peaks. Following \cite{DeGregorio2020}, the disk radius was matched to the tip vortex size, giving $r_\Gamma/R = 0.06$. This fixed radius biases the evaluation near boundaries, but the tip vortices lie well inside the measurement domain.

    Vortex candidates were identified as local maxima of $\vert \Gamma_2 \vert$ exceeding 0.85. Their surface area was computed as the 8-connected region satisfying $\vert \Gamma_2 \vert > 2 / \pi$ \cite{Graftieaux2001} and was used to compute the normalized circulation $\langle \Gamma_v \rangle / (\langle u \rangle_\textrm{tip}\,R) $ via the surface integral of the out-of-plane vorticity $\omega_z\,R / \langle u \rangle_\textrm{tip}$. To avoid detecting spurious peaks, valid vortices were finally required to have an average normalized vorticity larger than 3.

	\begin{figure*}[t]
		\centering
		\includegraphics[width=\linewidth]{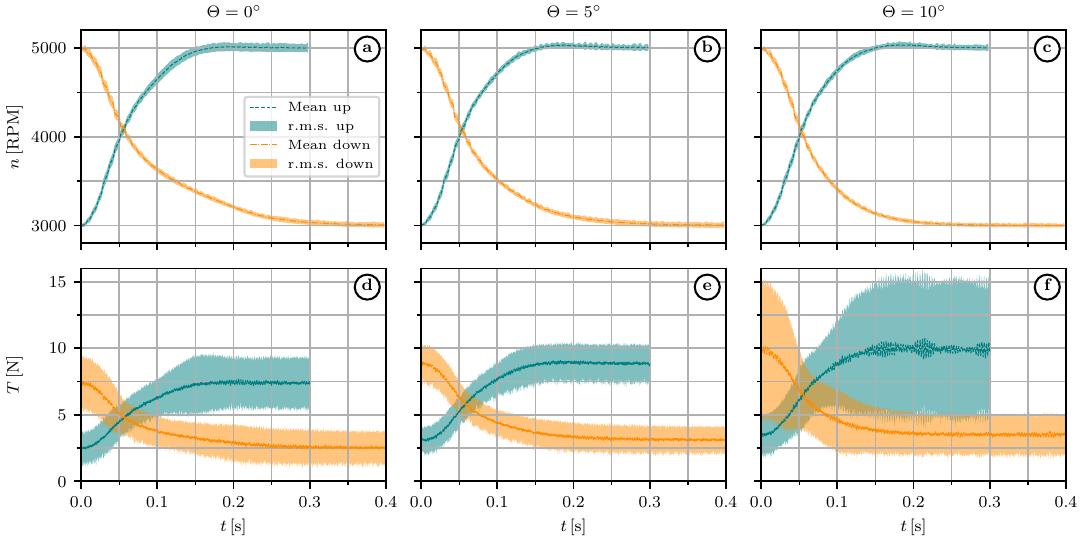}
		\caption{Load measurements: Raw, ensemble-averaged step up and down system response in the rotational speed $n$ (a--c) and thrust $T$ (d--f). The system responses are displayed with the r.m.s. of fluctuations as the shaded area. The columns correspond to different pitch offsets according to the title, and the legend in subfigure (a) applies to all other subfigures.}
		\label{fig:aero_steps}
	\end{figure*}
	
	\section{Results}
	\label{sec:5}

    The discussion starts with a steady-state BEMT analysis (Sec.~\ref{sec:5_0}) to estimate the effects of increasing the pitch angle and radius with the blade holders. Afterward, the experimental results are presented, starting with the thrust coefficient (Sec.~\ref{sec:5_1}) and blade deformation (Sec.~\ref{sec:5_2}). The section closes with the spanwise flow data (Sec.~\ref{sec:5_3}) and the slipstream data (Sec.~\ref{sec:5_4}).
	
	\subsection{Steady BEMT analysis}
	\label{sec:5_0}
	
	The blade holders take the propeller blade into off-design conditions as the sectional pitch and Reynolds number change. To estimate the effect of these changes, we conduct a BEMT analysis of the original and modified propellers. Aerodynamic coefficients are computed with XFOIL \cite{Drela1989}, using the $e^N$ method \cite{VanIngen2008} with $N = 9$.
	
	The BEMT formulation follows \cite{Ning2013}, using 201 spanwise elements. Lift coefficients account for delayed stall under rotation \cite{Himmelskamp1945} through a flat plate model \cite{Snel1992}, adapted for high tip speed ratios \cite{Marten2013}. The drag is unmodified, and the sectional lift is predicted as:
	\begin{align}
		\label{eq:3D_corr}
		C_{l,3D} = C_{l,2D} + \frac{3.1}{1 + \lambda(r)^2} \, \frac{c^2}{r^2} \,(2\pi(\alpha - \alpha_0) - C_{l,2D})\,,
	\end{align} with the local inflow ratio $\lambda(r) = v_i(r) / (\Omega r)$, where $C_{l,2D}$ is the lift coefficient of XFOIL, $v_i$ is the induced velocity, and $\alpha_0$ is the zero-lift angle of attack. Similar correction models resulted in better predictions for propellers at similar Reynolds numbers \cite{McCrink2015, MacNeill2017}.
    
	\begin{figure*}[t]
		\centering
		\includegraphics[width=\linewidth]{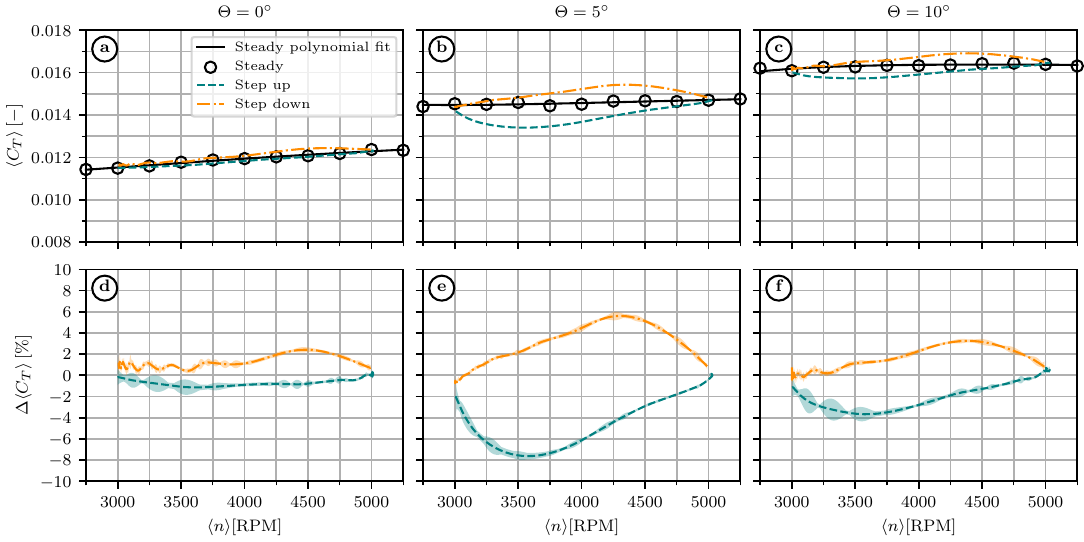}
		\caption{Load measurements: Thrust coefficient as a function of the rotational speed (a--c) and relative difference between transient and steady conditions (d--f) for each pitch offset according to the column title. The shaded area in subfigures (d--f) represents \SI{95}{\percent} confidence intervals, and the legend in subfigure (a) applies to all subfigures.}
		\label{fig:aero_cT}
	\end{figure*}

	\begin{figure*}[t]
		\centering
		\includegraphics[width=\linewidth]{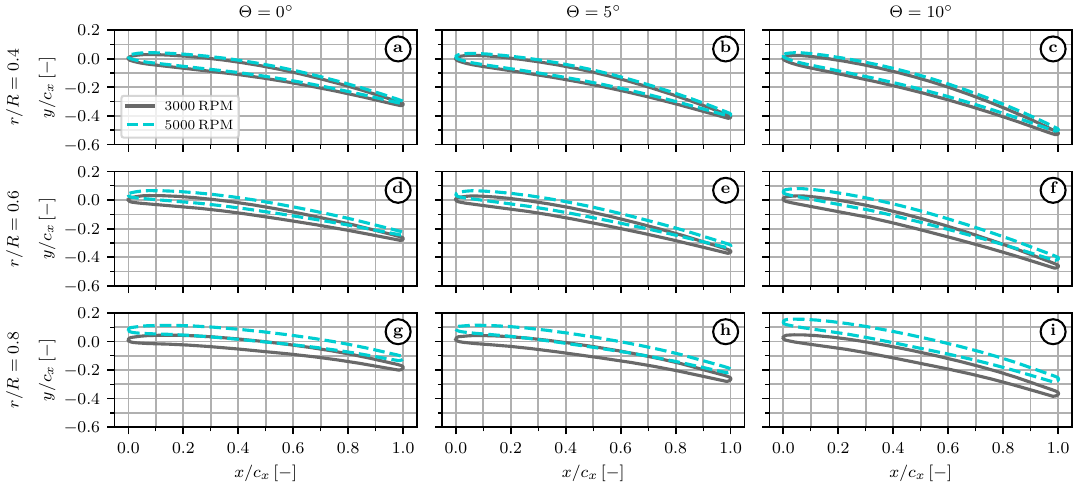}
		\caption{Spanwise measurements: Detected airfoil position at the minimum and maximum rotational speeds in the camera reference frame from the fitting of the local airfoil shape to the reflection of the mean image. The rows correspond to different radial sections according to each row label, and the columns correspond to different pitch offsets. The legend in subfigure (a) applies to all subfigures.}
		\label{fig:elast_air_shapes}
	\end{figure*}
	
	\begin{figure*}[t]
		\centering
		\includegraphics[width=\linewidth]{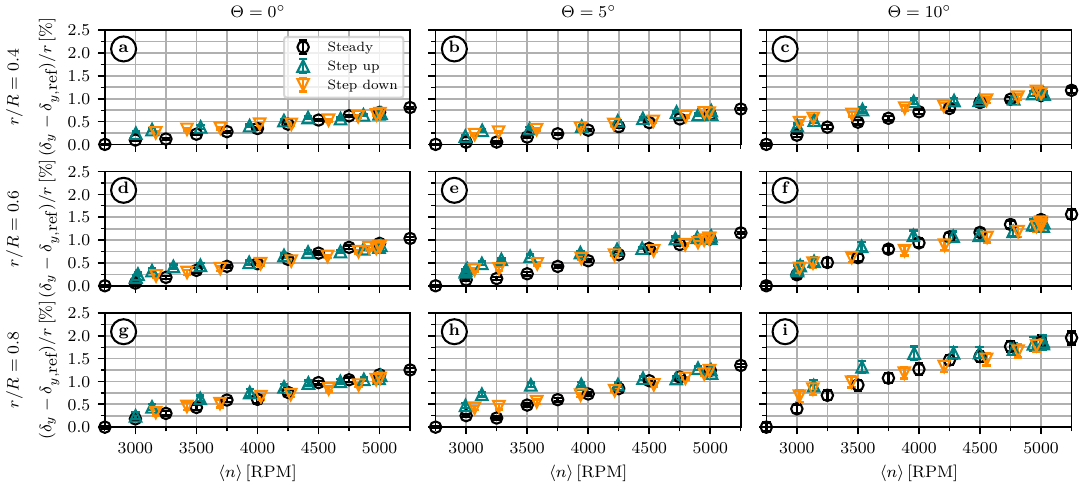}
		\caption{Spanwise measurements: Ensemble-averaged vertical bending of the trailing edge $\delta_y$ normalized by the local radius over the rotational speed for steady and transient conditions. The data are given with respect to the reference at \SI{2750}{\rpm} in steady conditions, and the error bars represent the r.m.s. of the bending fluctuations. The rows correspond to different radial sections according to each row label, and the columns correspond to different pitch offsets. The legend in subfigure (a) applies to all subfigures.}
		\label{fig:elast_disp_mean_blade}
	\end{figure*}
	
	\begin{figure*}[t]
		\centering
		\includegraphics[width=\linewidth]{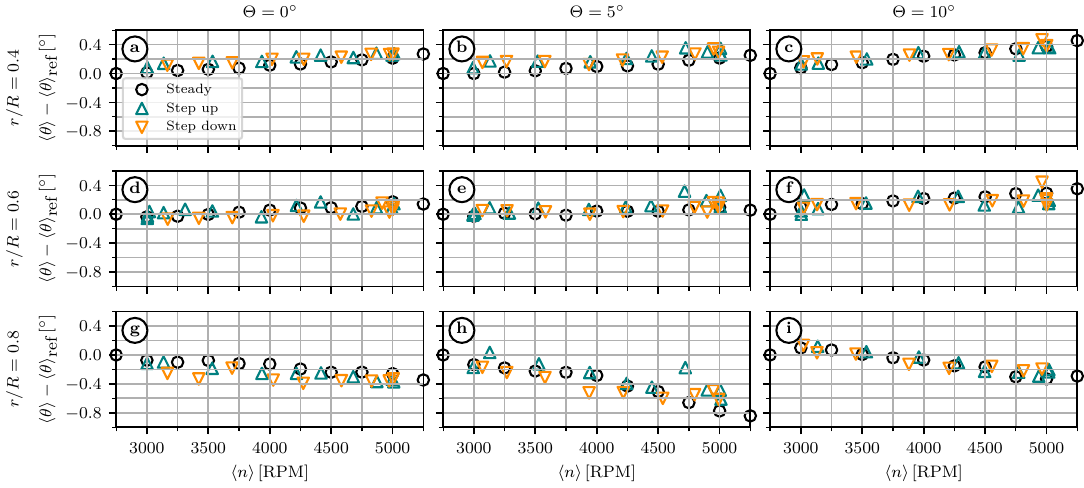}
		\caption{Spanwise measurements: Local twist angle due to blade loading $\langle \theta \rangle$ over the rotational speed. The data are given with respect to the reference at \SI{2750}{\rpm} in steady conditions. The rows correspond to different radial sections according to each row label, and the columns correspond to different pitch offsets. The legend in subfigure (a) applies to all subfigures.}
		\label{fig:elast_angles_blade}
	\end{figure*}

    Figure~\ref{fig:bemt} compares the sectional chord-based Reynolds number $\textrm{Re}_c$, angle of attack $\alpha$, and lift coefficient $C_l$ at 3000 and \SI{5000}{\rpm}. As expected, the larger radius leads to a larger Reynolds number, while differences among the pitch offsets remain small. Most sections operate above $\textrm{Re}_c = \num{40000}$, where the polar curves only show changes in the stall behavior, resulting in minor changes in the angle of attack and lift between minimum and maximum rotational speed.

    The pitch offset has a much stronger effect. At $\Theta = \SI{0}{\degree}$, the angle of attack at the mid-span increases by approximately \SI{2}{\degree} relative to the baseline propeller. At $\Theta = \SI{5}{\degree}$, the mid-span approaches two-dimensional stall, and for $\Theta = \SI{10}{\degree}$ a substantial portion of the blade stalls beyond what the simple flat plate model \eqref{eq:3D_corr} can correct (Fig.~\ref{fig:bemt}c,d). The lift correction is strongest toward the inner span and weakens as $c/r$ decreases, resulting in strongly varying lift coefficients (Fig.~\ref{fig:bemt}e,f).

    These predictions provide an initial map of the operating regimes, ranging from predominantly pre-stall conditions at $\Theta=\SI{0}{\degree}$ to extensive stall of the outer sections at $\Theta=\SI{10}{\degree}$. Their weak dependence on rotational speed is only partly observed in the PIV data, with the results in Sec.~\ref{sec:5_3} showing substantial changes between minimum and maximum speeds at $r/R=0.6$.
	
	\subsection{Aerodynamic loads}
	\label{sec:5_1}
	
	Figure~\ref{fig:aero_steps}(a--c) depicts the system response in rotational speed under transient conditions. The rise time is similar for all three pitch offsets, since inertia dominates the aerodynamic torque. The fall time decreases with increasing pitch, because the drag increases. Two distinct phases emerge during the fall time, particularly visible for $\Theta = \SI{0}{\degree}$ with a rapid decrease during the first \SI{0.06}{\second}, followed by an approximately linear decay between 0.1 and \SI{0.25}{\second}. This two-stage behavior is attributed to the ESC, since it was also observed without the propeller. Fluctuations remain comparable across all cases.

    The thrust (Fig.~\ref{fig:aero_steps}d--f), in contrast, fluctuates three times more for $\Theta = \SI{10}{\degree}$ than the other two pitch offsets. This behavior is consistent with increased unsteady loading due to flow separation at the outer span (see Sec.~\ref{sec:5_3}). Still, the mean response remains comparable and tracks the mean rotational speed. The fluctuations grow with rotational speed, in both steady and transient conditions, indicating that they are not specific to transient conditions.

    To check hysteresis effects, Fig.~\ref{fig:aero_cT}(a--c) depicts the thrust coefficients in steady and transient conditions for all three pitch offsets. For $\Theta = \SI{0}{\degree}$, the thrust coefficient increases linearly with rotational speed, with transient conditions almost perfectly matching steady ones. In contrast, the steady thrust coefficient remains almost constant at $\Theta = 5$ and $\SI{10}{\degree}$, and the acceleration (deceleration) produces lower (higher) thrust than the steady-state condition. The relative difference $\Delta \langle C_T \rangle$ \eqref{eq:rel_ct} quantifies this hysteresis in Fig.~\ref{fig:aero_cT}(d--f). At $\Theta = \SI{0}{\degree}$, maximum differences of \SI{-1}{\percent} (\SI{2.5}{\percent}) are observed during acceleration (deceleration), giving a quasi-steady trend. The differences grow to $-8$ and \SI{5}{\percent} at $\Theta = \SI{5}{\degree}$, with extrema near 3600 and \SI{4250}{\rpm}, respectively. At $\Theta = \SI{10}{\degree}$, the differences reduce to $-4$ and \SI{3}{\percent} at similar speeds.
	
	This trend opposes dynamic inflow theory \cite{Pitt1981, Peters1991, Peters1995}. A lag in the induced velocity causes higher (lower) sectional angles of attack during acceleration (deceleration), leading to higher (lower) thrust. Similar trends were also observed in numerical simulations \cite{Grava2026a} and for low-pitch propellers \cite{Ratz2026b}. The present experimental data indicate the opposite trend, and the inflow results in Sec.~\ref{sec:5_4_2} show no clear inflow effect.
	
	\subsection{Blade deformation}
	\label{sec:5_2}

	Figure~\ref{fig:elast_air_shapes} presents the detected airfoil positions in the camera reference frame at 3000 and \SI{5000}{\rpm}, for the three radial sections and pitch offsets. The vertical bending dominates the twist. The bending follows the expected structural trend, increasing from root (Fig.~\ref{fig:elast_air_shapes}a--c) to tip (Fig.~\ref{fig:elast_air_shapes}g--i), as the chord narrows and the blade velocity rises. Higher pitch offsets (Fig.~\ref{fig:elast_air_shapes}c,f,i) amplify bending further through the increased sectional lift.

    To relate these steady trends to the transient cases, Fig.~\ref{fig:elast_disp_mean_blade} compares the vertical bending $\delta_y $ with respect to the reference image at \SI{2750}{\rpm}. The steady data follow a linear increase with rotational speed. The step up matches this behavior, slightly exceeding steady values between 3000 and \SI{4000}{\rpm}, but differences stay below \SI{0.5}{\percent} and never surpass the maximum steady value of approximately \SI{2}{\percent}. The step down follows similar trends, with increased bending at low and quasi-steady trends at high rotational speed (Fig.~\ref{fig:elast_disp_mean_blade}a--c). This behavior appears consistent, with generally small r.m.s. fluctuations, and high-speed video analysis (not shown) resolved no higher-order bending modes. Large fluctuations appear only for $r/R=0.6$ and 0.8 at $\Theta=\SI{10}{\degree}$ (Fig.~\ref{fig:elast_disp_mean_blade}f,i), in both steady and transient conditions, consistent with flow separation of the outer sections (Sec.~\ref{sec:5_3_1}).
	
	\begin{figure*}[t]
		\centering
		\includegraphics[width=\linewidth]{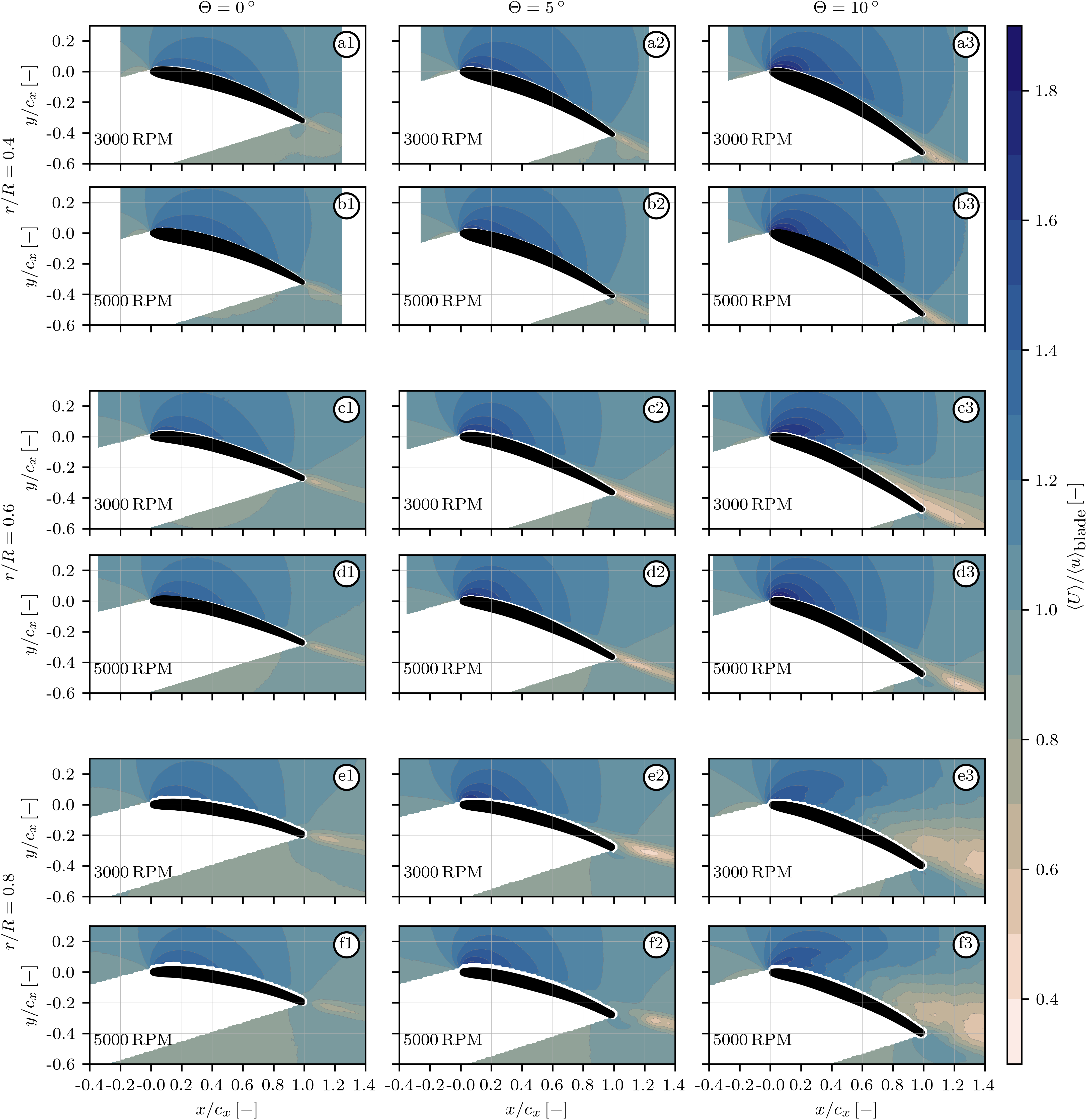}
		\caption{Spanwise measurements: Ensemble-averaged velocity magnitude $\langle U \rangle/\langle u \rangle_\textrm{blade}$ in steady conditions in the coordinate system centered at the leading edge. The columns correspond to different blade pitch offsets according to the titles, and the colorbar is shared among all subfigures. The row blocks correspond to different spanwise locations according to the ordinate label and give results at minimum and maximum rotational speeds, i.e., (a1--a3) correspond to \SI{3000}{\rpm} and (b1--b3) to \SI{5000}{\rpm} for $r/R=0.4$ and equivalent for (c1--c3) and (d1--d3) for $r/R=0.6$, and (e1--e3) and (f1--f3) for $r/R=0.8$.}
		\label{fig:span_min_max_vel}
	\end{figure*}
	
	\begin{figure*}[t]
		\centering
		\includegraphics[width=\linewidth]{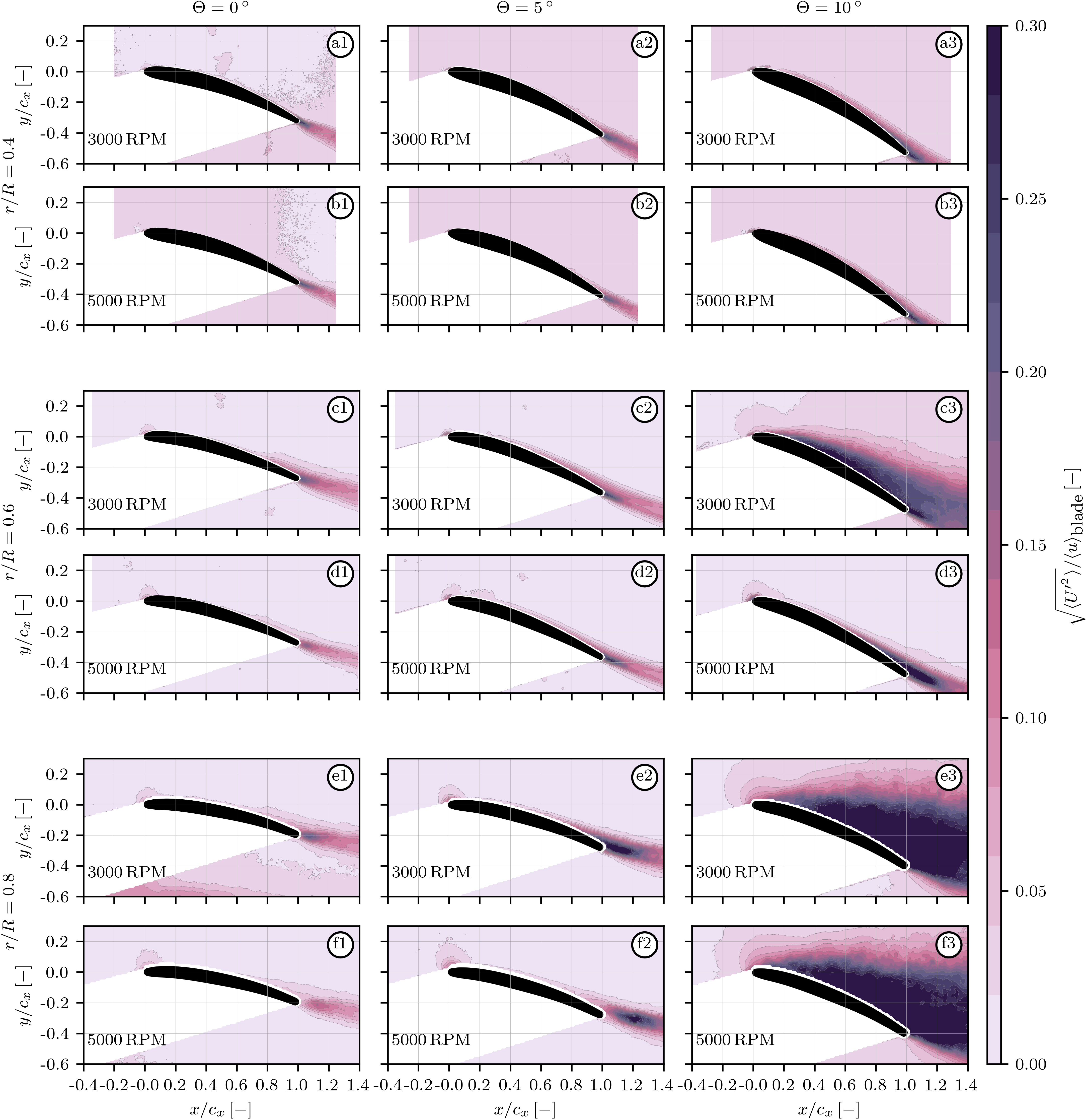}
		\caption{Same as Fig.~\ref{fig:span_min_max_vel} except for the r.m.s. of magnitude fluctuations $\sqrt{\langle U^{\prime^2} \rangle} / \langle u \rangle_\textrm{blade}$.}
		\label{fig:span_min_max_rms}
	\end{figure*}
    
    \begin{figure*}[t]
		\centering
		\includegraphics[width=\linewidth]{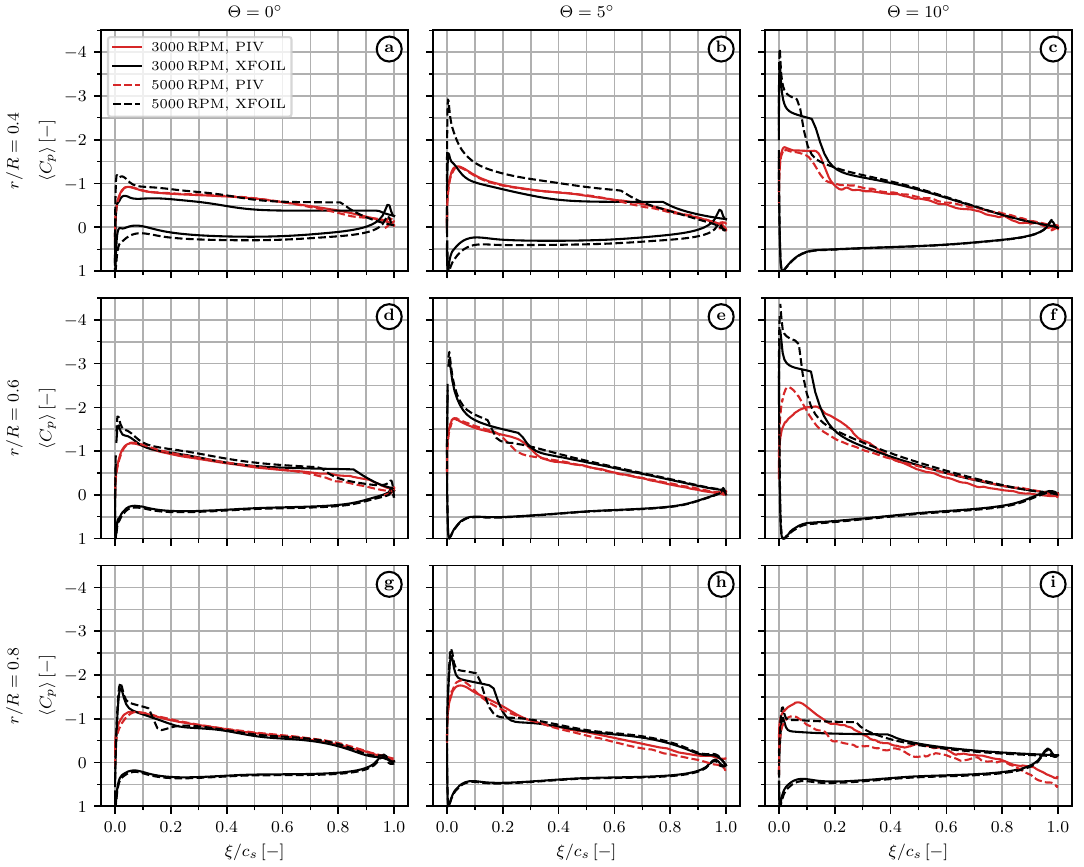}
		\caption{Spanwise measurements: Surface pressure coefficient $\langle C_p \rangle$ from PIV compared to XFOIL in steady conditions at minimum and maximum rotational speed. The rows correspond to different radial sections according to each row label, and the columns correspond to different pitch offsets. The legend in subfigure (a) applies to all subfigures.}
		\label{fig:span_pressure}
	\end{figure*}
	
	The twist angle $\langle \theta \rangle$, depicted in Fig.~\ref{fig:elast_angles_blade}, recovers similar linear trends. The inner section at $r/R=0.4$ (Fig.~\ref{fig:elast_angles_blade}a--c) shows larger twist during both transient conditions, consistent with the bending--twist coupling of a twisted blade. The twist reduces at $r/R=0.6$ and reverses sign at $r/R=0.8$, which appears linked to the blade tip geometry rather than pitch, as it is consistent across all pitch offsets. The largest difference occurs at $\Theta = \SI{5}{\degree}$, reaching steady twists of approximately \SI{-0.8}{\degree} compared to \SI{-0.4}{\degree} for the other pitch offsets. The reduced magnitude for $\Theta = \SI{10}{\degree}$ is consistent with reduced loading under flow separation.

    Thus, the blades exhibit measurable departures in transient conditions compared to steady ones, but both bending and twist remain small compared to rotor radius and flow angles. Simulations with comparable bending revealed no appreciable difference in thrust \cite{Russell2017}, which is why the blade bending is considered quasi-steady and unlikely to be the primary source of hysteresis.
	
	\subsection{Spanwise flow fields}
	\label{sec:5_3}

    The spanwise flow fields are characterized by comparing conditions at minimum and maximum rotational speed to establish the separation characteristics (Sec.~\ref{sec:5_3_1}). Afterward, the r.m.s. of magnitude fluctuations (Sec.~\ref{sec:5_3_2}), the sectional circulation (Sec.~\ref{sec:5_3_3}), and wake vorticity (Sec.~\ref{sec:5_3_4}) follow the flow from the blade surface into the wake.

	\begin{figure*}[t]
		\centering
		\includegraphics[width=\linewidth]{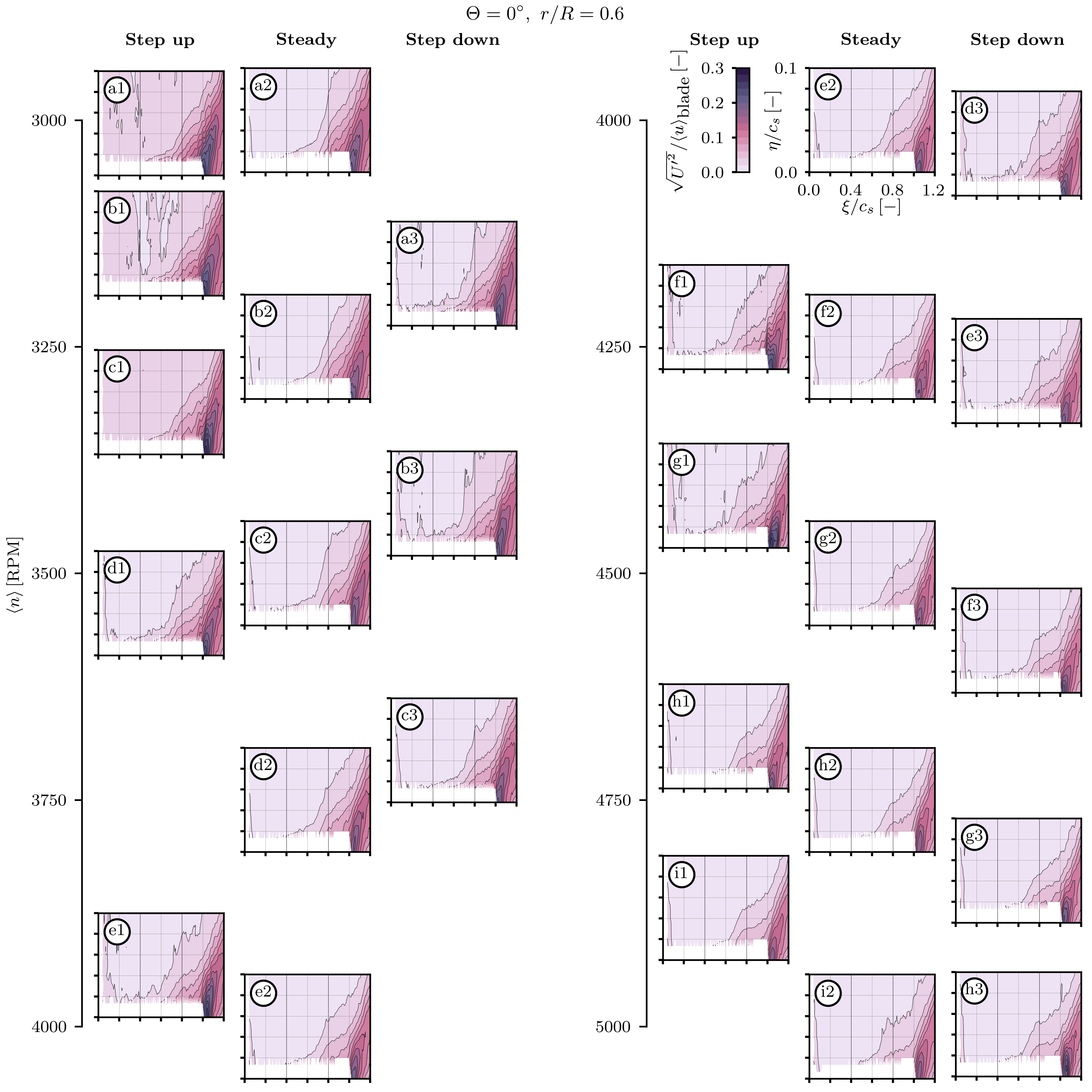}
		\caption{Spanwise measurements: ensemble-averaged velocity r.m.s. of magnitude fluctuations $\sqrt{\langle U^{\prime^2} \rangle}/\langle u \rangle_\textrm{blade}$ in steady and transient conditions for $\Theta = \SI{0}{\degree}$ at $r/R=0.6$. The three columns show data during step up (a1--i1), steady (a2--i2) and step down (a3--h3) where the subplots are spaced according to their rotational speed on the ordinate axis. The left column shows data between 3000 and \SI{4000}{\rpm} and the right one between 4000 and \SI{5000}{\rpm}, where the steady reference at \SI{4000}{\rpm} (e2) is shown in both columns for reference. Data are shown in surface-normal coordinates (see Fig.~\ref{fig:span_wake_surface_unwrapping}) in the domain $\xi/c_s \times \eta/c_s \in [0, 1.2] \times [0, 0.1]$, given in subfigure (e2) in the right column. The colorbar range is shared among all subfigures and matches Fig.~\ref{fig:span_min_max_rms}.}
		\label{fig:span_wake_rms_00}
	\end{figure*}
	
	\begin{figure*}[t]
		\centering
		\includegraphics[width=\linewidth]{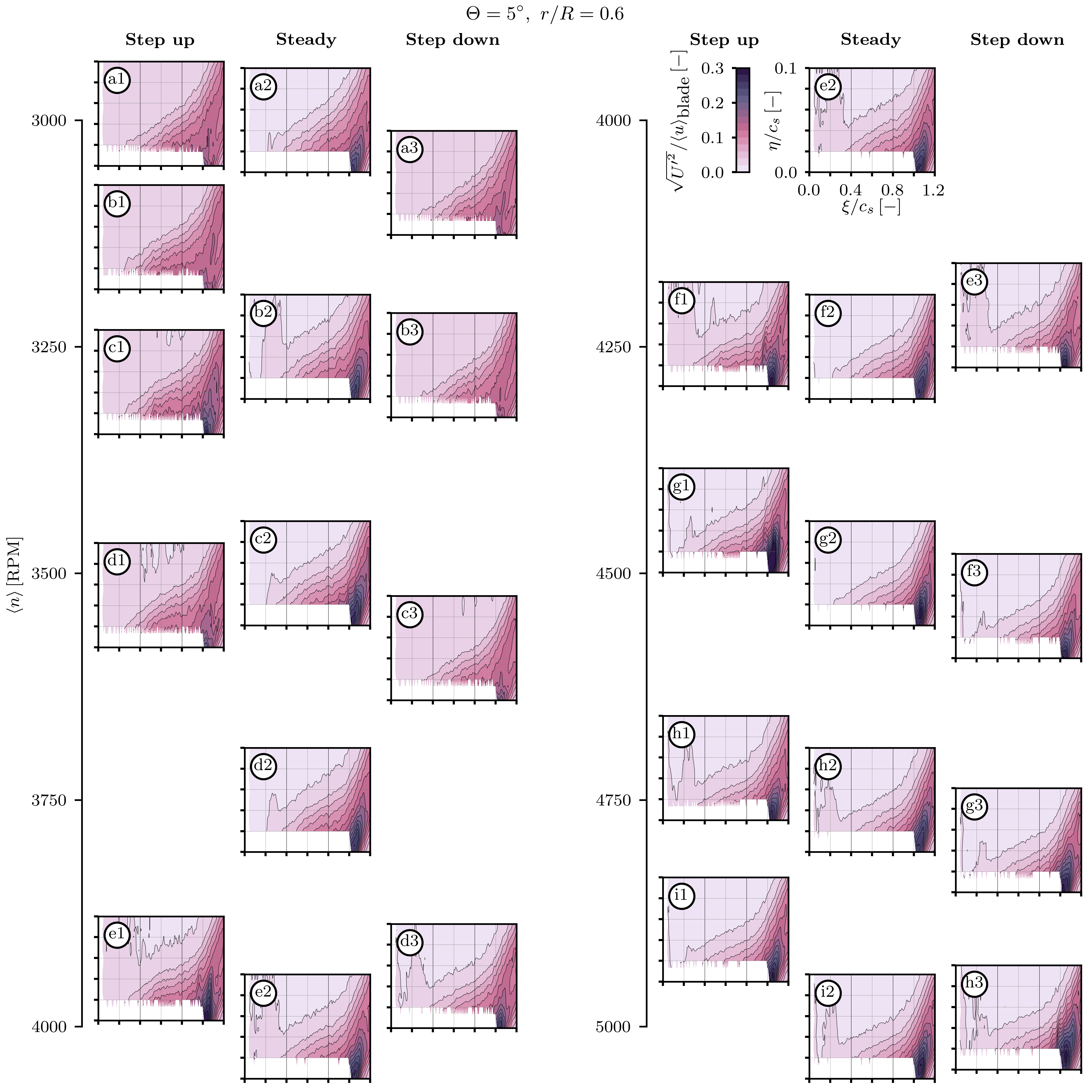}
		\caption{Spanwise measurements: Same as Fig.~\ref{fig:span_wake_rms_00} for $\Theta = \SI{5}{\degree}$ at $r/R=0.6$.}
		\label{fig:span_wake_rms_05}
	\end{figure*}
	
	\begin{figure*}[t]
		\centering
		\includegraphics[width=\linewidth]{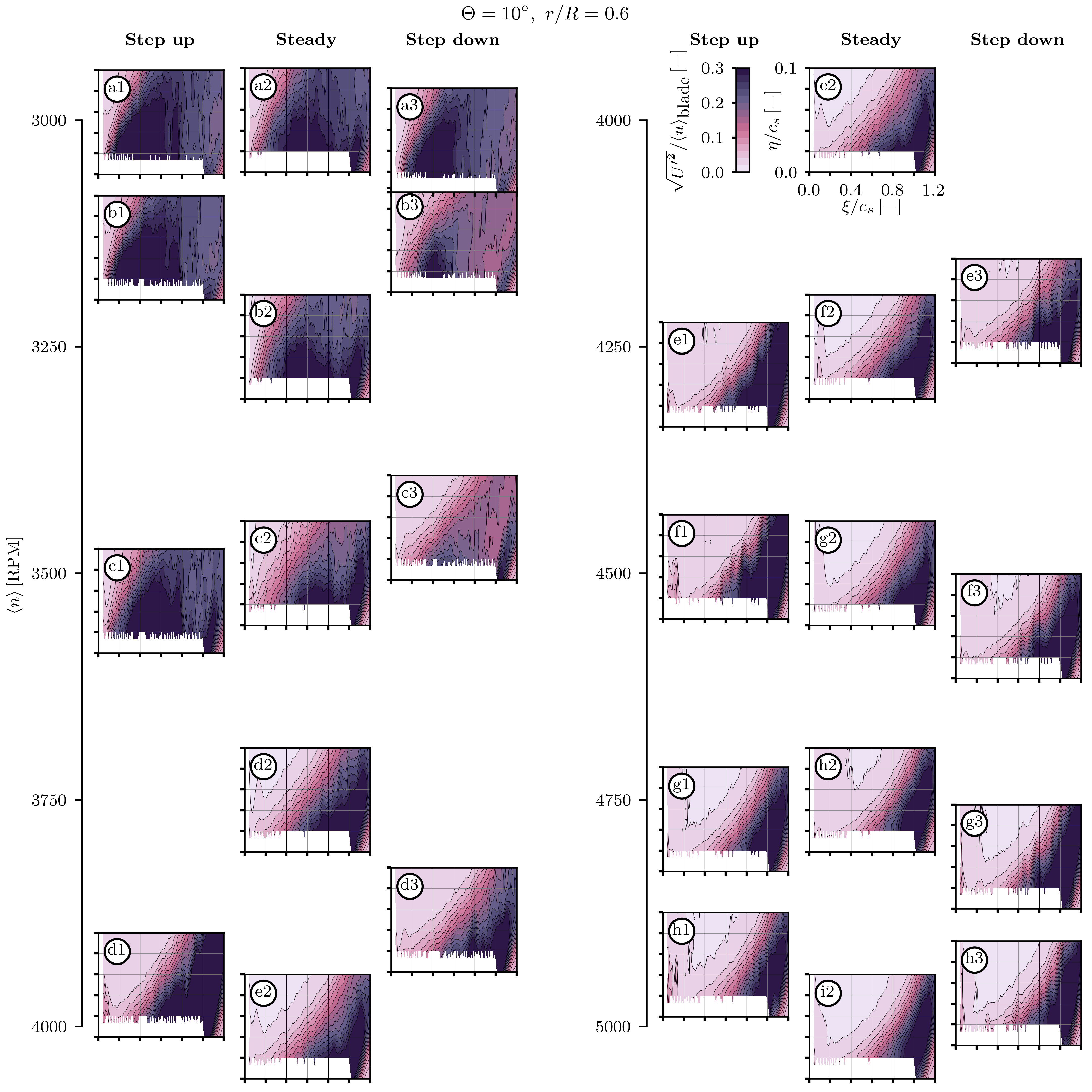}
		\caption{Spanwise measurements: Same as Fig.~\ref{fig:span_wake_rms_00} for $\Theta = \SI{10}{\degree}$ at $r/R=0.6$.}
		\label{fig:span_wake_rms_10}
	\end{figure*}
	
	\subsubsection{Steady velocity fields}
	\label{sec:5_3_1}

    Figures~\ref{fig:span_min_max_vel} and \ref{fig:span_min_max_rms} compare the velocity magnitude $\langle U \rangle / \langle u\rangle_\textrm{blade}$ and r.m.s. of magnitude fluctuations $\sqrt{\langle U^{\prime^2} \rangle} / \langle u \rangle_\textrm{blade}$ at 3000 and \SI{5000}{\rpm}. Figure~\ref{fig:span_pressure} complements the analysis with the pressure coefficient $\langle C_p \rangle$, derived both from PIV and 2D XFOIL simulations. The XFOIL pressure coefficients were computed by estimating the local flow angle and changing the angle of attack to match the slope or transition point. Together, these data indicate flow regimes ranging from rear-chord separation, through reattached separated flow forming an LSB, to fully separated flow. A pressure plateau followed by rapid recovery indicates a reattached LSB. Because the near-wall resolution limits precise identification of separation and reattachment, the flow states are interpreted qualitatively, using the first location with $\sqrt{\langle U'^2\rangle}/\langle u\rangle_\textrm{blade}\geq0.02$ as a proxy for transition onset. This threshold exceeds the estimated noise floor of \SI{1.24}{\percent} (see \ref{app:4_2}).

    At $r/R=0.4$, pressure plateaus are discernible across all three pitch offsets. The transition onset moves from $\xi/c_s=0.80$ to 0.60 and 0.15 as $\Theta$ increases from 0 to 5 and \SI{10}{\degree} (Figs.~\ref{fig:span_min_max_rms}a1--b3 and \ref{fig:span_pressure}a--c). The suction peak likewise strengthens at higher pitch offsets (Figs.~\ref{fig:span_min_max_vel}a1--b3 and \ref{fig:span_pressure}a--c) with no evidence of stall, consistent with rotational augmentation models \cite{Himmelskamp1945,Snel1992,Marten2013}. Differences between minimum and maximum rotational speed remain small, with maximum differences in the estimated transition location of less than \SI{10}{\percent} of the chord.

    At $r/R=0.6$, the $\Theta = 0$ and \SI{5}{\degree} cases show shorter LSBs with estimated transition at $\xi/c_s \approx 0.5$ and 0.3 (Figs.~\ref{fig:span_min_max_rms}c1--d2 and \ref{fig:span_pressure}d,e), with similar results at both speeds. For $\Theta = \SI{10}{\degree}$, clear changes from minimum to maximum speed are observed. At \SI{3000}{\rpm}, fluctuations exceed \SI{30}{\percent} of the blade velocity across almost the entire chord, while the flow resembles a reattached LSB at \SI{5000}{\rpm}, although the pressure profiles do not identify a clear pressure plateau (Figs.~\ref{fig:span_min_max_rms}c3,d3 and \ref{fig:span_pressure}f).

    At $r/R=0.8$, the $\Theta = \SI{10}{\degree}$ case remains fully separated at both speeds, which could explain why no clear LSB is observed at $r/R=0.6$. The other pitch offsets likewise show no LSB topology (Fig.~\ref{fig:span_pressure}g--i). Both radial flow and tip vortex effects may contribute to these observations, as both are affect LSB formation at the blade tip \cite{Li2024b,Grande2021,Toppings2021}.

    Reynolds number effects remain small throughout, except for $\Theta = \SI{10}{\degree}$ at $r/R=0.6$, where fluctuations become restricted to the aft portion of the blade. In the cases with LSB, the separation point moves upstream by less than \SI{10}{\percent} between minimum and maximum speed. The shear layer also widens at lower speeds, consistent with observations for airfoils (e.g., \cite{OMeara1987}) and rotating blades \cite{Fava2024a,Fava2024b}.
	
	\subsubsection{Transient velocity fluctuations}
	\label{sec:5_3_2}

    Figures~\ref{fig:span_wake_rms_00}--\ref{fig:span_wake_rms_10} compare the fluctuation fields at $r/R=0.6$ in surface-normal coordinates (Fig.~\ref{fig:span_wake_surface_unwrapping}). The other spans follow similar trends, and results are discussed where relevant. Across all cases, the transients primarily affect the shear layer and near-blade wake, without large-scale differences associated with dynamic stall (e.g., \cite{McCroskey1976,Mulleners2011}). Yet, the differences remain systematic, despite the small nondimensional accelerations (Table~\ref{tab:exp_params}).

    For $\Theta = \SI{0}{\degree}$, both transients show an earlier onset of elevated fluctuations below \SI{4000}{\rpm} (Fig. \ref{fig:span_wake_rms_00}a1--e1,a3--d3) with larger values around the trailing edge region. These similarities are notable because the nondimensional acceleration at this speed reaches 0.005 for the step up while the step down is five times smaller at 0.001. Above \SI{4000}{\rpm}, the transition onset recovers quasi-steady values around $\xi / c_s \approx 0.5$, but the r.m.s. values near the trailing edge at $\xi /c_s \approx 0.9$ remain elevated in transient conditions (Fig. \ref{fig:span_wake_rms_00}f1,g1,d3--f3). After approximately \SI{4600}{\rpm}, both recover quasi-steady trends in the entire domain even though the step down records the maximum nondimensional deceleration at this point. The near wake ($\xi /c_s \approx 1.1)$ responds differently, remaining quasi-steady for the step down throughout and elevated for the step up until approximately \SI{4400}{\rpm}, when quasi-steady values are recovered. The surface and wake fluctuations therefore distinguish aspects of the response that the thrust does not reveal.

    For $\Theta = \SI{5}{\degree}$ (Fig. \ref{fig:span_wake_rms_05}), the transition point moves upstream to $\xi / c_s \approx 0.3$, but the overall behavior remains similar. Both transients exhibit an earlier transition onset and larger fluctuations over the blade. The step down recovers quasi-steady conditions after \SI{4000}{\rpm}, while differences persist for the step up until \SI{4500}{\rpm}. The near-wake trend is reversed. Lower fluctuations are recovered in transient conditions, with both step up and step down matching steady values after \SI{4000}{\rpm}.
    
	\begin{figure*}[t]
		\centering
		\includegraphics[width=\linewidth]{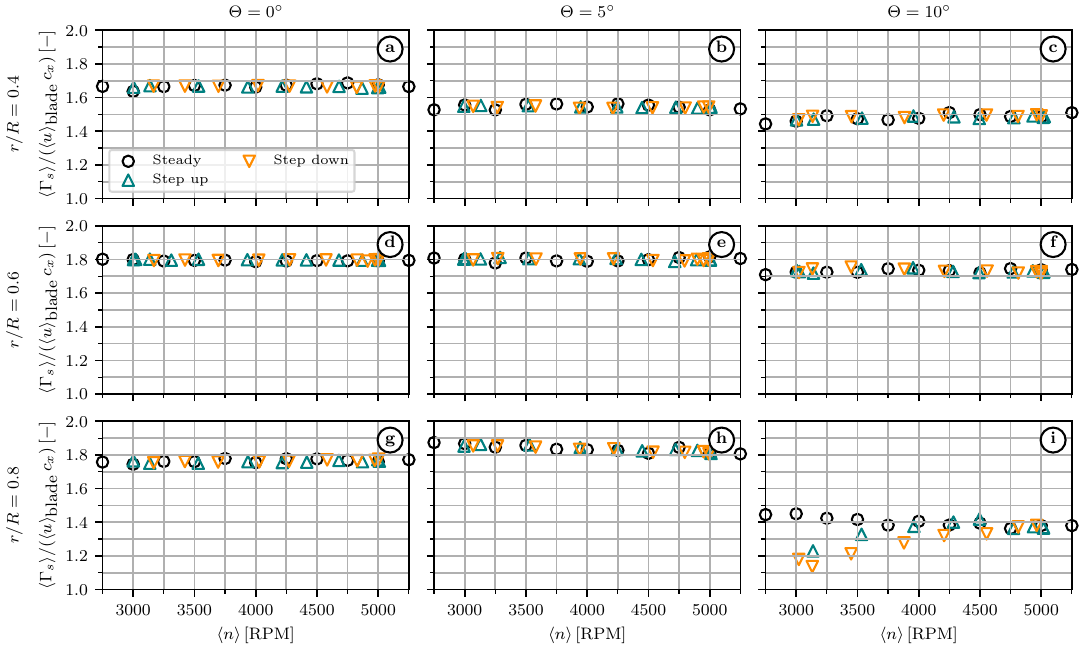}
		\caption{Estimation of the sectional circulation $\langle \Gamma_s \rangle $ around the local airfoil sections, normalized by the local projected blade chord $c_x$ and blade velocity $\langle u \rangle_\textrm{blade}$. The columns correspond to different blade pitch offsets and the rows to different spans, and the legend in subfigure (a) applies to all subfigures.}
		\label{fig:span_local_circ}
	\end{figure*}

    The $\Theta = \SI{10}{\degree}$ case most clearly shows a delayed change in separation state (Fig. \ref{fig:span_wake_rms_10}). At 3000 and \SI{3250}{\rpm} (Fig. \ref{fig:span_wake_rms_10}a2,b2), the steady fields contain a large region of strong fluctuations centered around $\xi/c_s \approx 0.6$, accompanied by high wake fluctuations. With increasing speed, this region moves downstream, and a coherent shear layer forms around \SI{3750}{\rpm} (Fig. \ref{fig:span_wake_rms_10}d2), with estimated transition around $\xi /c_s \approx 0.2$. During the start of the acceleration, the region of strong fluctuations is centered more upstream around $\xi/c_s \approx 0.4$ with smaller fluctuations than in steady conditions. Around \SI{3500}{\rpm}, a shear layer starts to develop in steady conditions (Fig. \ref{fig:span_wake_rms_10}c2), while the step up still shows elevated fluctuations in the fore portion of the blade. At \SI{4000}{\rpm}, the step up also exhibits a coherent shear layer, but wake fluctuations remain higher until \SI{4750}{\rpm}. The deceleration mirrors this behavior with quasi-steady values above \SI{4000}{\rpm}. Most pronounced differences occur around \SI{3450}{\rpm} (Fig. \ref{fig:span_wake_rms_10}c3), where both mid-chord and wake fluctuations are weaker than during steady conditions. At the lowest speeds (Fig. \ref{fig:span_wake_rms_10}a3,b3), the surface maximum lies again around $\xi /c_s \approx 0.4$ with a weaker wake than in steady conditions.

    Data at the other spans reinforce this picture. At $r/R=0.4$, the $\Theta = \SI{10}{\degree}$ case resembles the $\SI{5}{\degree}$ case, while the outer section remains fully separated. Across all pitch offsets and all spans, the largest differences occur at low speeds, near the start (end) of the acceleration (deceleration), even though the nondimensional acceleration differs by a factor of five between them. This indicates a dependence on rotational speed and Reynolds number rather than the acceleration magnitude.

    The behavior near \SI{3500}{\rpm} at $\Theta=\SI{10}{\degree}$ is particularly revealing, but its near-surface mechanism remains unresolved. Sensitivity is already apparent in the first acceleration frame, acquired only \SI{1}{\milli\second} after the step begins (Sec.~\ref{sec:2_4}); the wake also differs from the steady reference near the end of deceleration. Thus, a small acceleration number does not by itself ensure a quasi-steady response near a change in separation state, consistent with observations on low-Reynolds-number airfoils and wings \cite{Toppings2024,Toppings2025}.
	
	\subsubsection{Local airfoil circulation}
    \label{sec:5_3_3}

    The circulation $\langle \Gamma_s \rangle / (\langle u \rangle_\textrm{blade}\,c_x)$ in Fig.~\ref{fig:span_local_circ} remains almost perfectly constant for all spans and all pitch offsets, with steady and transient conditions separated by less than \SI{1}{\percent}. The one exception is $\Theta = \SI{10}{\degree}$ at $r/R=0.8$, where transient circulation drops by up to \SI{25}{\percent} at rotational speeds under \SI{4000}{\rpm}, with step up and step down closely following each other.

    For a time-varying oncoming velocity, thin-airfoil theory predicts quasi-steady circulation, plus unsteady corrections from shed wake and apparent mass, which grow with velocity amplitude and reduced frequency \cite{Isaacs1945,vanderWall1992}. At the present accelerations, this correction falls below the measurement scatter, so the agreement bounds the magnitude of the transient rather than validating the theory. These formulations assume attached flow, a condition the data violate to some degree at every section and comprehensively for $\Theta = \SI{10}{\degree}$ at $r/R=0.8$, where flow separation without reattachment also invalidates the Kutta condition \cite{Strangfeld2016}. This corresponds to the single case with considerable differences from steady conditions.
	
	\begin{figure*}[t]
		\centering
		\includegraphics[width=\linewidth]{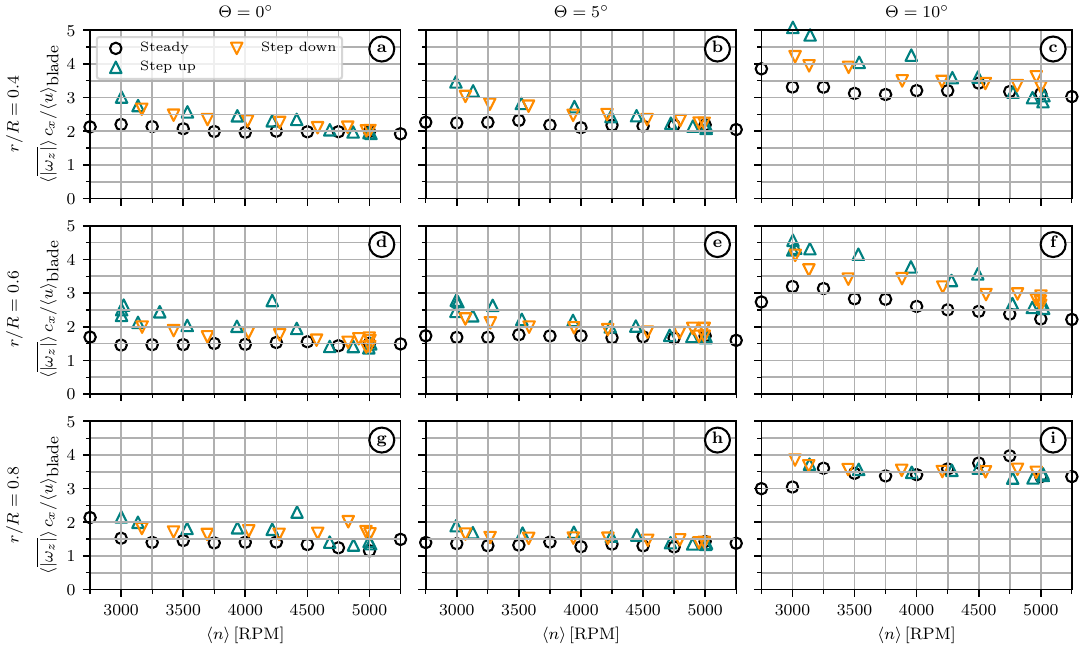}
		\caption{Spanwise measurements: Average norm of the vorticity  in the wake region $x/c_x > 0.05,\,y/c_x < 0.4$ in the trailing-edge coordinate system. The columns correspond to different blade pitch offsets and the rows to different spans. The vorticity is normalized by the local projected chord $c_x$ and the local blade velocity $\langle u \rangle_\textrm{blade}$, and the legend in subfigure (a) applies to all subfigures.}
		\label{fig:span_wake_vort}
	\end{figure*}

	\subsubsection{Airfoil wake vorticity}
	\label{sec:5_3_4}
    
    Circulation therefore remains constant while the near-wake r.m.s. does not, which motivates investigating the far-wake vorticity. The absolute vorticity $\langle \overline{\vert \omega_z \vert} \rangle c_x/\langle u \rangle_\textrm{blade}$ is depicted in Fig.~\ref{fig:span_wake_vort}. Transient effects are likewise visible, with vorticity values up to \SI{50}{\percent} above steady conditions at low speeds, with the difference diminishing at higher rotational speed. The discrepancy is largest at $r/R=0.4$ (Fig.~\ref{fig:span_wake_vort}a--c) and $r/R=0.6$ (Fig.~\ref{fig:span_wake_vort}d--f), while the outer span at $r/R=0.8$ (Fig.~\ref{fig:span_wake_vort}g--i) remains largely quasi-steady.

    This pattern follows the earlier onset and stronger fluctuations over the blade: the separated shear layer develops over a larger part of the chord and is accompanied by higher wake vorticity. The immediate near-wake fluctuations show a less direct correspondence. At $\Theta=\SI{10}{\degree}$ and $r/R=0.6$, for example, their reduction at the start of acceleration is not reflected in the wake-vorticity measure. The averaging region partly explains this difference: it emphasizes the broader shear layer and starts at $x/c_x>0.05$ downstream of the trailing edge, reducing the contribution of the thin near-blade wake.

    Both transients increase this absolute-vorticity measure at low speeds. The shared trend is consistent with changes in the separated shear layer, but it does not distinguish the sign of shed vorticity or exclude an unsteady circulatory contribution. The interpretation is therefore that separation-state changes contribute to the wake response, in line with studies of low-Reynolds-number airfoils \cite{Strangfeld2016,Greenblatt2023}.

	\subsection{Slipstream flow fields}
    \label{sec:5_4}

    The slipstream measurements extend the analysis beyond the blade sections. As before, mean velocity fields are analyzed to establish flow topology (Sec.~\ref{sec:5_4_1}), followed by the rotor-disk inflow (Sec.~\ref{sec:5_4_2}) and the trajectories and circulation of the tip vortices (Sec.~\ref{sec:5_4_3}).
	
	\subsubsection{Mean velocity fields}
	\label{sec:5_4_1}
	
	\begin{figure*}[t]
		\centering
		\includegraphics[width=\linewidth]{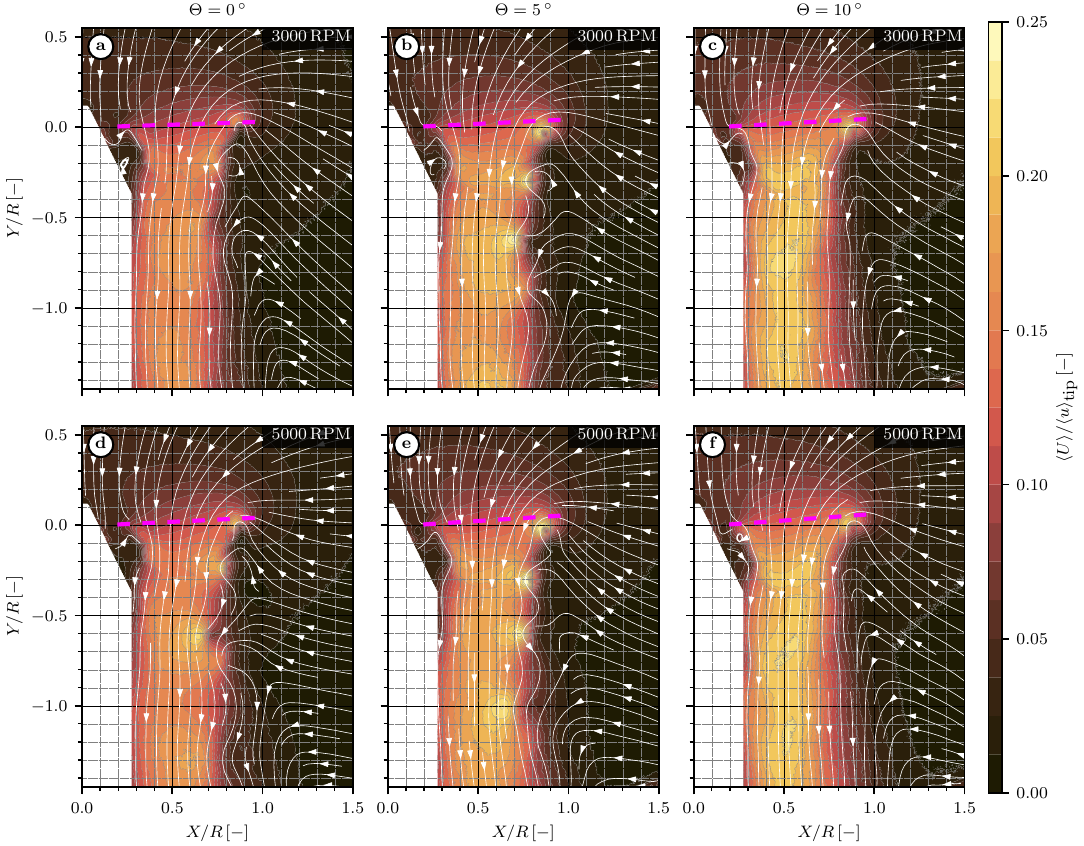}
		\caption{Slipstream measurements: ensemble-averaged velocity magnitude $\langle U \rangle / \langle u \rangle_\textrm{tip}$ at minimum (a--c) and maximum (d--f) rotational speed of the maneuver. The columns correspond to different blade pitch offsets according to the titles, and the colorbar is shared among all subfigures. The magenta line indicates the rotor disk, as depicted in Fig.~\ref{fig:inflow_methodology}.}
		\label{fig:wake_min_max_vel}
	\end{figure*}

    At 3000 and \SI{5000}{\rpm}, the mean fields show a similar downwash structure at all pitch offsets, with tip vortices visible through phase-locking (Fig.~\ref{fig:wake_min_max_vel}). In the mean fields, up to four vortices can be distinguished at $\Theta=\SI{5}{\degree}$, compared with three at \SI{0}{\degree} and one at \SI{10}{\degree}. The less coherent tip vortex at \SI{10}{\degree} is consistent with the extensive outer-span separation. Vortices also appear stronger at higher rotational speeds, but the instantaneous tracking reveals an important distinction: stronger meandering at low speeds spreads the cores in the ensemble average (Sec.~\ref{sec:5_4_3}). Their prominence in the mean field therefore cannot be read directly as vortex strength.
    
    The meandering tip vortex likewise smears the mean velocity fields, masking differences between steady and transient conditions, particularly in the shear layer. Thus, we forgo a detailed discussion of the slipstream profiles. The overall trend is that far from the blade ($Y/R < -0.25$), the slipstream is weaker (stronger) during acceleration (deceleration) as the fluid in this region was shed earlier at lower (higher) rotational speed. Near the blade, the slipstream adapts almost instantly. Both observations are consistent with other works \cite{Grava2026a,Chance2026}.
    	
	\subsubsection{Inflow analysis}
	\label{sec:5_4_2}
	
	\begin{figure*}[t]
		\centering
		\includegraphics[width=\linewidth]{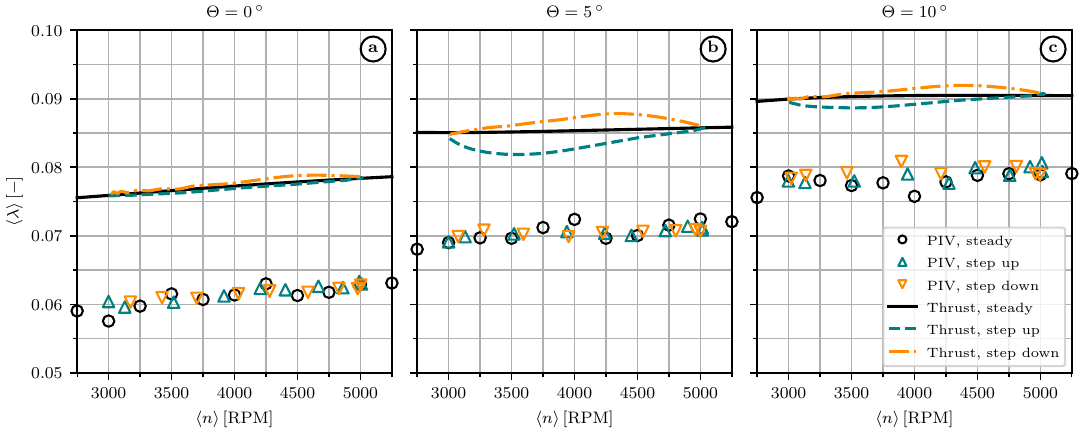}
		\caption{Slipstream measurements: Ensemble-averaged inflow ratio $\langle \lambda \rangle$ over the rotational speed in steady and transient conditions. Markers are given with uncertainties based on 250 samples. For comparison, the inflow estimated from momentum theory ($\langle \lambda \rangle= \sqrt{\langle C_T \rangle/2}$) is also shown in steady and transient conditions. The legend in subfigure (c) applies to all subfigures.}
		\label{fig:inflow_over_rpm}
	\end{figure*}
	
	The inflow analysis confirms rapid near-disk adjustment of the flow. The thrust measurements are converted to an equivalent uniform inflow through $\langle \lambda \rangle = \sqrt{\langle C_T \rangle/2}$ \cite{Leishman2016}, relating the scale of the thrust hysteresis to the inflow. The results in Fig.~\ref{fig:inflow_over_rpm} show offsets with respect to the inflow measured by the PIV. We relate these differences to the root cutout and tip vortices which are not accounted for in momentum theory.
    
    For $\Theta = \SI{0}{\degree}$ (Fig.~\ref{fig:inflow_over_rpm}a), both the thrust and the PIV data indicate quasi-steady behavior. For $\Theta = \SI{5}{\degree}$ (Fig.~\ref{fig:inflow_over_rpm}b), the PIV data again show quasi-steady trends, opposing the hysteresis of the thrust measurements. The pitch offset $\Theta = \SI{10}{\degree}$ (Fig.~\ref{fig:inflow_over_rpm}c) likewise shows no definitive hysteresis, although the larger scatter due to flow separation makes the comparison less conclusive.

    These results agree with the assessment made during the analysis of the thrust coefficient (Sec.~\ref{sec:5_1}): the hysteresis loop in $\langle C_T \rangle$ opposes classic dynamic inflow theory, which predicts smaller/larger inflow during acceleration/deceleration. The acceleration magnitudes further rule out dynamic inflow effects. For the given inflow, the first-order inflow constants stay below 0.3 revolutions \cite{Pitt1981}, an order of magnitude below the ramps spanning \numrange{6}{11} revolutions. Coupled with the associated accelerations, the predicted inflow lag is two orders of magnitude below the measured hysteresis and unlikely to be the dominant source of the observed thrust hysteresis.
    	
	\begin{figure*}[t]
		\centering
		\includegraphics[width=\linewidth]{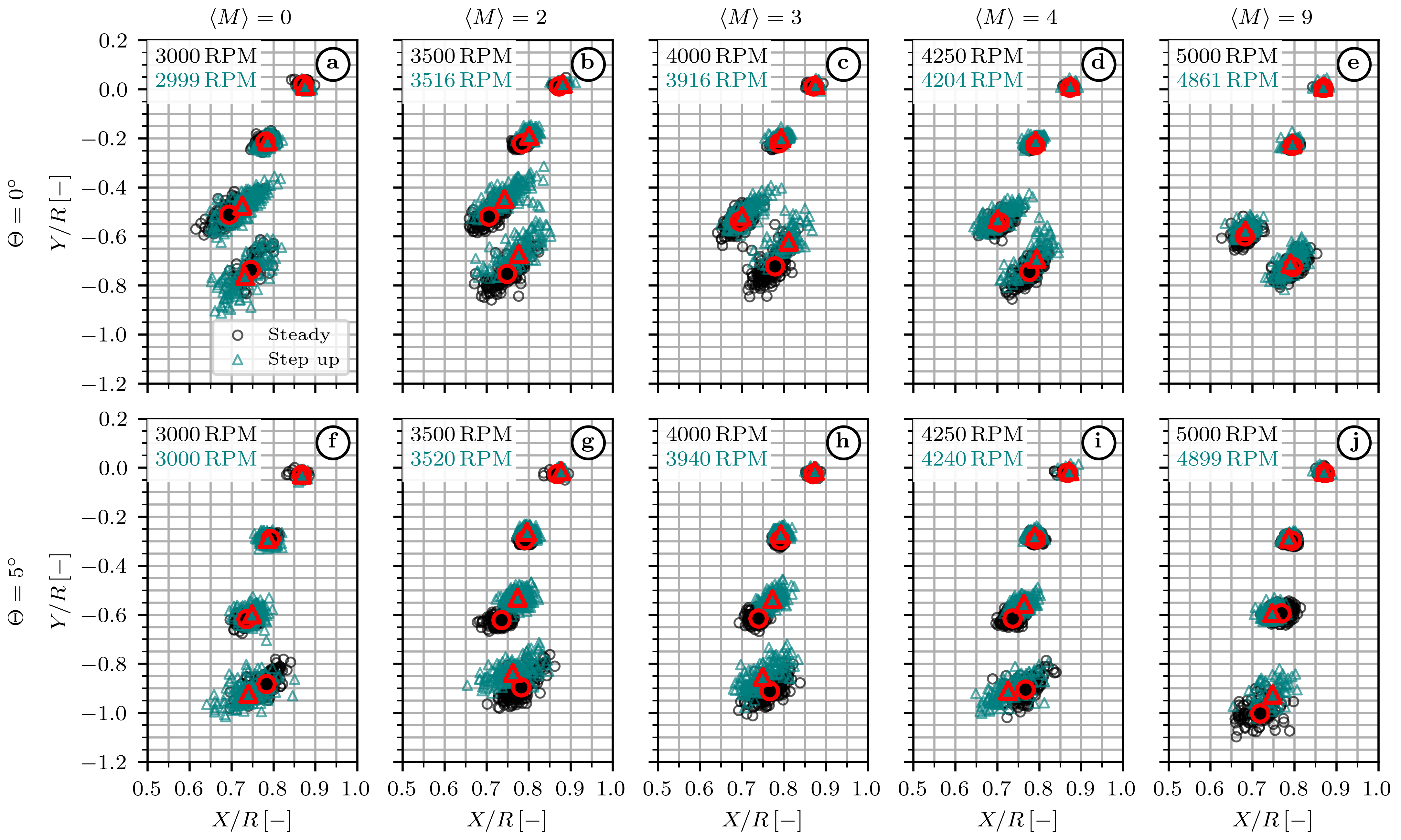}
		\caption{Slipstream measurements: Vortex positions of the first four tip vortices for every instantaneous snapshot during the step up for $\Theta = \SI{0}{\degree}$ (a--e) and \SI{5}{\degree} (f--j). The column titles correspond to the number of revolutions $\langle M \rangle$ since $t=0$. Shown are the tip vortices in transient conditions (teal triangles) together with the closest corresponding steady measurement (black circles). The corresponding median positions are given by the corresponding red marker. The legend in subfigure (a) applies to all subfigures.}
		\label{fig:tip_vort_pos_Up}
	\end{figure*}
	
	\begin{figure*}[t]
		\centering
		\includegraphics[width=\linewidth]{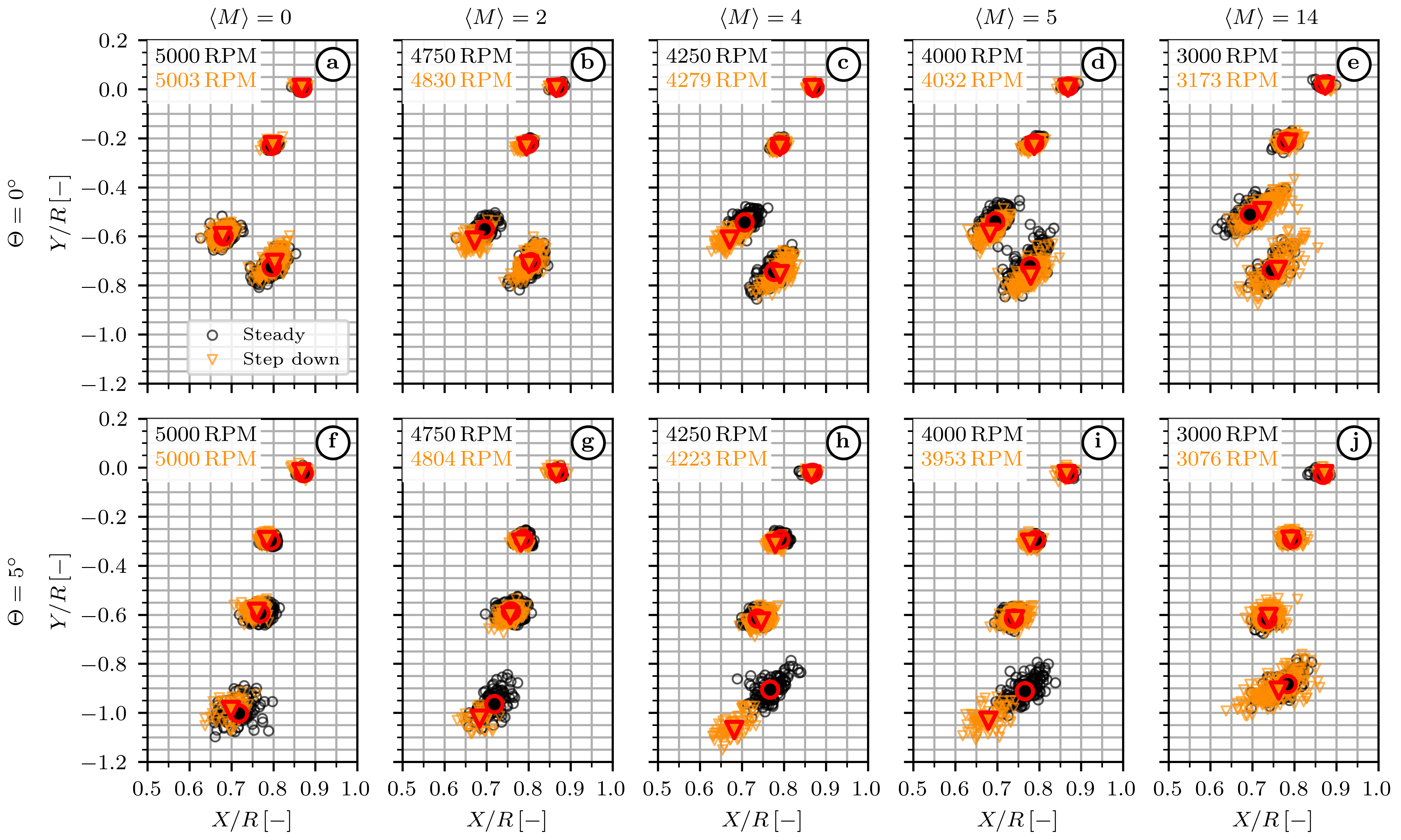}
		\caption{Slipstream measurements: Same as Fig.~\ref{fig:tip_vort_pos_Up}, except for the step down with markers as orange downward triangles.}
		\label{fig:tip_vort_pos_Do}
	\end{figure*}
	
	\begin{figure*}[t]
		\centering
		\includegraphics[width=\linewidth]{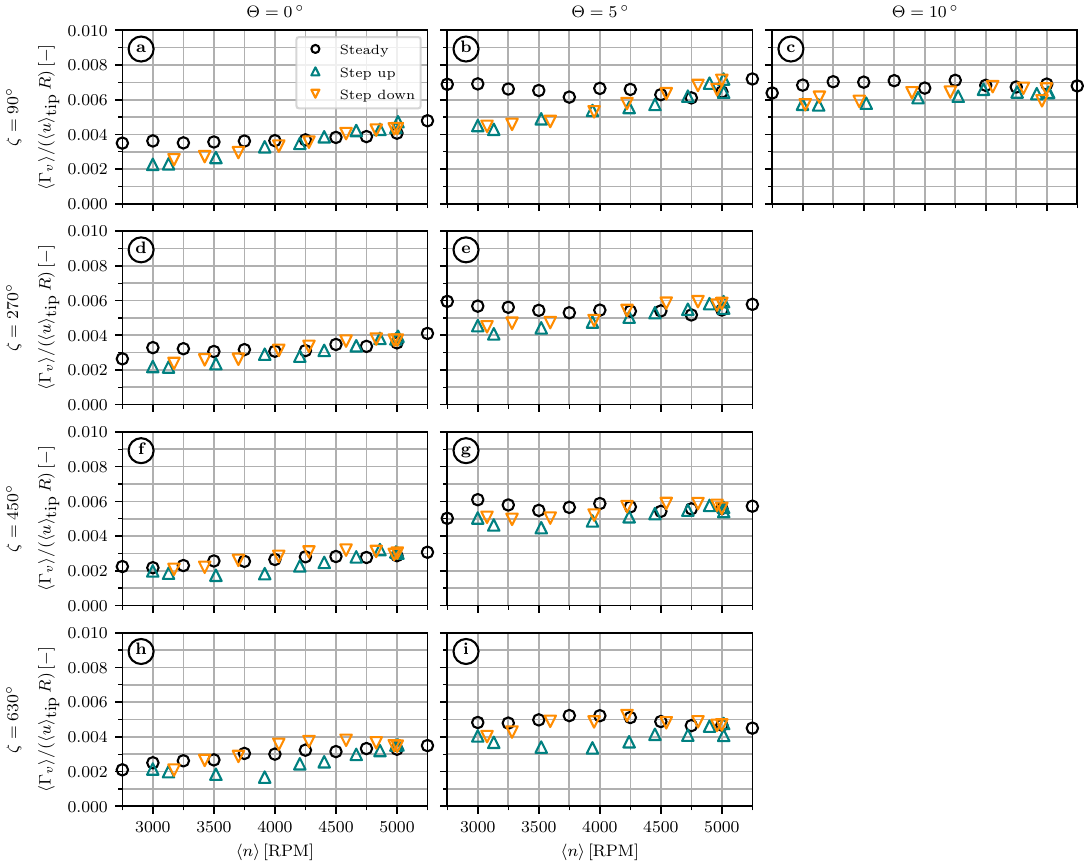}
		\caption{Slipstream measurements: ensemble-averaged circulation $\langle \Gamma_v \rangle / (\langle u \rangle_\textrm{tip}\,R)$ of the tip vortex based on the vorticity surface integral over the rotational speed in steady and transient conditions. The columns correspond to different propeller pitch offsets according to the title, and the rows correspond to different vortex ages $\zeta$ (see Fig.~\ref{fig:inflow_methodology}) according to the labels. For $\Theta = 0$ and \SI{5}{\degree}, four tip vortices were identified, while only one could be reliably found for $\Theta = \SI{10}{\degree}$. The legend in subfigure (a) applies to all subfigures.}
		\label{fig:tip_vort_circ}
	\end{figure*}
	
	\subsubsection{Tip vortex analysis}
	\label{sec:5_4_3}

    The slipstream description closes with the instantaneous tip vortices, beginning with their centers. The first four vortex positions are depicted in Fig.~\ref{fig:tip_vort_pos_Up} for $\Theta = 0$ and \SI{5}{\degree} since only one vortex could be reliably tracked for $\Theta = \SI{10}{\degree}$. The two youngest vortices ($\zeta = 90$ and \SI{270}{\degree}) closely match between steady and step up conditions with minimal spatial scatter, consistent with the rapid flow adjustment near the rotor. The older vortices depart from this pattern and show leapfrogging for $\Theta = \SI{0}{\degree}$. Their spread decreases from approximately $\pm 0.1R$ at 3000 to $\pm0.05R$ at \SI{5000}{\rpm}, both in steady conditions and during the step up. At $\Theta = \SI{5}{\degree}$, no leapfrogging is observed, and the spread of the third vortex remains at $\pm 0.05R$, but its mean position shifts from $(X/R,Y/R) = (0.73,-0.63)$ in steady to $(0.77, -0.55)$ in transient conditions. As the position is almost constant in steady conditions across all rotational speeds, this shift cannot be explained by slipstream adaptation.

    During the step down, the youngest two vortices likewise closely follow the steady positions for both pitch offsets (Fig.~\ref{fig:tip_vort_pos_Do}). Leapfrogging persists for $\Theta = \SI{0}{\degree}$, but the spread is approximately half that of the step up. At the moment of strongest deceleration (Fig.~\ref{fig:tip_vort_pos_Do}b,c) the vortices shift toward smaller $X$ and $Y$, opposing the shift of the step up. For $\Theta = \SI{5}{\degree}$, the third vortex matches steady conditions, while the fourth vortex shifts toward smaller $X$ and $Y$ values. This trend of displaced older vortices is consistent with observations of numerical simulations \cite{Grava2026a}.

    The youngest vortices reveal a different aspect during transients: their position matches steady conditions, but their circulation does not. At $\zeta = \SI{90}{\degree}$, the normalized circulation $\langle \Gamma_v \rangle / (\langle u \rangle_\textrm{tip}\,R)$, depicted in Fig.~\ref{fig:tip_vort_circ}, remains almost constant across all rotational speeds for each pitch offset, with $\Theta = 5$ and \SI{10}{\degree} having nearly double the circulation of $\Theta = \SI{0}{\degree}$. In transient conditions, circulation reduces by up to \SI{50}{\percent} with largest differences around \SI{3000}{\rpm}, while steady values are recovered around \SI{4500}{\rpm}. The behavior is similar across all pitch offsets with largest departures at \SI{5}{\degree} pitch.

    At $\Theta = 0$ and \SI{5}{\degree}, the second vortex ($\zeta = \SI{270}{\degree})$ shows similar but smaller reductions. For the third and fourth vortices ($\zeta = 450$ and \SI{630}{\degree}), the step down closely follows steady conditions, while the step up still exhibits a weaker circulation. This difference can be explained by the normalization, as the vortices were shed at a different blade velocity. The step up has larger accelerations (Table~\ref{tab:exp_params}), leading to slower slipstream adaptation than the step down and hence larger differences. Such downstream history effects were also reported in numerical simulations \cite{Grava2026a}.

    The youngest vortex is more distinctive. Where numerical simulations predicted quasi-steady changes \cite{Grava2026a}, the experiments show the biggest differences. The reduced circulation occurs during both transients and at low rotational speeds, matching the behavior of the spanwise data, where differences were largest at low speeds, both for the fluctuations over the blade and in the wake. This correspondence suggests a role of the evolving separation state in the tip-vortex formation. Steady sectional polars as used in \cite{Grava2026a} do not account for the history of the local flow state; establishing a clear connection requires resolving the near-surface separation dynamics.
    
	\section{Conclusions and perspectives}
	\label{sec:6}

    This study investigated the transient aerodynamics of a small propeller operating at low Reynolds numbers during acceleration and deceleration between 3000 and \SI{5000}{\rpm}. A propeller assembled from commercial propeller blades was tested with additional pitch offsets of $\Theta=0$, 5, and \SI{10}{\degree}. Thrust measurements were combined with phase-locked particle image velocimetry (PIV) at 40, 60, and \SI{80}{\percent} span and in the meridian plane, providing ensemble-resolved flow statistics at selected instants during the transients.

At $\Theta=\SI{0}{\degree}$, the thrust coefficient remains close to its steady trend, whereas clear hysteresis appears at the higher pitch offsets. The largest deviations occur at $\Theta=\SI{5}{\degree}$, reaching approximately \num{-8} and \SI{5}{\percent} during acceleration and deceleration, respectively, compared with approximately \num{-4} and \SI{3}{\percent} at $\Theta=\SI{10}{\degree}$. The direction of this hysteresis opposes classic dynamic-inflow theory, where a lag in the induced velocity causes higher (lower) sectional angles of attack during acceleration (deceleration), leading to higher (lower) thrust.

The spanwise measurements reveal markedly different separation states with pitch offset and radial position, ranging from rear-chord separation to reattached separated flow forming a laminar separation bubble (LSB) and fully separated flow. During the transients, the largest differences from steady conditions occur at low rotational speeds, near the beginning of acceleration and the end of deceleration, rather than where the instantaneous acceleration magnitude is largest. These departures appear primarily in the development of the separated shear layer and in the blade wake. In particular, the onset of elevated velocity fluctuations moves upstream, and the wake vorticity increases relative to steady conditions, while the sectional circulation remains nearly unchanged in most cases.

The case $\Theta=\SI{10}{\degree}$ at $r/R=0.6$ is particularly revealing. The steady flow evolves from an extended separated state with downstream reattachment at low rotational speed toward a more LSB-like state at higher rotational speed. During acceleration and deceleration, this evolution follows different paths: the extended separated state persists to higher rotational speed during acceleration, whereas the more compact reattached state persists to lower rotational speed during deceleration. The separation state therefore retains a dependence on the preceding flow history even though the nondimensional acceleration remains quasi-steady.

The meridian-plane measurements show that the mean inflow through the rotor disk remains close to its steady trend during the transients. Within the present operating range, this provides little evidence for a dynamic-inflow contribution of sufficient magnitude to explain the measured thrust hysteresis. Further downstream, however, transient differences remain visible in the slipstream and tip vortices. The tip-vortex circulation can decrease by up to approximately \SI{50}{\percent} relative to the corresponding steady condition, with differences reducing downstream.

Taken together, the measurements indicate that the dominant transient effects are associated less with rotor-scale inflow or blade deformation than with the local state and history of the separated flow over the blade. Reduced-order models for this regime should therefore account not only for the Reynolds-number dependence of the sectional aerodynamics, but also for changes and possible history dependence of the separation state. Resolving the underlying mechanism will require measurements with improved near-wall and temporal resolution, together with denser spanwise coverage. Transition-resolving numerical simulations could complement these measurements by resolving the separation, transition, and reattachment dynamics that remain inaccessible to the present PIV measurements.

    \section*{CRediT authorship contribution statement} \textbf{M. Ratz:} Conceptualization, Methodology, Software, Validation, Formal analysis, Investigation, Resources, Data curation, Writing – original draft, Writing – review \& editing, Visualization. \textbf{C. Schram:} Writing – original draft, Writing – review \& editing, Supervision. \textbf{A. Sciacchitano:} Formal analysis, Investigation, Resources, Writing – original draft, Writing – review \& editing, Supervision, Project administration. \textbf{M. A. Mendez:} Conceptualization, Methodology, Formal analysis, Investigation, Resources, Writing – original draft, Writing – review \& editing, Supervision, Project administration, Funding acquisition.

    \section*{Declaration of competing interest} The authors declare that they have no known competing financial interests or personal relationships that could have appeared to influence the work reported in this paper.
    
    \section*{Declaration of generative AI and AI-assisted technologies in the manuscript preparation process}
    During the preparation of this work, the authors used Claude (Anthropic), ChatGPT (OpenAI), and Grammarly to improve the readability and language of the manuscript. After using these tools/service, the authors reviewed and edited the content as needed and took full responsibility for the publication’s content.

    \section*{Data availability}
    The data underlying the results presented in this study are publicly available on Zenodo \cite{Ratz2026e}.
	
	\section*{Acknowledgments}
	The authors kindly acknowledge Aren Vardanyan for his help in designing the propeller test rig. M.R. is funded by the Fonds de la Recherche Scientifique (F.R.S.-FNRS) through the FRIA grant No. FC57471. M.A.M.'s contributions are supported by the European Research Council (ERC) under the Horizon Europe programme (Starting Grant No. 101165479, RE-TWIST). The data are visualized using perceptually uniform colormaps to prevent distortion of the data and exclusion of readers with color-vision deficiencies \cite{Crameri2020, Crameri2023}.

    \def\bibfont{\footnotesize}
    \setlength{\bibsep}{0pt plus 0.8ex}
    \bibliographystyle{elsarticle-num-names}
    \bibliography{bibliography}

@Misc{Crameri2023,
  author    = {Crameri, Fabio},
  title     = {Scientific colour maps},
  year      = {2023},
  copyright = {MIT License},
  doi       = {10.5281/ZENODO.8409685},
  language  = {en},
  publisher = {Zenodo},
}

@Article{Crameri2020,
  author    = {Crameri, Fabio and Shephard, Grace E. and Heron, Philip J.},
  journal   = {Nature Communications},
  title     = {The misuse of colour in science communication},
  year      = {2020},
  issn      = {2041-1723},
  number    = {5444},
  volume    = {11},
  comment   = {My favorite reading about colormaps. Proposes perceptually uniform colormaps for all sorts of applications.
Also an excellent guide on how to choose the colormaps},
  doi       = {10.1038/s41467-020-19160-7},
  publisher = {Springer Science and Business Media LLC},
}

@Article{Scarano2000,
  author  = {F. Scarano and M. Riethmüller},
  journal = {Experiments in Fluids},
  title   = {Advances in iterative multigrid {PIV} image processing},
  year    = {2000},
  pages   = {S051--S060},
  volume  = {29},
  comment = {Original description of the window deformation technique.},
  doi     = {10.1007/s003480070007},
}

@Article{Westerweel2005,
  author  = {Westerweel, Jerry and Scarano, Fulvio},
  journal = {Experiments in Fluids},
  title   = {Universal outlier detection for {PIV} data},
  year    = {2005},
  pages   = {1096-1100},
  volume  = {39},
  doi     = {10.1007/s00348-005-0016-6},
}

@Article{Sciacchitano2016,
  author     = {Sciacchitano, Andrea and Wieneke, Bernhard},
  journal    = {Measurement Science and Technology},
  title      = {{PIV} uncertainty propagation},
  year       = {2016},
  month      = {08},
  pages      = {084006},
  volume     = {27},
  comment    = {Derived quantity uncertainty quantificiation for PIV applications, also Reynolds stresses},
  doi        = {10.1088/0957-0233/27/8/084006},
  readstatus = {read},
}

@Book{Raffel2018,
  author    = {Raffel, M. and Willert, C. E. and Scarano, F. and Kähler, C. and Wereley, S. T. and Kompenhans, J.},
  publisher = {Springer International Publishing AG},
  title     = {Particle {I}mage {V}elocimetry - {A} {P}ractical {G}uide},
  year      = {2018},
  edition   = {Third},
  doi       = {10.1007/978-3-319-68852-7},
}

@Article{Hassanalian2017,
  author   = {M. Hassanalian and A. Abdelkefi},
  journal  = {Progress in Aerospace Sciences},
  title    = {Classifications, applications, and design challenges of drones: A review},
  year     = {2017},
  issn     = {0376-0421},
  pages    = {99--131},
  volume   = {91},
  doi      = {10.1016/j.paerosci.2017.04.003},
}

@Article{Ayamga2021,
  author   = {Matthew Ayamga and Selorm Akaba and Albert Apotele Nyaaba},
  journal  = {Technological Forecasting and Social Change},
  title    = {Multifaceted applicability of drones: A review},
  year     = {2021},
  issn     = {0040-1625},
  pages    = {120677},
  volume   = {167},
  doi      = {10.1016/j.techfore.2021.120677},
}

@Article{Faical2014,
  author   = {Bruno S. Faiçal and Fausto G. Costa and Gustavo Pessin and Jó Ueyama and Heitor Freitas and Alexandre Colombo and Pedro H. Fini and Leandro Villas and Fernando S. Osório and Patrícia A. Vargas and Torsten Braun},
  journal  = {Journal of Systems Architecture},
  title    = {The use of unmanned aerial vehicles and wireless sensor networks for spraying pesticides},
  year     = {2014},
  issn     = {1383-7621},
  number   = {4},
  pages    = {393-404},
  volume   = {60},
  comment  = {Drone Motivation in pesticide control},
  doi      = {10.1016/j.sysarc.2014.01.004},
}

@Article{Goebel2015,
  author  = {Goebel, Michael E. and Perryman, Wayne L. and Hinke, Jefferson T. and Krause, Douglas J. and Hann, Nancy A. and Gardner, Steve and LeRoi, Donald J.},
  journal = {Polar Biology},
  title   = {A small unmanned aerial system for estimating abundance and size of {A}ntarctic predators},
  year    = {2015},
  pages   = {619--630},
  volume  = {38},
  comment = {Use propeller UAVs to assess population size in Antarctica
Drone Motivation in scientific application},
  doi     = {10.1007/s00300-014-1625-4},
}

@Article{Greenwood2019,
  author  = {Greenwood, William W. and Lynch, Jerome P. and Zekkos, Dimitrios},
  journal = {Journal of Infrastructure Systems},
  title   = {Applications of {UAV}s in civil infrastructure},
  year    = {2019},
  number  = {2},
  pages   = {04019002},
  volume  = {25},
  comment = {Drone Motivation in civil engineering},
  doi     = {10.1061/(ASCE)IS.1943-555X.0000464},
}

@Article{Hodgson2016,
  author    = {Hodgson, Jarrod C. and Baylis, Shane M. and Mott, Rowan and Herrod, Ashley and Clarke, Rohan H.},
  journal   = {Scientific Reports},
  title     = {Precision wildlife monitoring using unmanned aerial vehicles},
  year      = {2016},
  number    = {1},
  pages     = {22574},
  volume    = {6},
  comment   = {Drone Motivation in wildlife monitoring application},
  doi       = {10.1038/srep22574},
  publisher = {Nature Publishing Group UK London},
}

@Article{Lucieer2014,
  author  = {Lucieer, Arko and Turner, Darren and King, Diana H. and Robinson, Sharon A.},
  journal = {International Journal of Applied Earth Observation and Geoinformation},
  title   = {Using an {U}nmanned {A}erial {V}ehicle {(UAV)} to capture micro-topography of {A}ntarctic moss beds},
  year    = {2014},
  pages   = {53--62},
  volume  = {27},
  comment = {Use fly-by of drones to map 3D structures in antarctica through a model from motion approach from computer vision.
Drone Motivation in scientific application},
  doi     = {10.1016/j.jag.2013.05.011},
}

@Article{Restas2015,
  author  = {Restas, Agoston},
  journal = {World Journal of Engineering and Technology},
  title   = {Drone applications for supporting disaster management},
  year    = {2015},
  number  = {3},
  pages   = {316--321},
  volume  = {3},
  comment = {Drone Motivation in disaster management
nuclear accidents, dangerous material releases, floods, earthquakes andforest fires},
  doi     = {10.4236/wjet.2015.33C047},
}

@Article{Shouji2021,
  author   = {Cheng Shouji and Reza {Alidoost Dafsari} and Seung-Hwa Yu and Yong Choi and Jeekeun Lee},
  journal  = {Computers and Electronics in Agriculture},
  title    = {Mean and turbulent flow characteristics of downwash air flow generated by a single rotor blade in agricultural drones},
  year     = {2021},
  issn     = {0168-1699},
  pages    = {106471},
  volume   = {190},
  comment  = {Perform hotwire measurements in the downwash of a rotor blade for agriculture
Drone Motivation in agriculture},
  doi      = {10.1016/j.compag.2021.106471},
}

@Article{Tatum2017,
  author   = {Mark C. Tatum and Junshan Liu},
  journal  = {Procedia Engineering},
  title    = {Unmanned {A}ircraft {S}ystem {A}pplications in {C}onstruction},
  year     = {2017},
  issn     = {1877-7058},
  pages    = {167--175},
  volume   = {196},
  comment  = {Drone Motivation in civil engineering},
  doi      = {10.1016/j.proeng.2017.07.187},
}

@Article{Tavana2017,
  author   = {Madjid Tavana and Kaveh Khalili-Damghani and Francisco J. Santos-Arteaga and Mohammad-Hossein Zandi},
  journal  = {Expert Systems with Applications},
  title    = {Drone shipping versus truck delivery in a cross-docking system with multiple fleets and products},
  year     = {2017},
  issn     = {0957-4174},
  pages    = {93-107},
  volume   = {72},
  comment  = {Drone Motivation in delivery applications},
  doi      = {10.1016/j.eswa.2016.12.014},
}

@Article{Viegas2022,
  author  = {Viegas, Carlos and B. Chehreh, Babak and Andrade, José and Lourenço, João},
  journal = {Journal of Intelligent \& Robotic Systems},
  title   = {Tethered {UAV} with {C}ombined {M}ulti-rotor and {W}ater {J}et {P}ropulsion for {F}orest {F}ire {F}ighting},
  year    = {2022},
  number  = {21},
  volume  = {104},
  comment = {Drone Motivation in fighting forest fires},
  doi     = {10.1007/s10846-021-01532-w},
}

@Article{Lissaman1983,
  author  = {Lissaman, P. B. S.},
  journal = {Annual Review of Fluid Mechanics},
  title   = {Low-{R}eynolds-{N}umber {A}irfoils},
  year    = {1983},
  number  = {1},
  pages   = {223--239},
  volume  = {15},
  comment = {Very nice citation to give some references on laminar separation bubbles which only occur at low Reynolds number},
  doi     = {10.1146/annurev.fl.15.010183.001255},
}

@Article{Burgmann2008,
  author  = {S. Burgmann and W. Schröder},
  journal = {Experiments in Fluids},
  title   = {Investigation of the vortex induced unsteadiness of a separation bubble via time-resolved and scanning {PIV} measurements},
  year    = {2008},
  pages   = {675--691},
  volume  = {45},
  doi     = {10.1007/s00348-008-0548-7},
}

@InProceedings{Pisharoti2023,
  author    = {Naina Pisharoti and Jeremiah Whelchel and W. Nathan Alexander and Stefano Brizzolara},
  booktitle = {Vertical Flight Society’s 79th Annual Forum \& Technology Display},
  title     = {Experimental and {N}umerical {I}nvestigation of {I}solated {S}wept {R}otors for {UAV}s},
  year      = {2023},
  comment   = {Compare swept and unswept propellers.
They conduct noise measurements and the swept propeller has much less broadband noise which is confiemd by URANS simulations with SSG/LRR. They attribute this to lower levels of blade wake interaction},
  doi       = {10.4050/F-0079-2023-17946},
}

@Article{Grande2021,
  author  = {Grande, Edoardo and Romani, Gianluca and Ragni, Daniele and Avallone, Francesco and Casalino, Damiano},
  journal = {AIAA Journal},
  title   = {Aeroacoustic {I}nvestigation of a {P}ropeller {O}perating at {L}ow {R}eynolds {N}umbers},
  year    = {2022},
  number  = {2},
  pages   = {860--871},
  volume  = {60},
  comment = {Extremely comprehensive experimental investigation of the flow field around a propeller; pressure integration based on Oudheusden (PIV based). They use stereo PIV. They neglect the Reynolds stresses because they found them to not contribute a lot.

Low Reynolds number},
  doi     = {10.2514/1.j060611},
}

@Article{Grande2022,
  author  = {Grande, Edoardo and Ragni, Daniele and Avallone, Francesco and Casalino, Damiano},
  journal = {AIAA Journal},
  title   = {Laminar {S}eparation {B}ubble {N}oise on a {P}ropeller {O}perating at {L}ow {R}eynolds {N}umbers},
  year    = {2022},
  number  = {9},
  pages   = {5324--5335},
  volume  = {60},
  comment = {Followup to Grande2021 where they do a bit more analysis and used more measurements, also of the wake. They compare measurements with a semi-empirical model.

They compute the shedding frequency through the 2D auto correlation function of the instantaneous vorticity field in the wake region. Based on frozen turbulence hypothesis, the shedding frequency is computed with the wavelength from the correlation and the advection velocity from the PIV mean flow.},
  doi     = {10.2514/1.J061691},
}

@InProceedings{Dierl2024,
  author    = {Dierl, Wolfgang and Yarusevych, Serhiy and Hain, Rainer and Kähler, Christian J.},
  booktitle = {21st {I}nternational {S}ymposium on {A}pplications of {L}aser and {I}maging {T}echniques to {F}luid {M}echanics},
  title     = {Flow development over an {SD}7003 airfoil at different accelerations from rest},
  year      = {2024},
  comment   = {Accelerate an airfoil from rest with different accelerations
They mention Guerra2023, Nati2015 and Toppings2023 as the most relevant papers
Chord 250mm, towing tank. Measurements were repeated with 400 frames each
When the strong acceleration stops, it takes some time for the reattachment point to reach quasi-steady values. Above a certain threshold, this is independent of the acceleration and depends only on the distance traveled
It takes 8 convective time units (i.e. a distance of 8 airfoil chords) to reach steady state values.
The roll-up location moves up-stream with initial high velocities and then decreases with traveled distance.
The vortex shedding of the LSB during the acceleration phase remains at almost constant frequency},
  doi       = {10.55037/lxlaser.21st.240},
}

@Article{Cutler2015,
  author  = {Cutler, Mark and How, Jonathan P.},
  journal = {Journal of Dynamic Systems, Measurement, and Control},
  title   = {Analysis and {C}ontrol of a {V}ariable-{P}itch {Q}uadrotor for {A}gile {F}light},
  year    = {2015},
  number  = {10},
  pages   = {101002},
  volume  = {137},
  comment = {Full Journal paper of Cutler2011
They demonstrate the flipping capabilities by video
The blades are symmetric to allow inverted flight
There are still discrepancies between the reference command and the position, but the difference is small},
  doi     = {10.1115/1.4030676},
}

@Article{Porter2016,
  author  = {Robert Porter and Bijan Shirinzadeh and Man Ho Choi},
  journal = {Journal of Intelligent \& Robotic Systems},
  title   = {Experimental {A}nalysis of {V}ariable {C}ollective-pitch {R}otor {S}ystems for {M}ultirotor {H}elicopter {A}pplications},
  year    = {2016},
  pages   = {271--288},
  volume  = {83},
  doi     = {10.1007/s10846-015-0311-2},
}

@InProceedings{Asper2024,
  author    = {Matthew S. Asper and Jayant Sirohi},
  booktitle = {AIAA SCITECH 2024 Forum},
  title     = {Testing and {L}ow {O}rder {M}odeling of a {S}peed-{C}ontrolled {R}otor in {H}over},
  year      = {2024},
  number    = {AIAA 2024-2475},
  doi       = {10.2514/6.2024-2475},
}

@InProceedings{Paolillo2024,
  author    = {Paolillo, Gerardo and Astarita, Tommaso},
  booktitle = {21st {I}nternational {S}ymposium on {A}pplications of {L}aser and {I}maging {T}echniques to {F}luid {M}echanics},
  title     = {{PaIRS-UniNa}: a robust and accurate free tool for digital particle image velocimetry and optical camera calibration},
  year      = {2024},
  doi       = {10.55037/lxlaser.21st.120},
}

@Article{Menter2006,
  author  = {F. R. Menter and R. Langtry and S. Völker},
  journal = {Flow, Turbulence and Combustion},
  title   = {Transition {M}odelling for {G}eneral {P}urpose {CFD} {C}odes},
  year    = {2006},
  pages   = {277--303},
  volume  = {77},
  doi     = {10.1007/s10494-006-9047-1},
}

@Article{Ellsworth1991,
  author  = {Ellsworth, R. H. and Mueller, T. J.},
  journal = {Experiments in Fluids},
  title   = {Airfoil boundary layer measurements at low {R}e in an accelerating flow from a nonzero velocity},
  year    = {1991},
  pages   = {368--374},
  volume  = {11},
  comment = {Perform hotwire measurements over a wing between Re 100.000-150.000
Use a variable frequency drive motor with different blockages to get different accelerations
Free-stream is changed by 1-2 m/s (6m/s reference) over 150 ms
The separation bubble shifts opposite in dynamic conditions than in static ones},
  doi     = {10.1007/BF00211791},
}

@Article{Li2024b,
  author   = {Xiangtian Li and Han Wu and Yuhong Li and Peng Zhou and Siyang Zhong and Xin Zhang and Zhaokai Ma},
  journal  = {AIAA Journal},
  title    = {Experimental {S}tudy of the {E}ffect of {B}oundary-{L}ayer {T}ransition on {R}otor {A}eroacoustics},
  year     = {2024},
  number   = {8},
  pages    = {1974--1981},
  volume   = {62},
  comment  = {Use temperature sensitive paint (TSP) to study boundary layer transition over propelers, coupled with noise measurements
Give nice references on how broadband noise is caused by the boundary layer features
Very nice analysis of where the transition happens over the blade, at different profiles.
The blade is investigated untripped and with zigzag strip
Rotor diameter 1m, rpm 600 to 1800, Reynolds 10**5 - 3.7*10**5
Validate their transition identification technique with the flow over a normal wing},
  doi      = {10.2514/1.J063785},
}

@InProceedings{Sekula2022,
  author    = {Sekula, Martin and Russell, Carl},
  booktitle = {Vertical Flight Society’s 78th Annual Forum \& Technology Display},
  title     = {Time-{F}requency {A}nalysis of {E}xperimental and {A}nalytical {H}ub {L}oads of a {R}otor {U}ndergoing a {R}otor {S}peed {C}hange},
  year      = {2022},
  month     = {05},
  comment   = {Perform RPM variations from 1500 to 2500 for 600mm diameter blades
They compare with numerical data of a commercial solver. They need to include different wake models to well-capture the experimental data},
  doi       = {10.4050/F-0078-2022-17526},
}

@Book{Leishman2016,
  author    = {J. Gordon Leishman},
  publisher = {Cambridge University Press},
  title     = {Principles of {H}elicopter {A}erodynamics},
  year      = {2016},
  edition   = {Third},
  isbn      = {9781107013353},
}

@Article{Mortimer2025,
  author    = {Mortimer, Patrick and Sirohi, Jayant},
  journal   = {Experiments in Fluids},
  title     = {Experimental characterization of a rotor to dynamic collective pitch inputs in hover},
  year      = {2025},
  number    = {6},
  volume    = {66},
  comment   = {Perform variable pitch inputs with sinusodials for a large scale rotor
They derive a transfer function for the dynamic inflow state to model the delays
Dynamic inflow state is extracted from PIV measurements.},
  doi       = {10.1007/s00348-025-04037-6},
}

@Article{Boutilier2012,
  author   = {Boutilier, Michael and Yarusevych, Serhiy},
  journal  = {Experiments in Fluids},
  title    = {Parametric study of separation and transition characteristics over an airfoil at low {R}eynolds numbers},
  year     = {2012},
  pages    = {1491--1506},
  volume   = {52},
  comment  = {Perform Timeresolve surface pressure measurements for Reynolds between 50k and 250k at AoA between 0 and 21
They show that the lift does not increase beyond 6 degree AoA because the LSB forms near the leading edge and flattens the lift curve. Above stall, the shear layer does not reattach and significantly reduces lift},
  doi      = {10.1007/s00348-012-1270-z},
}

@Article{Toppings2025,
  author    = {Toppings, Connor E. and Yarusevych, Serhiy},
  journal   = {Journal of Fluid Mechanics},
  title     = {Transient dynamics of stall and reattachment at low {R}eynolds number},
  year      = {2025},
  pages     = {A38},
  volume    = {1011},
  doi       = {10.1017/jfm.2025.348},
  publisher = {Cambridge University Press (CUP)},
}

@TechReport{Carmichael1981,
  author      = {B. H. Carmichael},
  institution = {NASA CR-165803. Low Energy Transportation Systems},
  title       = {Low {R}eynolds number airfoil survey},
  year        = {1981},
  comment     = {Nice fundamental paper showing that the BL typically transitions downstream of the separation},
}

@Article{Toppings2024,
  author    = {Toppings, Connor E. and Yarusevych, Serhiy},
  journal   = {Journal of Fluid Mechanics},
  title     = {Laminar separation bubble formation and bursting on a finite wing},
  year      = {2024},
  pages     = {A26},
  volume    = {986},
  comment   = {Transients on an airfoil take longer than on a finite wing
They produce ensemble averages based off of 20 runs only.
They conduct slices and span-wise measurements of both the wing and the airfoil.
When the flow reattaches, it is not visible in the span-wise measurements of the re-attached flow state (high Reynolds) since the bubble becomes too thin.
When the flow reattaches, the uv Reynolds stress becomes larger},
  doi       = {10.1017/jfm.2024.321},
  publisher = {Cambridge University Press (CUP)},
}

@Article{Khedr2025,
  author    = {Khedr, Amr and Castellani, Francesco},
  journal   = {Physics of Fluids},
  title     = {Large eddy simulation of the effect of blade rotation on laminar separation bubbles in horizontal axis wind turbines},
  year      = {2025},
  pages     = {044109},
  volume    = {37},
  comment   = {Perform LES of a wind turbine blade. Compared to a wing, the bubble appears shorter and flatter which reduces drag and lift, enhancing efficiency
Very nice visualization of the Reynolds stresses compared to a wing. The typical peak in the Reynolds stress is not visible. Instead, a smeared out Reynolds stress is visible and the peak stress is much lower
The Cp is also very very different between a 2D and 3D case with the 2D flow having a more distinct pressure plateau},
  doi       = {10.1063/5.0261925},
  publisher = {AIP Publishing},
}

@PhdThesis{Himmelskamp1945,
  author  = {Himmelskamp, H.},
  school  = {Göttingen},
  title   = {Profile investigations on a rotating airscrew},
  year    = {1945},
  comment = {The first work to describe that lift is much higher for propellers than airfoils because stall is delayed},
}

@InProceedings{Snel1992,
  author    = {Snel, H. and Houwink, R. and Piers, W. J.},
  booktitle = {18th European Rotorcraft Forum},
  title     = {Sectional prediction of {3D} effects for separated flow on rotating blades},
  year      = {1992},
  comment   = {Describe that 3D effects such as radial velocities and radial pressure gradient affect the polars for rotating flows (Himmelskamp effect)
Derive a correction which can be implemented in BEMT simulations},
}

@Article{MacNeill2017,
  author    = {MacNeill, R. and Verstraete, D.},
  journal   = {The Aeronautical Journal},
  title     = {Blade element momentum theory extended to model low {R}eynolds number propeller performance},
  year      = {2017},
  issn      = {2059-6464},
  month     = may,
  number    = {1240},
  pages     = {835--857},
  volume    = {121},
  comment   = {Enhance BEMT computations with rotation correction and stall delay models to get a better agreement to experimental data
Very nice, clean BEMT derivation
Nice description of Viterna1982 stall model
Pretty much all models perform in the same way},
  doi       = {10.1017/aer.2017.32},
  publisher = {Cambridge University Press (CUP)},
}

@InProceedings{McCrink2015,
  author    = {McCrink, Matthew and Gregory, James},
  booktitle = {33rd AIAA Applied Aerodynamics Conference},
  title     = {Blade {E}lement {M}omentum {M}odeling of {L}ow-{R}e {S}mall {UAS} {E}lectric {P}ropulsion {S}ystems},
  year      = {2015},
  number    = {AIAA 2015-3296},
  comment   = {Focus on describing the scaling effect due to Reynolds number
They discuss tip-loss, Mach number correction and 3D flow correction; the latter is due to locally separated regions 
They get the aerodynamic coefficients from XFOIL and augment it with experimental data 
Thrust decreases parabolically with advance ratio, which is typical for small scale propellers},
  doi       = {10.2514/6.2015-3296},
}

@InCollection{Drela1989,
  author    = {Drela, Mark},
  booktitle = {Low Reynolds Number Aerodynamics},
  publisher = {Springer Berlin Heidelberg},
  title     = {{XFOIL}: {A}n {A}nalysis and {D}esign {S}ystem for {L}ow {R}eynolds {N}umber {A}irfoils},
  year      = {1989},
  edition   = {First},
  editor    = {Mueller, Thomas J.},
  isbn      = {978-3-642-84010-4},
  pages     = {1--12},
  series    = {Lecture Notes in Engineering 54},
  doi       = {10.1007/978-3-642-84010-4},
}

@Article{Morgado2016,
  author     = {J. Morgado and R. Vizinho and M. A. R. Silvestre and J. C. Páscoa},
  journal    = {Aerospace Science and Technology},
  title      = {{XFOIL} vs {CFD} performance predictions for high lift low {R}eynolds number airfoils},
  year       = {2016},
  issn       = {1270-9638},
  pages      = {207-214},
  volume     = {52},
  comment    = {Very detailed overview about how each CFD model works. Compare XFOIL vs CFD for airfoils at low Re. Standard k omega is not better than xfoil.

They conclude that modeling the transition is highly important for accurate solutions.},
  doi        = {10.1016/j.ast.2016.02.031},
  readstatus = {skimmed},
}

@Article{Schwermer2019,
  author    = {Schwermer, Till and Gardner, Anthony D. and Raffel, Markus},
  journal   = {Journal of the American Helicopter Society},
  title     = {A {N}ovel {E}xperiment to {U}nderstand the {D}ynamic {S}tall {P}henomenon in {R}otor {A}xial {F}light},
  year      = {2019},
  issn      = {2161-6027},
  month     = jan,
  number    = {1},
  pages     = {1--11},
  volume    = {64},
  comment   = {Describe a facility to measure unsteady pitching to analyze dynamic stall of rotors.},
  doi       = {10.4050/jahs.64.012004},
  publisher = {American Helicopter Society},
}

@InProceedings{Moore2010,
  author    = {Moore, Mark},
  booktitle = {10th AIAA Aviation Technology, Integration, and Operations (ATIO) Conference},
  title     = {{NASA} {P}uffin {E}lectric {T}ailsitter {VTOL} {C}oncept},
  year      = {2010},
  comment   = {Main takeaway: Electric propulsion is strong since the efficiency does not decrease while varying the RPM},
  doi       = {10.2514/6.2010-9345},
  number    = {AIAA 2010-9345},
}

@InProceedings{McDonald2014,
  author    = {McDonald, Robert A.},
  booktitle = {52nd Aerospace Sciences Meeting},
  title     = {Electric {P}ropulsion {M}odeling for {C}onceptual {A}ircraft {D}esign},
  year      = {2014},
  number    = {AIAA 2014-0536},
  comment   = {Review different motors and their efficiency (which is quite high over a large range)},
  doi       = {10.2514/6.2014-0536},
}

@MastersThesis{Nowicki2016,
  author  = {Nathalie Nowicki},
  school  = {KTH - Royal Institute of Technology},
  title   = {Measurement and modelling of multicopter {UAS} rotor blade deflection in hover},
  year    = {2016},
  comment = {Measure blade deflection and coning angle via two cameras, photogrammetry and dots painted on the propeller},
}

@InProceedings{Russell2017,
  author    = {Russell, Carl R. and Sekula, Martin K.},
  booktitle = {Proceedings of the AHS International Annual Forum and Technology Display},
  title     = {Comprehensive {A}nalysis {M}odeling of {S}mall-{S}cale {UAS} {R}otors},
  year      = {2017},
  comment   = {Use the measurements from Nowicki (blade deflection) to model a propeller in detail
They use CAMRAD II to account for deflection, twisting, vibration, etc.
Thrust is barely affected whether the blade is rigid or elastic (simulations), the power coefficient differs by around 10% at 8000 RPM},
  doi       = {10.4050/F-0073-2017-12196},
}

@Article{Sicard2013,
  author    = {Sicard, Jérôme and Sirohi, Jayant},
  journal   = {Measurement Science and Technology},
  title     = {Measurement of the deformation of an extremely flexible rotor blade using digital image correlation},
  year      = {2013},
  issn      = {1361-6501},
  month     = may,
  number    = {6},
  pages     = {065203},
  volume    = {24},
  comment   = {Follow up journal paper to Lawson2011
Not much more information gained, RPM is still quite low around 1000},
  doi       = {10.1088/0957-0233/24/6/065203},
  publisher = {IOP Publishing},
}

@Article{Mendez2017,
  author     = {Mendez, Miguel A. and Raiola, Marco and Masullo, Alessandro and Discetti, Stefano and Ianiro, Andrea and Theunissen, R. and Buchlin, Jean-Marie},
  journal    = {Experimental Thermal and Fluid Science},
  title      = {{POD}-based {B}ackground {R}emoval for {P}article {I}mage {V}elocimetry},
  year       = {2017},
  pages      = {181--192},
  volume     = {80},
  comment    = {POD background removal paper from Miguel},
  doi        = {10.1016/j.expthermflusci.2016.08.021},
  readstatus = {read},
}

@Article{Marten2013,
  author  = {D. Marten and J. Wendler and G. Pechlivanoglou and C. N. Nayeri and C. O. Paschereit},
  journal = {International Journal of Emerging Technology and Advanced Engineering},
  title   = {{QB}lade: an {O}pen {S}ource {T}ool for {D}esign and {S}imulation of {H}orizontal and {V}ertical {A}xis {W}ind {T}urbines},
  year    = {2013},
  pages   = {264--269},
  volume  = {3},
}

@Article{Ning2013,
  author    = {Ning, S. Andrew},
  journal   = {Wind Energy},
  title     = {A simple solution method for the blade element momentum equations with guaranteed convergence},
  year      = {2014},
  number    = {9},
  pages     = {1327--1345},
  volume    = {17},
  doi       = {10.1002/we.1636},
  publisher = {Wiley},
}

@InProceedings{Ratz2026b,
  author    = {Manuel Ratz and Christophe Schram and Andrea Sciacchitano and Miguel Alfonso Mendez},
  booktitle = {Vertical Flight Society’s 82nd Annual Forum \& Technology Display},
  title     = {Experimental {C}haracterization of an {A}ccelerating and {D}ecelerating {P}ropeller at {L}ow {R}eynolds {N}umbers},
  year      = {2026},
  doi       = {10.4050/F-0082-2026-0088},
}

@Article{Grava2026a,
  author    = {Grava, Alessandro and Serpieri, Jacopo and Bernardos, Luis and Cafiero, Gioacchino},
  journal   = {AIAA Journal},
  title     = {Aerodynamics of {S}mall-{S}cale {P}ropellers {U}ndergoing {D}ynamic {V}ariations of {R}otational {S}peed},
  year      = {2026},
  number    = {8},
  pages     = {4535--4548},
  volume    = {64},
  doi       = {10.2514/1.j066036},
  publisher = {American Institute of Aeronautics and Astronautics (AIAA)},
}

@InProceedings{Chance2026,
  author    = {Yannick G. A. Chance and Lourenco Tercio Lima Pereira and Riccardo Zamponi and Daniele Ragni},
  booktitle = {32nd AIAA/CEAS Aeroacoustics 2026 Conference},
  title     = {Aeroacoustics of a {S}caled e{VTOL} {R}otor {U}ndergoing {H}armonic {R}otational {V}elocity {C}hanges},
  year      = {2026},
  number    = {AIAA 2026-3515},
  comment   = {Perform sinusodial variations up to a maximum k of 0.01 (similar to us).
Thrust shows almost no hysteresis in hovering conditions.
The non-phasel-locked wake during acceleration is weaker than during acceleration. The time-average lies in the middle},
  doi       = {10.2514/6.2026-3515},
}

@Article{Wang2014,
  author  = {Wang, S. and Zhou, Y. and Alam, Md. Mahbub and Yang, H.},
  journal = {Physics of Fluids},
  title   = {Turbulent intensity and {R}eynolds number effects on an airfoil at low {R}eynolds numbers},
  year    = {2014},
  issn    = {1089-7666},
  pages   = {115107},
  volume  = {26},
  comment = {Perform measurements of an airfoil at different turbulence levels and Reynolds.
Identify the ultra-low, low, medium and high Reynolds regime (UAVs are low)
The stall characteristics change significantly among them.},
  doi     = {10.1063/1.4901969},
}

@Article{Winslow2018,
  author   = {Winslow, Justin and Otsuka, Hikaru and Govindarajan, Bharath and Chopra, Inderjit},
  journal  = {Journal of Aircraft},
  title    = {Basic {U}nderstanding of {A}irfoil {C}haracteristics at {L}ow {R}eynolds {N}umbers ({$10^4$}–{$10^5$})},
  year     = {2018},
  number   = {3},
  pages    = {1050--1061},
  volume   = {55},
  comment  = {Study low Reynolds airfoils with URANS, spalart allmaras and an intermittency tracking model.
They compare lift and drag with experimental measurements and reach a good agreement.
They want to promote their solver and show that it can well predict the transition and LSB
Very nice overview of what changes for low Reynolds airfoils.
Laminar to turbulent transition and the like happens.
They use an inhouse code TURNS2D to characterize the separation},
  doi      = {10.2514/1.C034415},
}

@Article{Mueller2003,
  author  = {Mueller, Thomas J. and DeLaurier, James D.},
  journal = {Annual Review of Fluid Mechanics},
  title   = {Aerodynamics of {S}mall {V}ehicles},
  year    = {2003},
  pages   = {89--111},
  volume  = {35},
  comment = {Review mostly focused on wings at low Reynolds.
Above Re=50.000, the boundary layer can reattach.
Between Re=70.000-200.000, the flow is mostly laminar.
Very thorough overviw on wings, airfoils, and flapping applications.},
  doi     = {10.1146/annurev.fluid.35.101101.161102},
}

@InProceedings{Moehren2022,
  author    = {Möhren, Felix and Bergmann, Ole and Braun, Carsten and Janser, Frank},
  booktitle = {AIAA SCITECH 2022 Forum},
  title     = {Prediction and {C}omparison of {D}ynamic {L}oads of {P}ropellers},
  year      = {2022},
  number    = {AIAA 2022-2267},
  comment   = {Build an elastic model for a large scale (1.6-4m) propeller.
Transient loads are significant compared to steady loads.
Swept propellers do not bend as much. as unswept ones.
Large propellers bend less then small ones (in terms of moments)
They build the elastic and aerodynamic model using section-wise discretization},
  doi       = {10.2514/6.2022-2267},
}

@Article{Paulos2018,
  author    = {Paulos, James and Yim, Mark},
  journal   = {Journal of Guidance, Control, and Dynamics},
  title     = {Cyclic {B}lade {P}itch {C}ontrol {W}ithout a {S}washplate for {S}mall {H}elicopters},
  year      = {2018},
  issn      = {1533-3884},
  month     = Mar,
  number    = {3},
  pages     = {689--700},
  volume    = {41},
  comment   = {Describe the swash-plate less rotor that acts through torque modulation
One of the first to do it, final journal publication after a few conferences.},
  doi       = {10.2514/1.g002683},
}

@Article{Bernardes2023,
  author    = {Bernardes, Evandro and Boyer, Frédéric and Viollet, Stéphane},
  journal   = {Aerospace Science and Technology},
  title     = {Modelling, control and simulation of a single rotor {UAV} with swashplateless torque modulation},
  year      = {2023},
  issn      = {1270-9638},
  pages     = {108433},
  volume    = {140},
  comment   = {Full description of the equations of motion for the variable rpm control leading to variable pitch
Swashplate-less rotor},
  doi       = {10.1016/j.ast.2023.108433},
  publisher = {Elsevier BV},
}

@Article{Flores2024,
  author    = {Flores, Jonathan and Salazar, Sergio and Gonzalez-Hernandez, Ivan and Rosales, Yukio and Lozano, Rogelio and Salazar, Eduardo and Nicolas, Benjamin},
  journal   = {Drones},
  title     = {Control of {H}elicopter {U}sing {V}irtual {S}washplate},
  year      = {2024},
  issn      = {2504-446X},
  number    = {7},
  pages     = {327},
  volume    = {8},
  comment   = {Full description of the equations of motion for the variable rpm control leading to variable pitch
Swashplate-less rotor},
  doi       = {10.3390/drones8070327},
  publisher = {MDPI AG},
}

@Article{Mulleners2012b,
  author  = {Mulleners, K. and Kindler, K. and Raffel, M.},
  journal = {Aerospace Science and Technology},
  title   = {Dynamic stall on a fully equipped helicopter model},
  year    = {2012},
  number  = {1},
  pages   = {72--76},
  volume  = {19},
  comment = {Perform phase-locked PIV measurements on a rotating helicopter blade.
Intro clearly states that 3D measurements are important because a 2D polar does not tell the full story.},
  doi     = {10.1016/j.ast.2011.03.013},
}

@Article{Pavel2022,
  author  = {Pavel, Marilena D.},
  journal = {Aerospace Science and Technology},
  title   = {Understanding the control characteristics of electric vertical take-off and landing (e{VTOL}) aircraft for urban air mobility},
  year    = {2022},
  pages   = {107143},
  volume  = {125},
  comment = {Mention that acting on pitch and rpm is the best option for controlling a quadcopter (and its angle in particular).
Very very thorough modeling of the dynamics of the quadcopter.
If we act on pitch, we overshoot the roll rate
If we act on velocity, we are very slow.
If we do both, we have a nice system.},
  doi     = {10.1016/j.ast.2021.107143},
}

@InProceedings{Mortimer2020a,
  author    = {Mortimer, Patrick and Johnson, Chloe and Sirohi, Jayant and Platzer, Stefan and Rauleder, Juergen},
  booktitle = {AIAA Aviation 2020 Forum},
  title     = {Experimental and {N}umerical {I}nvestigation of a {V}ariable-{S}peed {R}otor for {T}hrust {C}ontrol},
  year      = {2020},
  number    = {AIAA 2020-2791},
  comment   = {Perform a ramp up of a low speed (900RPM) large (1m radius) propeller and compare it with CFD simulations
Use time-resolved PIV in the wake due to the large rotor speed
Reynolds is 500.000, so super large
CFD data comes from DLR TAU URANS simulations
Their ramp also slightly fluctuates and is rounded off
Torque overshoots, thrust doesn't. They calculate the inertial torque contribution based on the RPM.
They had some issues with their HPC cluster, so not all CFD data is presented.
Their torque overshoot is bananas. For low blade loadings, the torque overshoots by 400\%},
  doi       = {10.2514/6.2020-2791},
}

@Article{Zhang2024,
  author    = {Zhang, Kong and Pakrashi, Vikram and Murphy, Jimmy and Hao, Guangbo},
  journal   = {Sensors},
  title     = {Inspection of {F}loating {O}ffshore {W}ind {T}urbines {U}sing {M}ulti-{R}otor {U}nmanned {A}erial {V}ehicles: Literature {R}eview and {T}rends},
  year      = {2024},
  issn      = {1424-8220},
  number    = {3},
  pages     = {911},
  volume    = {24},
  comment   = {Motivate drones for remote wind turbine inspections},
  doi       = {10.3390/s24030911},
  publisher = {MDPI AG},
}

@Article{Park2023,
  author    = {Park, SunHoo and Lee, Sihun and Im, Byeonguk and Lee, Dongyeol and Shin, SangJoon},
  journal   = {Aerospace Science and Technology},
  title     = {Improvement of a multi-rotor {UAV} flight response simulation influenced by gust},
  year      = {2023},
  issn      = {1270-9638},
  month     = mar,
  pages     = {108156},
  volume    = {134},
  comment   = {Estimate gusts based on the drone trajectory},
  doi       = {10.1016/j.ast.2023.108156},
  publisher = {Elsevier BV},
}

@Article{Sheng2016,
  author         = {Sheng, Shouzhao and Sun, Chenwu},
  journal        = {Energies},
  title          = {Control and {O}ptimization of a {V}ariable-{P}itch {Q}uadrotor with {M}inimum {P}ower {C}onsumption},
  year           = {2016},
  number         = {4},
  pages          = {232},
  volume         = {9},
  article-number = {232},
  comment        = {perform contorl of a variable pitch quadcopter/drone

they have simple models for the torques for the three-axis motions and use similar motor equations to Porter2016.

They then sample the equation in steady state at multiple points and compute the coefficients via a regression. They have five params, so they need at least 5 data points for this

efficiency for the vpp is better than for the fpp},
  doi            = {10.3390/en9040232},
}

@Article{Peters1995,
  author    = {Peters, David A. and He, Cheng Jian},
  journal   = {Journal of Aircraft},
  title     = {Finite state induced flow models. {II} - {T}hree-dimensional rotor disk},
  year      = {1995},
  number    = {2},
  pages     = {323--333},
  volume    = {32},
  comment   = {Build an inflow model based on potential theory.},
  doi       = {10.2514/3.46719},
}

@Article{Pitt1981,
  author  = {Pitt, D. M. and Peters, D. A.},
  journal = {Vertica},
  title   = {Theoretical {P}rediction of {D}ynamic-{I}nflow {D}erivatives},
  year    = {1981},
  number  = {1},
  pages   = {21--34},
  volume  = {5},
}

@Article{Peters1991,
  author    = {Peters, David A. and He, Cheng Jian},
  journal   = {Journal of the American Helicopter Society},
  title     = {Correlation of {M}easured {I}nduced {V}elocities with a {F}inite‐{S}tate {W}ake {M}odel},
  year      = {1991},
  number    = {3},
  pages     = {59--70},
  volume    = {36},
  doi       = {10.4050/jahs.36.59},
  publisher = {American Helicopter Society},
}

@Book{Bramwell2001,
  editor    = {Anthony R. S. Bramwell and George Done and David Balmford},
  publisher = {American Institute of Aeronautics and Astronautics},
  title     = {Bramwell's{ H}elicopter {D}ynamics},
  year      = {2001},
  address   = {Reston, VA},
  edition   = {Second},
  isbn      = {0080528309},
}

@Article{Graftieaux2001,
  author    = {Graftieaux, Laurent and Michard, Marc and Grosjean, Nathalie},
  journal   = {Measurement Science and Technology},
  title     = {Combining {PIV}, {POD} and vortex identification algorithms for the study of unsteady turbulent swirling flows},
  year      = {2001},
  issn      = {1361-6501},
  month     = Aug,
  number    = {9},
  pages     = {1422--1429},
  volume    = {12},
  comment   = {Gamma 2 criterion paper},
  doi       = {10.1088/0957-0233/12/9/307},
  publisher = {IOP Publishing},
}

@Article{DeGregorio2020,
  author  = {De Gregorio, Fabrizio and Visingardi, Antonio},
  journal = {Experiments in Fluids},
  title   = {Vortex detection criteria assessment for {PIV} data in rotorcraft applications},
  year    = {2020},
  number  = {179},
  volume  = {61},
  comment = {Propose to use the Gamma2 criterion to identify vortices in rotor applications where noise can be large and gradients suffer. Their method does not require gradients and just uses a summation with a normal projection and the discrete data.},
  doi     = {10.1007/s00348-020-03012-7},
}

@Book{Sutton2009,
  author    = {Sutton, Michael A. and Orteu, Jean-José and Schreier, Hubert},
  publisher = {Springer New York, NY},
  title     = {Image {C}orrelation for {S}hape, {M}otion and {D}eformation {M}easurements: {B}asic {C}oncepts, {T}heory and {A}pplications},
  year      = {2009},
  edition   = {First},
  isbn      = {9780387787473},
  doi       = {10.1007/978-0-387-78747-3},
}

@InProceedings{Yu2026,
  author    = {Daniel Yu and Jayant Sirohi},
  booktitle = {Vertical Flight Society’s 82nd Annual Forum \& Technology Display},
  title     = {Forward {F}light {D}ynamic {I}nflow {M}odel {I}dentification from {E}xperimental {W}ake {M}easurements},
  year      = {2026},
  comment   = {Followup to Mortimer2024, where they now perform identification in forward flight.
PIV system can be rotated around the rotor to allow phase-locking at different azimuths.
The inflow change as a function of time looks very clean.
Uncertainties in the fitted model are high because first order might not be good enough.},
  doi       = {10.4050/F-0082-2026-0061},
}

@Article{Toppings2021,
  author  = {Toppings, Connor E. and Yarusevych, Serhiy},
  journal = {Journal of Fluid Mechanics},
  title   = {Structure and dynamics of a laminar separation bubble near a wingtip},
  year    = {2021},
  pages   = {A39},
  volume  = {929},
  doi     = {10.1017/jfm.2021.881},
}

@Article{McCroskey1976,
  author    = {McCroskey, W. J. and Carr, L. W. and McAlister, K. W.},
  journal   = {AIAA Journal},
  title     = {Dynamic {S}tall {E}xperiments on {O}scillating {A}irfoils},
  year      = {1976},
  number    = {1},
  pages     = {57--63},
  volume    = {14},
  comment   = {Seminal work on dynamic stall},
  doi       = {10.2514/3.61332},
}

@Article{Mulleners2011,
  author    = {Mulleners, Karen and Raffel, Markus},
  journal   = {Experiments in Fluids},
  title     = {The onset of dynamic stall revisited},
  year      = {2012},
  pages     = {779--793},
  volume    = {52},
  comment   = {PIV analysis of dynamic stall for a helicopter blade},
  doi       = {10.1007/s00348-011-1118-y},
  publisher = {Springer Science and Business Media LLC},
}

@Article{Zhang2020a,
  author  = {K. Zhang and D. E. Rival},
  journal = {Experiments in Fluids},
  title   = {Direct {L}agrangian method to characterize entrainment dynamics using particle residence time: a case study on a laminar separation bubble},
  year    = {2020},
  number  = {243},
  volume  = {61},
  doi     = {10.1007/s00348-020-03081-8},
}

@InProceedings{Valentin2024,
  author    = {Valentin, Johan and Bernardos, Luis F. and Pinon, Gregory and Rivoalen, Elie},
  booktitle = {AIAA Aviation Forum and Ascend 2024},
  title     = {Hybrid {E}ulerian-{L}agrangian {M}ethod for {C}omplex {3D} {V}iscous {F}lows},
  year      = {2024},
  number    = {AIAA 2024-3865},
  comment   = {VPM citation with lifting line theory},
  doi       = {10.2514/6.2024-3865},
}

@Article{OMeara1987,
  author  = {O’Meara, M. M. and Mueller, T. J.},
  journal = {AIAA Journal},
  title   = {Laminar separation bubble characteristics on an airfoil at low {R}eynolds numbers},
  year    = {1987},
  number  = {8},
  pages   = {1033--1041},
  volume  = {25},
  comment = {As Reynolds number increases, LSB length/thickness decrease. Transition happens earlier and the separation angle decreases},
  doi     = {10.2514/3.9739},
}

@Article{Jaroslawski2022,
  author  = {Jaroslawski, Thomas and Forte, Maxime and Moschetta, Jean-Marc and Delattre, Gregory and Gowree, Erwin R.},
  journal = {Experimental Thermal and Fluid Science},
  title   = {Characterisation of boundary layer transition over a low {R}eynolds number rotor},
  year    = {2022},
  pages   = {110485},
  volume  = {130},
  comment = {Phase-locked IR thermography plus thrust/torque on a three-bladed untwisted NACA0012 rotor in hover (R = 0.125 m, c = 0.025 m, AR = 5, α = 10°, Re_tip = 6.2×10³--1.44×10⁵), smooth and with 2D/3D roughness trips at 0.05c and 0.28c. On the smooth blade the elevated-I_n region is identified as an LSB, and the extent of the laminar separation zone shrinks with increasing RPM --- the rotor equivalent of the 2D Re trend. Two rotor-specific observations matter here: (i) the LSB is strongly three-dimensional, with spanwise (radial) non-uniformity that becomes more pronounced with rotational speed, unlike the spanwise-uniform bubble on a 2D wing; (ii) radial tilting of roughness wakes indicates spanwise flow driven by rotational forces and radial pressure gradients. The relevant scaling is the Rossby number, which for a rotor in hover reduces to Ro = AR = r/c (Lentink \& Dickinson) and is therefore set by blade geometry rather than Ω. At AR ≈ 5 rotational and convective accelerations are the same order; Weiss et al. found no rotational effect on transition or chordwise laminar extent at local AR = 4.76, while Hernandez found a stabilizing effect only as Ro → 1. The authors conclude that sectional BEMT-type analysis may be insufficient for transition on low-Re rotors.},
  doi     = {10.1016/j.expthermflusci.2021.110485},
}

@Article{Fava2024a,
  author  = {Fava, T. C. L. and Henningson, D. S. and Hanifi, A.},
  journal = {Physics of Fluids},
  title   = {Boundary layer stability on a rotating wind turbine blade section},
  year    = {2024},
  pages   = {094128},
  volume  = {36},
  comment = {Blade section at r/R = 0.68, AoA = 12.8/4.2/1.2 deg, rotating vs non-rotating. At 4.2 deg the LSB is long and open (x = 0.34-0.99), h_max/c = 2.7e-2, h_max/delta\* = 0.56, splits into three cells with flapping. Rotation thickens the bubble 74.7% at 12.8 deg, raises reverse flow from 7% to 16% U_inf, delays transition and reattachment by 3-4% chord via Coriolis counteracting the APG, then accelerates breakdown through an oblique mechanism. No unstable cusp at 12.8 deg: convectively unstable, so noise-sensitive. Tip flow only 5-6% U_inf, too weak to trigger transition.
Complement to Fava2024b},
  doi     = {10.1063/5.0223207},
}

@Article{Fava2024b,
  author    = {Fava, T. C. L. and Massaro, D. and Schlatter, P. and Henningson, D. S. and Hanifi, A.},
  journal   = {Journal of Fluid Mechanics},
  title     = {Transition to turbulence on a rotating wind turbine blade at {$Re_c=3 \times 10^5$}},
  year      = {2024},
  pages    = {A54},
  volume    = {999},
  comment   = {Same geometry, Ro_c swept 0.052-0.314. At matched AoA = 4.2 deg the bubble is 3.6x thinner than at 1e5 (h_max/c = 7.6e-3) and closes at x = 0.52 instead of running to the trailing edge, while x_tr shifts only 0.50 to 0.467 --- Re controls closure, not onset. h_max/delta\* rises to 0.82, so the bubble does not scale with the BL. Thin bubbles (h_max/delta\* = 0.27) switch from KH roll-up to a wall-influenced TS/KH blend with subharmonic lambda vortices. Rotation effect nearly vanishes at this Re (4% thickness change). Hover maps to Ro_r = 1, Ro_c = c/r, i.e. cases 4C/4D, where crossflow reaches 28-56% U_inf and moves transition to the LE. Open: nothing below 1e5, no imposed transients, spanwise-periodic slice, no load/C_T link.},
  doi       = {10.1017/jfm.2024.913},
  publisher = {Cambridge University Press (CUP)},
}

@Article{Isaacs1945,
  author    = {Isaacs, Rufus},
  journal   = {Journal of the Aeronautical Sciences},
  title     = {Airfoil {T}heory for {F}lows of {V}ariable {V}elocity},
  year      = {1945},
  number    = {1},
  pages     = {113--117},
  volume    = {12},
  doi       = {10.2514/8.11202},
}

@InProceedings{vanderWall1992,
  author    = {van der Wall, Berend G. and Leishman, J. Gordon},
  booktitle = {18th European Rotorcraft Forum},
  title     = {The {I}nfluence of {V}ariable {F}low {V}elocity on {U}nsteady {A}irfoil {B}ehavior},
  year      = {1992},
  comment   = {Summary of Isaacs1945 and related theories.},
}

@Article{Strangfeld2016,
  author  = {Strangfeld, C. and Müller-Vahl, H. and Nayeri, C. N. and Paschereit, C. O. and Greenblatt, D.},
  journal = {Journal of Fluid Mechanics},
  title   = {Airfoil in a high amplitude oscillating stream},
  year    = {2016},
  pages   = {79--108},
  volume  = {793},
  comment = {Followup of Isaacs1945},
  doi     = {10.1017/jfm.2016.126},
}

@Article{Greenblatt2023,
  author    = {Greenblatt, David and Müller-Vahl, Hanns and Strangfeld, Christoph},
  journal   = {Physical Review Fluids},
  title     = {Laminar separation bubble bursting in a surging stream},
  year      = {2023},
  pages     = {L012102},
  volume    = {8},
  comment   = {Not the journal paper but the preprint. Followup of Isaacs1945},
  doi       = {10.1103/physrevfluids.8.l012102},
  publisher = {American Physical Society (APS)},
}

@Article{Yu2024b,
  author    = {Yu, Meilin and Hrynuk, John T. and Booth, David T. and Poudel, Naresh},
  journal   = {Aerospace Science and Technology},
  title     = {Freestream turbulence effects on low {R}eynolds number {NACA} 0012 airfoil laminar separation bubble and lift generation},
  year      = {2024},
  pages     = {109145},
  volume    = {149},
  doi       = {10.1016/j.ast.2024.109145},
  publisher = {Elsevier BV},
}

@InProceedings{VanIngen2008,
  author    = {van Ingen, J.},
  booktitle = {38th Fluid Dynamics Conference and Exhibit},
  title     = {The e{N} {M}ethod for {T}ransition {P}rediction. {H}istorical {R}eview of {W}ork at {TU} {D}elft},
  year      = {2008},
  number    = {AIAA 2008-3830},
  doi       = {10.2514/6.2008-3830},
}

@Article{Oudheusden2013,
  author  = {van Oudheusden, B. W.},
  journal = {Measurement Science and Technology},
  title   = {{PIV}-based pressure measurement},
  year    = {2013},
  pages   = {032001},
  volume  = {24},
  comment = {Very nice overview and introduction to pressure reconstruction in 2013. They also mention that the boundary conditions can be time dependent when solving the pressure poisson equation.},
  doi     = {10.1088/0957-0233/24/3/032001},
}

@Misc{Ratz2026e,
  author    = {Ratz, Manuel and Schram, Christophe and Sciacchitano, Andrea and Mendez, Miguel Alfonso},
  title     = {Dataset of: "{O}n the aerodynamics of an accelerating and decelerating propeller in hover at low {R}eynolds numbers"},
  year      = {2026},
  copyright = {Creative Commons Attribution 4.0 International},
  doi       = {10.5281/zenodo.23083730},
  language  = {en},
  howpublished = {Zenodo},
  note         = {(Version 1.0.0) [dataset]},
}

    \appendix

    \section{Repeatability and uncertainty}
	\label{app:4}
	
	\begin{figure*}[t]
		\centering
		\includegraphics[width=\linewidth]{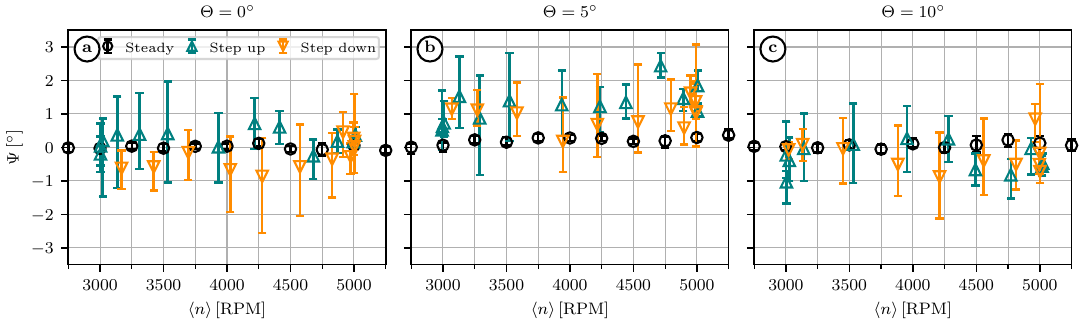}
		\caption{Offset in the phase angle $\Psi$ with respect to the reference image for each pitch offset at $r/R=0.6$. Shown are both the mean (markers) and the range (error bars). For the steady measurements, the variations are of the order of the marker size. The legend in subfigure (a) applies to all subfigures.}
		\label{fig:rep_phase_angle}
	\end{figure*}
	
	\begin{figure}[t]
		\centering
		\includegraphics[width=\linewidth]{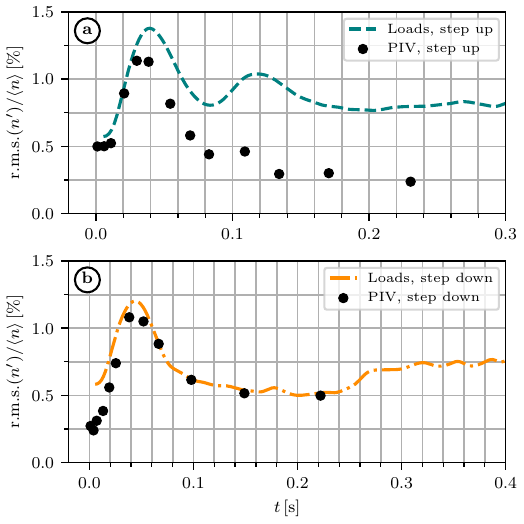}
		\caption{Normalized r.m.s. of the fluctuation $n^\prime = n - \langle n \rangle$ over the mean rotational speed for the step up (a) and step down (b) for $\Theta = \SI{10}{\degree}$ (loads) at $r/R=0.6$ (PIV). Uncertainty bounds for the PIV timing are not depicted since they are not visible at this scale.}
		\label{fig:rep_rpms_rpm}
	\end{figure}
    
    Each test case extended over several hours, so steady and transient data were acquired at different times of the day. This section quantifies the repeatability of the operating conditions and the main uncertainties affecting the load and PIV measurements.
	
	\subsection{Repeatability}
	\label{app:4_1}
	
	The laboratory room temperature ranged from 16.5 to \SI{21.5}{\celsius} over the entire campaign. For each pitch offset, steady and transient loads were collected within \SI{30}{\minute} with temperature changes of less than \SI{1}{\kelvin}, leading to negligible density changes. The PIV acquisitions required more time, but the Reynolds number variation remained below \SI{2}{\percent}, which is an order of magnitude below the \SI{67}{\percent} change imposed by the step responses.

    Thermal drift of the load cell was mitigated through compressed air (Sec.~\ref{sec:2_1}) and by subtracting the thrust offset from each acquisition. In transient conditions, five batches of 100 step cycles were acquired, and the temperature was allowed to settle after each batch. With these measures, the load cell temperature changed by \SIrange{0.9}{2.1}{\kelvin} across each case.

    The speed fluctuations reached a relative r.m.s. between 0.6 and \SI{0.7}{\percent} of the mean across both loads and PIV measurements in steady conditions. In transient conditions, they peaked at about \SI{1.4}{\percent} in acceleration and \SI{1.2}{\percent} in deceleration, as is illustrated in Fig.~ \ref{fig:rep_rpms_rpm}, which shows the r.m.s. values of both PIV and load measurements during the step responses for $\Theta = \SI{10}{\degree}$. The timing jitter of the PIV reached maximum standard deviations of \SI{0.7}{\milli\second} (\SI{1.2}{\milli\second}) for the step up (step down), corresponding to \SI{1}{\percent} of the rise/fall times. Similar fluctuation values across all pitch offsets and spans link these variations to the ESC response and not the blade aerodynamics.
    
	The jitter of the rotational speed degrades the phase-locking, as it is based on the mean response (Sec.~\ref{sec:2_4}). The jitter is quantified in Fig.~\ref{fig:rep_phase_angle} for $r/R=0.6$ based on the horizontal position of the trailing edge reflection (Sec.~ \ref{sec:3_2_1}). The mean bias is smaller than \SI{0.5}{\degree} with spread smaller than \SI{0.3}{\degree} in steady conditions. Both bias and spread increase in transient conditions, with a maximum bias of \SI{2.5}{\degree} and a maximum offset of \SI{3}{\degree}; the other span locations show similar trends. Pre-processing corrects the jitter, and three-dimensional effects are neglected because the angles are small across all cases.

    The jitter likewise degrades the slipstream measurements, but cannot be quantified as the images do not contain the blade contour reflection. Because the phase-locking principle remains the same, the spread is expected to be similar to the spanwise measurements, while the bias increases. Similar measurements (e.g., \cite{Mortimer2025}) suggest limited impact on inflow, downwash, and tip vortices since all three were extracted far from the blade passage.
	
	\subsection{Measurement uncertainty}
	\label{app:4_2}

    \noindent\textbf{Rotational speed.} Rotor vibrations due to imbalance occasionally caused the encoder to miss pulses. Based on preliminary tests with and without a propeller, this behavior is intermittent, with less than \SI{0.5}{\percent} of data being affected by it. The ensemble averaging is expected to reduce this effect.

    \noindent\textbf{Thrust coefficient.} The thrust analysis rests on the relative difference $\Delta \langle C_T \rangle$ \eqref{eq:rel_ct}. Since data were taken within \SI{30}{\minute}, temperature effects on the load cell sensitivity and density were neglected. The uncertainty quantification therefore reduces to random errors of the steady and transient averages and the sensitivity to the low-pass filter.

    The random error in steady conditions was estimated from non-overlapping segments of ten revolutions, yielding 0.02 to \SI{0.07}{\percent} relative to the mean. Inverse variance weighting of the second-order fit resulted in a standard error of 0.01 to \SI{0.03}{\percent}. In transient conditions, the standard error of the ensemble average ranged from 0.031 to \SI{0.07}{\percent} across all cases. The uncertainty of the low-pass filter was estimated as the difference between the chosen \SI{30}{\hertz} cut-off and frequencies between 20 and \SI{70}{\hertz}, giving standard errors between 0.01 and \SI{0.79}{\percent}. Combining all contributions yields an uncertainty between 0.07 and \SI{1.58}{\percent} on $\Delta \langle C_T \rangle$ with a \SI{95}{\percent} confidence level.
    
    \noindent\textbf{PIV.} The PIV uncertainty is dominated by velocity fluctuations \cite{Sciacchitano2016} even when considering the constant pulse separation in transient conditions (Table~\ref{tab:piv_params}), as the typical cross-correlation uncertainty is less than \SI{0.1}{\pixel} \cite{Raffel2018}. For the slipstream measurements, the maximum uncertainty relative to the mean is \SIrange{0.9}{1.3}{\percent} near the rotor disk mid-span, \SIrange{0.5}{1.9}{\percent} at the mid-span downstream, and \SIrange{7.2}{17.2}{\percent} in the shear layer, across all cases, with a \SI{95}{\percent} confidence level. For the spanwise data, uncertainties range between 1.1 and \SI{5.3}{\percent} in the shear layer and 6.6 to \SI{12.1}{\percent} in the immediate blade wake, where velocity fluctuations are larger. 

    The normalized uncertainties of the r.m.s. of magnitude fluctuations are bounded by the number of samples and range from 6.2 to \SI{10.5}{\percent}. Moreover, the r.m.s. values are affected by the instantaneous velocity uncertainty \cite{Sciacchitano2016}, and a correction was applied to all r.m.s. values, assuming \SI{0.1}{\pixel} uncertainty in the cross-correlation. Furthermore, the cross-correlation uncertainty and rotational speed fluctuations were used to estimate a noise floor of \SI{1.24}{\percent} of the local blade velocity; fluctuations below were excluded from the analysis.

\end{document}